\documentclass[12pt]{article}
\usepackage{amsfonts}
\usepackage[utf8]{inputenc}
\usepackage[margin=1in]{geometry}
\usepackage{setspace}
\usepackage{amsmath}
\usepackage{amssymb}
\usepackage{bbm}
\usepackage{graphicx}
\usepackage{xcolor}
\usepackage{enumerate}
\usepackage[inline, shortlabels]{enumitem}
\usepackage[multiple]{footmisc}
\usepackage[authoryear]{natbib}
\usepackage[colorlinks,linkcolor=blue,citecolor=blue]{hyperref}
\usepackage{algorithm,algorithmic}
\usepackage{titlesec}
\newcommand{\norm}[1]{\left\lVert#1\right\rVert}
\newcommand{\wh}{\widehat}
\newcommand{\wt}{\widetilde}
\newcommand{\E}{\mathbb{E}}
\newcommand{\R}{\mathbb{R}}
\newcommand{\Op}{O_p}
\newcommand{\op}{o_p}
\newcommand{\mw}{\wedge}
\DeclareMathOperator{\tr}{tr}
\newcommand{\fro}[1]{\left\langle #1\right\rangle_F}

\newtheorem{theorem}{Theorem}
\newtheorem{proposition}{Proposition}
\newtheorem{remark}{Remark}
\newtheorem{lemma}{Lemma}
\newtheorem{corollary}{Corollary}

\newtheorem{example}{Example}
\newtheorem{assumption}{Assumption}
\newenvironment{proof}[1][Proof]{\noindent \textbf{#1.} }{\hfill \rule{0.5em}{0.5em}}

\title{Linear Regression with 
Common Unobserved Heterogeneity}
\author{Kaicheng Chen\thanks{
                   School of Economics,
                   Shanghai University of Finance and Economics, Shanghai, China.
                   }
\and 
Hugo Freeman\thanks{
                   Department of Economics,
                   Michigan State University,
                   East Lansing,
                   USA.
                   }
}\date{\today}

\begin{document}
\maketitle

\begin{abstract}

We study inference on common slope parameters in linear panel models with two-way
unobserved heterogeneity. In linear regression, bias arises from unobserved
heterogeneity that is shared between the residual and the covariates. Our common unobserved
heterogeneity (CUH) estimator targets this common variation directly: 
it searches over functionals of the residualised outcome for variation that also 
resides in the covariates, and seeks to further project out only bias-inducing variation.
This makes the estimator less sensitive to weak
outcome factors while avoiding the complete removal of exogenous variation in covariates 
that is unrelated to the confounding heterogeneity. We develop a
weak-factor theory based on a singular-value strength index and establish
$\sqrt{NT}$-asymptotic normality of the cross-fitted CUH estimator. Under
homoskedastic errors, its asymptotic variance is no larger than that of a
common correlated effects factor estimator. Simulations and empirical
applications illustrate the resulting bias--variance trade-off and the gains from selectively removing common variation.
\end{abstract}

\section{Introduction}\label{sec:intro}

This paper studies estimation of common slope parameters in linear panel
regressions with two-way unobserved heterogeneity. We observe $(y_{it},x_{it})$
for $i=1,\dots,N$ and $t=1,\dots,T$, with $x_{it}\in\R^{d}$ and $d$ fixed,
generated as
\begin{equation}\label{eq:y}
  y_{it}=x_{it}'\beta_0+g(\alpha_i,\gamma_t)+\varepsilon_{it},
\end{equation}
where $\alpha_i\in\Omega_\alpha\subset\R^{d_\alpha}$ and
$\gamma_t\in\Omega_\gamma\subset\R^{d_\gamma}$ are unobserved individual and
time effects, $g$ is unknown and $\varepsilon_{it}$ is an idiosyncratic shock.
The covariates are exogenous with respect to $\varepsilon_{it}$ but correlated
with $(\alpha_i,\gamma_t)$. They obey a \emph{two-block} structure that
separates heterogeneity common to both equations from heterogeneity present
only in $x_{it}$,
\begin{equation}\label{eq:x}
  x_{it}=\underbrace{g_x(\alpha_i,\gamma_t)}_{\text{common (endogenous)}}
        +\underbrace{\wt g_x(\wt\alpha_i,\wt\gamma_t)}_{\text{$x$-only (exogenous)}}
        +\eta_{it},
  \qquad
  \E[\eta_{it}\mid\alpha_i,\gamma_t,\wt\alpha_i,\wt\gamma_t]=0,
\end{equation}
where $(\wt\alpha_i,\wt\gamma_t)$ are latent effects absent from \eqref{eq:y}.
Equation~\eqref{eq:x} is a construction, not an assumption. Set
$g_x(\alpha_i,\gamma_t):=\E[x_{it}\mid \alpha_i,\gamma_t]$,
$\wt g_x(\wt\alpha_i,\wt\gamma_t):=\E[x_{it}\mid\alpha_i,\gamma_t,\wt\alpha_i,\wt\gamma_t]-g_x(\alpha_i,\gamma_t)$
and $\eta_{it}=x_{it} - \E[x_{it}\mid\alpha_i,\gamma_t,\wt\alpha_i,\wt\gamma_t]$.
Then $\E[\wt g_x(\wt\alpha_i,\wt\gamma_t)\mid \alpha_i,\gamma_t]=0$, and this
normalisation assigns to $g_x$ the \emph{entire} dependence of $x_{it}$ on
$(\alpha_i,\gamma_t)$. The correlation of $g_x$ with $g(\alpha_i,\gamma_t)$ is
the source of endogeneity, while $\wt g_x(\wt\alpha_i,\wt\gamma_t)$ and $\eta_{it}$ are
exogenous variation that identifies $\beta_0$ and should be preserved. Our estimator is therefore designed to remove from $x_{it}$ only the block $g_x$ \emph{common} to $y_{it}$ and $x_{it}$, together with the heterogeneity in the residualised
outcome.

Three leading approaches to interactive unobserved heterogeneity frame our
contribution.

\begin{example}[Factor model]\label{ex:factor}
The factor model identifies $g(\alpha_i,\gamma_t)$ from
$y_{it}-x_{it}'\beta_0=g(\alpha_i,\gamma_t)+\varepsilon_{it}$.
Assume $g(\alpha_i,\gamma_t)=\varphi(\alpha_i)'\psi(\gamma_t)$ for
$\varphi(\alpha_i),\psi(\gamma_t)\in\R^{R_0}$ with $R_0<\infty$ and the
strong-factor conditions
\begin{align}\label{eqn:strongfactor}
  \frac1N\varphi(\alpha)'\varphi(\alpha)>0,
  &&
  \frac1T\psi(\gamma)'\psi(\gamma)>0 .
\end{align}
Then, under regularity conditions, factor-model estimates of $\beta$ are
$\sqrt{NT}$-consistent and asymptotically normal. When
\eqref{eqn:strongfactor} fails they are in general slower than $\sqrt{NT}$.
The approach residualises only through $y_{it}-x_{it}'\beta$, so the $x$-only
block $\wt g_x(\wt\alpha_i,\wt\gamma_t)$ of \eqref{eq:x} is never projected
out, however strong or low-rank, and keeps identifying $\beta_0$.
\end{example}

We call the next model the common correlated effects factor (CCEF) model to
distinguish it from the CCE model of \citet{Pesaran2006}, whose conditions are
tied to projection on cross-sectional means. CCEF lets the interactive effects
enter $x_{it}$ by assumption and admits generic factor-model estimators. {Given a preliminary
$\wt\beta$, estimate factor spaces jointly from $Y-X\wt\beta$ and the
covariates $X_j$. The CCEF estimator is then OLS after residualising both $Y$
and $X_j$ on these spaces.}

\begin{example}[Common correlated effects factor model (CCEF)]\label{ex:cce}
The CCEF model takes \eqref{eq:y}--\eqref{eq:x} and identifies functionals of
$(\alpha_i,\gamma_t)$ from functionals of $y_{it}$ and $x_{it}$ jointly. Assume
$g(\alpha_i,\gamma_t)=\varphi(\alpha_i)'\psi(\gamma_t)$ and
$g_x(\alpha_i,\gamma_t)=\varphi_x(\alpha_i)'\psi_x(\gamma_t)$ with
$\varphi,\psi,\varphi_x,\psi_x\in\R^{R_0}$, $R_0<\infty$. If either
\eqref{eqn:strongfactor} or
\begin{align}\label{eqn:CCEXstrongfactor}
  \frac1N\varphi_x(\alpha)'\varphi_x(\alpha)>0,
  &&
  \frac1T\psi_x(\gamma)'\psi_x(\gamma)>0,
  &&
  \varphi(\alpha)\in{\rm span}(\varphi_x(\alpha)),\
  \psi(\gamma)\in{\rm span}(\psi_x(\gamma)),
\end{align}
holds, CCEF estimates of $\beta$ are $\sqrt{NT}$-consistent and asymptotically
normal. Requiring only one of the two is weaker than
\eqref{eqn:strongfactor} alone. The price is higher variance, and in some cases
complete loss of identification. A functional
of $x_{it}$ cannot distinguish $g_x$ from $\wt g_x$, so
with $\wt g_x(\wt\alpha_i,\wt\gamma_t)=\wt\varphi_x(\wt\alpha_i)'
\wt\psi_x(\wt\gamma_t)$ the projected span is that of the \emph{combined}
loadings $(\varphi_x,\wt\varphi_x)$ and $(\psi_x,\wt\psi_x)$, and the exogenous
$x$-only variation is projected out with the common block.
\end{example}

\begin{example}[Bias-aware confidence intervals]\label{ex:ba}
The bias-aware approach allows preliminary estimates of $g(\alpha_i,\gamma_t)$
to be biased and widens the confidence interval for $\beta$ accordingly. For an
estimate $\wh\beta^{(\rm ba)}$ and an upper bound $\bar B$ on its absolute bias,
the interval is
\begin{align*}
  \big\{\wh\beta^{(\rm ba)}\pm[\bar B+z_{1-\alpha/2}\cdot se(\wh\beta^{(\rm ba)})]\big\},
\end{align*}
where $z_{1-\alpha/2}$ is the $1-\alpha/2$ Gaussian quantile. Any $\bar B>0$
costs power. The bound must cover the worst-case alignment of the unremoved
heterogeneity with the regressors. The method cannot tell the endogenous $g_x$
from the exogenous $\wt g_x$ in the common structure of $x_{it}$, so a strong
$x$-only block inflates the worst case and hence the width, although it
carries no bias.
\end{example}

Our common unobserved heterogeneity (CUH) estimator takes three steps. Start from a preliminary estimate
$\wt\beta$, for example the interactive-effects estimator of
\citet{Bai2009}, and the residuals $y_{it}-x_{it}'\wt\beta$.
\begin{enumerate}[label=\arabic*., leftmargin=*]
  \item Compute the leading $R$ principal components of the residuals, unit
  loadings and time factors, as in a factor model.
  \item Regress each regressor on every principal component of the residuals,
  unit by unit on the time factors and period by period on the unit loadings.
  On each side keep the $L$ coefficient vectors that explain the most
  variation in that regressor.
  \item Project the leading components of steps 1 and 2 from both $y_{it}$ and
  $x_{it}$. With $\wt y_{it}$ and $\wt x_{it}$ the projected data, the estimator is,
  \[
    \wh\beta=\Big(\sum_{i,t}\wt x_{it}\wt x_{it}'\Big)^{-1}
    \sum_{i,t}\wt x_{it}\wt y_{it}.
  \]
\end{enumerate}
Step 2 sets CUH apart. Its covariate projections are built from outcome
directions, so the part of $x_{it}$ that the outcome does not share survives
step 3 and identifies $\beta_0$. Without step 2 this is the factor model of
Example~\ref{ex:factor}. CCEF replaces step 2 by principal components of the
regressors, which removes exogenous variation as well (Proposition~\ref{prop:eff}).
Section~\ref{sec:cuh-estimator} states the steps formally, and
Algorithm~\ref{alg:cf} gives the cross-fitted version used for inference.

The factor model is efficient but can be asymptotically biased under weak
factors. The CCEF approach also uses common structure in the covariates
to recover the latent factor spaces, at a cost in variance, and bias-aware
intervals keep coverage at a cost in power. CUH combines features of the first
two approaches, exploiting directions that are weak in the outcome but strong in
the covariates. Under homoskedastic errors its asymptotic variance is no larger
than that of CCEF, and the weak-factor theory quantifies how weak the common
factors may be while retaining $\sqrt{NT}$ inference without widening the interval.

Two smoothing variants extend the projection estimator
(Section~\ref{sec:smoothing}). NPCUH replaces projection by an additive kernel
smoother over the estimated proxies, and BCUH applies the smoother after the
CUH projection. Appendix~\ref{app:npcuh} gives conditions under which NPCUH is
asymptotically normal.

\begin{figure}[!ht]
\centering
\includegraphics[width=\textwidth]{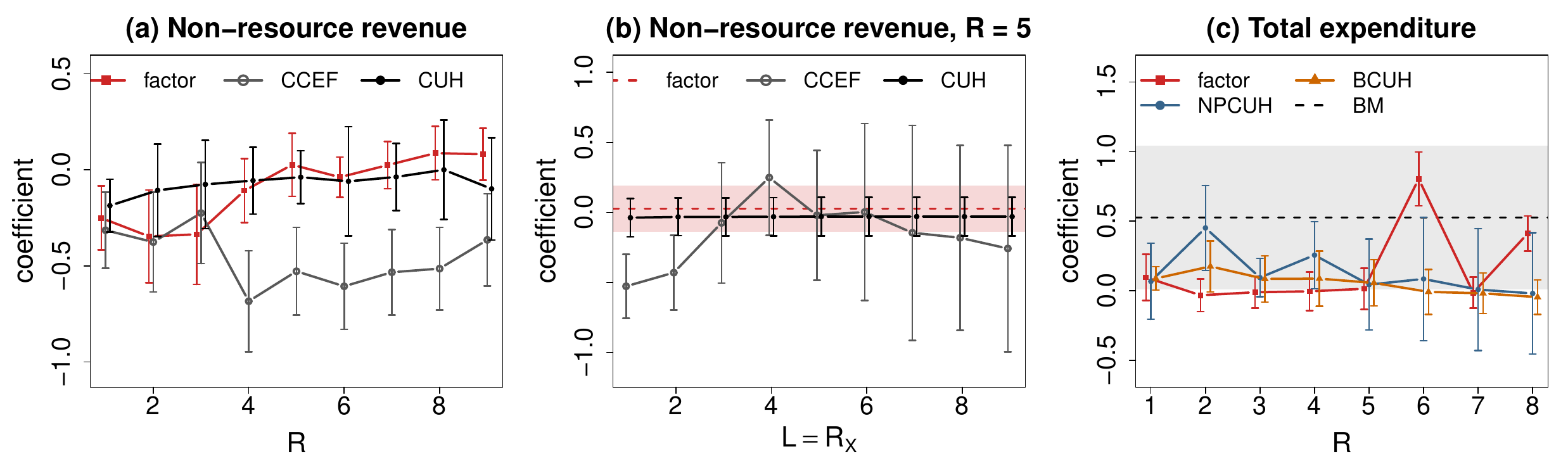}
\caption{Fiscal application of Section~\ref{sec:apps}. (a) Factor, CCEF and
CUH across the number $R$ of purged outcome directions. (b) Same outcome at $R=5$,
across $L=R_X$ covariate components. (c) Factor, NPCUH, BCUH and the two-way grouped
estimator of \citet{beyhum2025inference} (BM) on total expenditure. Bars are
95\% intervals, shaded bands the factor (b) and BM (c) intervals.}
\label{fig:intro-apps}
\end{figure}

Figure~\ref{fig:intro-apps} previews the empirical applications.
{It reports point estimates for two
outcomes from the fiscal application as the number of purged outcome directions
$R$ varies, and, for one outcome, as the number of purged covariate components
varies.}  In Section~\ref{sec:app-overlap} we show that when the residualised covariates 
across different estimators have high correlation, they unsurprisingly have similar estimates,
and large gaps arise only where the correlation is low. 

\begin{figure}[!ht]
\centering
\includegraphics[width=0.96\textwidth]{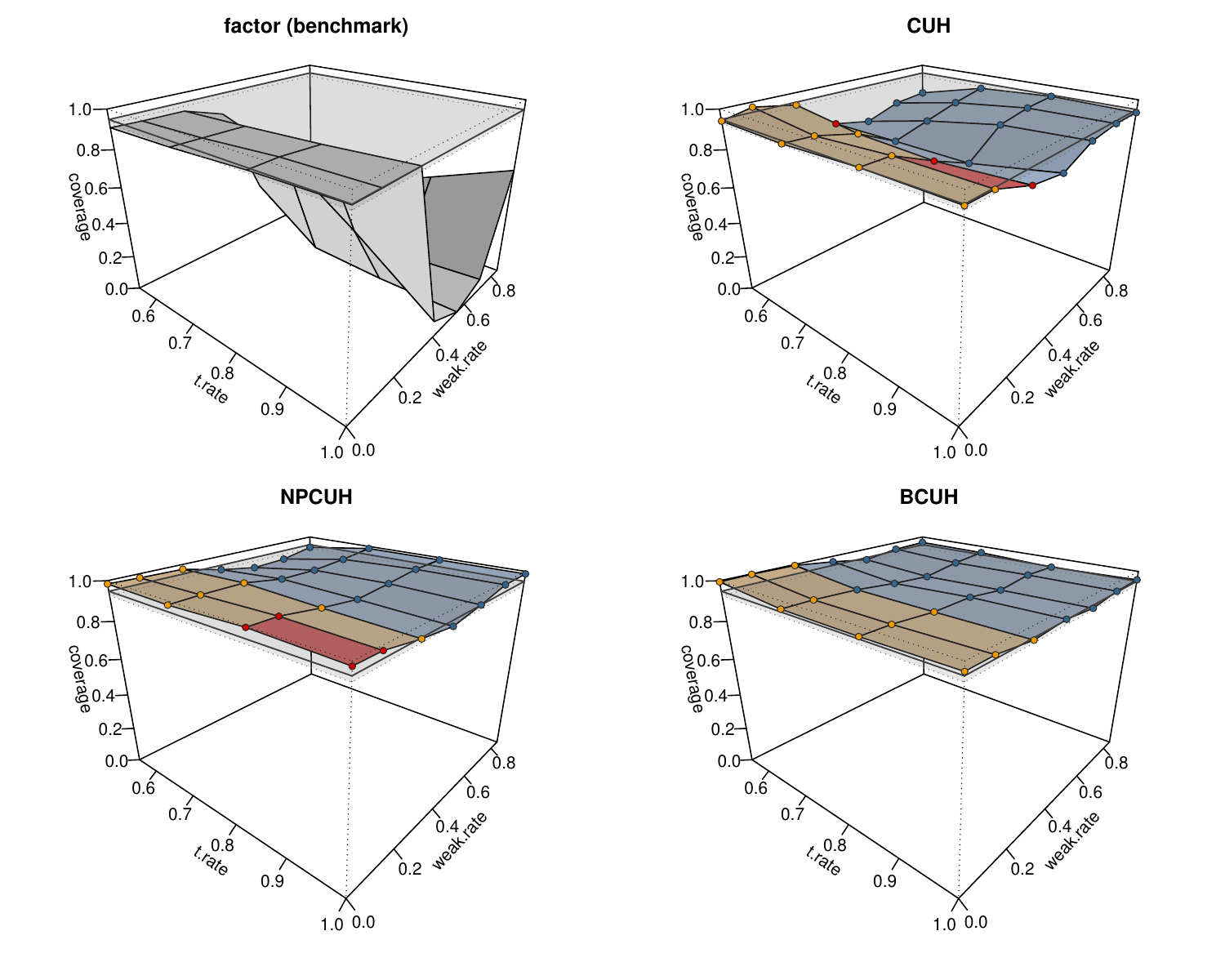}
\caption{Simulated coverage of nominal 95\% intervals at $N=200$, over $T=N^{t}$ and
factor strength $\min(N,T)^{-w}$. The grey factor benchmark fails along the
weak-factor boundary ($w\geq 0.2$). In the CUH, NPCUH and BCUH panels, blue
cells are closer to nominal than the benchmark beyond Monte Carlo noise, amber
within noise, red farther. Details in Section~\ref{sec:sim-manifold}.}
\label{fig:headline}
\end{figure}

Figure~\ref{fig:headline} previews the main simulation finding over a design
manifold of factor strength and the relative size of $T$ and $N$. Coverage of
the factor benchmark collapses along the weak-factor boundary. The CUH class
restores it, and BCUH and NPCUH have at least near-nominal coverage at almost all combinations of 
strength and panel length. 
Section~\ref{sec:sims} gives the full design with bias and variance surfaces.

\subsection*{Related literature}

CUH belongs to the literature on large-panel regression with interactive
fixed effects. Under strong factors, the least-squares estimator of
\citet{Bai2009} iterates between slope estimates and principal components of
the residualised outcome and is $\sqrt{NT}$-normal.
\citet{MoonWeidner2015,MoonWeidner2017} develop its asymptotic expansions and
its robustness to overstating the number of factors, and
\citet{moon2018nuclear} target the low-rank component by nuclear-norm
regularisation. The common correlated effects approach of \citet{Pesaran2006}
instead proxies the factor space by cross-sectional averages of outcomes
\emph{and} covariates, with extensions to non-stationary factors
\citep{KapetaniosPesaranYamagata2011}, spatial correlation
\citep{PesaranTosetti2011} and nonlinear models \citep{chen2023common}.
\citet{westerlund2015cross} compare the averages-based and principal-components
routes, and \citet{BaiNg2002,Onatski2005,ahn2013eigenvalue} select the number
of factors.

When strong-factor conditions fail, factor-based slope estimates are generally biased. \citet{onatski2012}
shows that principal-component estimates are inconsistent once factor eigenvalues fail to
dominate the upper edge of the noise spectrum, and characterises the bias
through the Marchenko--Pastur law. \citet{bai2023approximate} establish
consistency and asymptotic normality of principal components when loadings have
strength $N^{\alpha}$ with $\alpha\in(0,1]$. \citet{uematsuyamagata2022}
estimate weak-factor structures with heterogeneous, consistently estimable
strength exponents, and \citet{ChudikPesaranTosetti2011} classify weak,
semi-strong and strong cross-section dependence in the CCE setting.
\citet{armstrong2026robust} develop the bias-aware intervals of
Example~\ref{ex:ba} for slope inference under weak factors. We
measure strength by the singular values of $g(\alpha_i,\gamma_t)$, an
index symmetric in loadings and factors (Remark~\ref{rem:alpha}).

We derive separate detection, selection and inference thresholds and their
dependence on the relative dimension $N\asymp T^\kappa$. Detection is necessary
to estimate the common directions. Below the detection threshold
the bias-aware construction is the natural fallback.

A third strand treats the heterogeneity nonparametrically.
\citet{freeman2023linear} approximate a smooth $g(\alpha_i,\gamma_t)$ by a
growing-rank factor structure, under which the least-squares slope estimator is
slower than $\sqrt{NT}$. \citet{fernandez2021low} study low-rank approximations
of nonseparable panel models,
\citet{chen2021nonlinear,zeleneev2020identification} nonlinear factor
structures in panel and network data, and \citet{freeman2026multidimensional}
multidimensional effects. Proxies for latent effects can be built by
discretisation and clustering \citep{bonhomme2021discretizing,beyhum2025inference},
from latent similarities \citep{deaner2025inferring}, or from leading
eigenfunctions. \citet{freeman2026nonparametric} develop
the identification theory for these eigenfunction proxies, and
\citet{freeman2026inference} combine them with Neyman-orthogonal moments and
two-way nonparametric regression to restore $\sqrt{NT}$-normality. Our
smooth-function regime (Section~\ref{sec:smooth}) uses the same spectral-decay
calibration, and Appendix~\ref{app:npcuh} adapts cross-fitted additive
smoothing to the CUH proxies.

The inference theory draws on orthogonal moments and bias corrections.
The moment condition is the partially linear score of
\citet{robinson1988root}.
Incidental-parameter bias in panels and its analytical and jackknife
corrections are studied by
\citet{HahnKuersteiner2002,HahnNewey2004,ArellanoHahn2007,FernandezValWeidner2013},
and the two-way sample split of \citet{freeman2023linear} is closest to our
construction. Cross-fitting centres the stochastic errors, so inference is
valid under primitive conditions without a separate bias correction.

Section~\ref{sec:cuh-estimator} defines the estimators.
Sections~\ref{sec:weak} and~\ref{sec:inference} develop the weak-factor and
inference theory, and Sections~\ref{sec:sims} and~\ref{sec:apps} report
simulations and empirical applications. The appendices contain proofs,
implementation details and the theory for NPCUH.

\section{The CUH estimator}\label{sec:cuh-estimator}

{This section defines the CUH estimator and the objects used in the weak-factor theory.}
Throughout, the data follow \eqref{eq:y}--\eqref{eq:x}.

{
Under exogeneity, $\mathbb{E}[x_{it}\varepsilon_{it}]=0$, the population
least-squares coefficient is
\begin{align*}
    \mathbb{E}[x_{it} x_{it}']^{-1} \mathbb{E}[x_{it} y_{it}]
    =
    \beta +
    \mathbb{E}[x_{it} x_{it}']^{-1} \mathbb{E}[x_{it} g(\alpha_i, \gamma_t)],
\end{align*}
so the bias comes from $\mathbb{E}[x_{it} g(\alpha_i,\gamma_t)]$.
Factor models and CCE identify $g(\alpha_i,\gamma_t)$ and partial it out by a
Frisch--Waugh--Lovell reduction. This is sufficient but not necessary. What
must be controlled is only the \emph{common} variation that makes
$\mathbb{E}[x_{it} g(\alpha_i,\gamma_t)]$ non-zero.

Projecting the covariates on the outcome directions strengthens the estimates
of factors that bias the slope but are hard to detect against the noise in
$y_{it}$. In this sense the covariate projections act as an additional sieve
that selects which outcome components induce bias.
CUH therefore needs far less factor strength than the strong-factor
conditions require.
}

\subsection{Notation and asymptotic regime}\label{sec:notation}\label{sec:regime}
Stack the data into $N\times T$ matrices $Y=(y_{it})$, $X_j=(x_{it,j})$ for
$j=1,\dots,d$, $G=(g(\alpha_i,\gamma_t))$ and $E=(\varepsilon_{it})$. Every
singular value decomposition (SVD) uses the statistical normalisation, so a
generic $N\times T$ matrix $M=\sum_r s_r u_r v_r'$ has $u_r'u_r/N=1$ and
$v_r'v_r/T=1$. This matches the population expansion
\begin{align*}
     g(a,c)=\sum_{r\ge1}\sigma_r u_r(a)v_r(c), \quad  \E[u_r(\alpha)u_s(\alpha)]
  =\E[v_r(\gamma)v_s(\gamma)]=\mathbf 1\{r=s\},
\end{align*}
which, evaluated at the sample, gives
\begin{equation}\label{eq:Gsvd}
  G=\sum_{r\ge1}\sigma_r\,u_r^{(g)}\,{v_r^{(g)}}',
  \qquad
  u_r^{(g)}=(u_r(\alpha_1),\dots,u_r(\alpha_N))',\quad
  v_r^{(g)}=(v_r(\gamma_1),\dots,v_r(\gamma_T))',
\end{equation}
so the strength of common direction $r$ is carried entirely by $\sigma_r$. We
write $\norm{\cdot}_{\rm op}$ and $\norm{\cdot}_F$ for the operator and
Frobenius norms, $a\mw b=\min\{a,b\}$, $a\vee b=\max\{a,b\}$ and
$\fro{A,B}=\tr(A'B)$. For a frame $U=(u_1,\ldots,u_K)$ and
$\mathcal I\subseteq\{1,\ldots,K\}$, let $U_{\mathcal I}=\{u_r:r\in\mathcal I\}$.

We work in the regime
\begin{equation}\label{eq:regime}
  N\asymp T^{\kappa},\qquad \kappa\in(0,\infty),
\end{equation}
where $\kappa=1$ is the square case and $N$ grows faster than $T$ when
$\kappa>1$. Then $N\mw T\asymp T^{\min\{1,\kappa\}}$,
$N\vee T\asymp T^{\max\{1,\kappa\}}$ and $NT\asymp T^{1+\kappa}$. Most results
are stated in $N\mw T$, $N\vee T$ and $NT$, and so hold for every $\kappa>0$.

\subsection{The CUH construction}\label{sec:cuh-alg}

Given a preliminary estimator $\wt\beta$, the residualised outcome matrix is
\begin{equation}\label{eq:W}
  W:=Y-\sum_{j=1}^d X_j\wt\beta_j
   =X(\beta_0-\wt\beta)+G+E,
\end{equation}
where $X(\beta_0-\wt\beta):=\sum_jX_j(\beta_{0,j}-\wt\beta_j)$. Let
$W=\wh U\wh\Sigma\wh V'$ be its SVD, whose singular vectors are the outcome
directions, and let $R$ be the number of detectable directions
(Assumption~\ref{ass:strength}). Write $\mathcal I_R=\{1,\ldots,R\}$,
$\wh U_R=\wh U_{\mathcal I_R}$ and $\wh V_R=\wh V_{\mathcal I_R}$. For
$j\le d$ and $r\le N\wedge T$, the covariate projections and their projection
energies are
\begin{align*}
  \wh\lambda_{j,r}(\alpha_i)&:=\frac1T\sum_t x_{it,j}\wh v_{tr},
  &
  \wh f_{j,r}(\gamma_t)&:=\frac1N\sum_i x_{it,j}\wh u_{ir},\\
  \wh\sigma_{\alpha,j,r}^2&:=\frac1N\sum_i\wh\lambda_{j,r}^2(\alpha_i),
  &
  \wh\sigma_{\gamma,j,r}^2&:=\frac1T\sum_t\wh f_{j,r}^2(\gamma_t).
\end{align*}
For each regressor, rank the outcome directions separately by the two
energies, let $\mathcal J_j^\alpha$ and $\mathcal J_j^\gamma$ index the
leading $L_j$ in each ranking, and set
\[
  \wh\Lambda_j
  :=\frac1T X_j\wh V_{\mathcal J_j^\alpha},
  \qquad
  \wh F_j
  :=\frac1N X_j'\wh U_{\mathcal J_j^\gamma}.
\]
The stacked projections $\wh\Lambda=(\wh\Lambda_1,\ldots,\wh\Lambda_d)$ and
$\wh F=(\wh F_1,\ldots,\wh F_d)$, with $\wh U_R$ and $\wh V_R$, determine the
CUH transforms \eqref{eq:transforms}.

Section~\ref{sec:weak} shows that the outcome directions, covariate
projections and energy rankings are consistent for the detectable candidates
$r\in\mathcal I_R$ under the detectability condition of
Assumption~\ref{ass:strength}. The implementation
screens all estimated outcome directions, which can help near the detection
boundary, while the inference results are established for ranking restricted
to $\mathcal I_R$. {The smoothing variants of Section~\ref{sec:smoothing} accommodate the undetectable tail by smoothing rather than projecting.}

Our main proposal is the cross-fitted CUH estimator. Cross-fitting builds the
transforms on training subsamples disjoint from the block they residualise,
which limits their dependence on the evaluation errors.
Algorithm~\ref{alg:cf} in Appendix~\ref{app:crossfit} states the procedure.

\subsection{Transforms and scores}\label{sec:transforms}
Let $P_C:=C(C'C)^{+}C'$ project onto the column span of $C$, with
$(\cdot)^{+}$ the Moore--Penrose inverse, and let $M_C:=I-P_C$ be the residual
maker. The CUH transforms are
\begin{equation}\label{eq:transforms}
  \wh A=M_{\wh\Lambda }M_{\wh U_R},\qquad
  \wh B=M_{\wh V_R}M_{\wh F },
\end{equation}
each a product of two residual makers that need not commute. The oracle
transforms $A=M_{\Lambda}M_{U_R}$ and $B=M_{V_R}M_{F}$ use the population
analogues $U_R,V_R,\Lambda,F$.

For any $N\times T$ matrix $H$, the feasible and oracle operators are
\[
  \widehat{\mathcal R}[H]
  :=\widehat AH\widehat B,
  \qquad
  \mathcal R^{0}[H]
  :=AHB .
\]
The CUH estimator is $\wh\beta=\wh Q^{-1}\wh S$, where, for
$j,\ell=1,\dots,d$,
\begin{equation}
    [\widehat Q]_{j\ell}
  =\frac1{NT}\fro{\widehat{\mathcal R}[X_j],\widehat{\mathcal R}[X_\ell]},
  \qquad
  [\widehat S]_j
  =\frac1{NT}\fro{\widehat{\mathcal R}[X_j],\widehat{\mathcal R}[Y]}. \label{eq:gram_score}
\end{equation}
Substituting $Y=\sum_{q=1}^d X_q\beta_{0,q}+G+E$ gives
\begin{equation}
  \widehat S
  =\widehat Q\beta_0 +\frac1{NT}
  \left(
    \left\langle
      \widehat{\mathcal R}[X_j],
      \widehat{\mathcal R}[G]
    \right\rangle_F
  \right)_{j=1}^d +\frac1{NT}
  \left(
    \left\langle
      \widehat{\mathcal R}[X_j],
      \widehat{\mathcal R}[E]
    \right\rangle_F
  \right)_{j=1}^d .
  \label{eq:score-decomp}
\end{equation}
The $\widehat{\mathcal R}[G]$ term is the heterogeneity left after
projection, and the $\widehat{\mathcal R}[E]$ term carries the score of the
central limit theorem.

\subsection{Smoothing variants, NPCUH and BCUH}\label{sec:smoothing}
The transforms \eqref{eq:transforms} remove the span of the estimated
directions, so they need those directions estimated accurately. Two variants
smooth over the same directions instead. A smoother removes any function of
the proxies, so it needs only injectivity of the proxy collection, not
accurate estimates of the population directions. Collect the outcome
directions and the selected covariate projections of
Section~\ref{sec:cuh-alg} into the unit-side and time-side proxy vectors
\[
 \wh\pi_i^{(\alpha)}
 :=\big(\{\wh u_{ir}\}_{r\in\mathcal I_R},\,
 \{\wh\lambda_{j,r}(\alpha_i)\}
 _{j\le d,\,r\in\mathcal J_j^\alpha}\big),
 \qquad
 \wh\pi_t^{(\gamma)}
 :=\big(\{\wh v_{tr}\}_{r\in\mathcal I_R},\,
 \{\wh f_{j,r}(\gamma_t)\}
 _{j\le d,\,r\in\mathcal J_j^\gamma}\big),
\]
each with $Q_\pi:=R+\sum_jL_j$ coordinates. Write
$\wh\pi_{i,q}^{(\alpha)}$ and $\wh\pi_{t,q}^{(\gamma)}$ for the $q$th
coordinates and $K_h(u):=h^{-1}K(u/h)$ for a univariate kernel $K$. For each
coordinate $q$ the row-normalised kernel smoothers are
\begin{equation}\label{eq:np-smoother}
 \big[S^{(\alpha)}_{q,h_\alpha}\big]_{ij}
 :=\frac{K_{h_\alpha}\big(\wh\pi^{(\alpha)}_{j,q}-\wh\pi^{(\alpha)}_{i,q}\big)}
        {\sum_{\ell\le N}K_{h_\alpha}\big(\wh\pi^{(\alpha)}_{\ell,q}-\wh\pi^{(\alpha)}_{i,q}\big)},
 \qquad
 \big[S^{(\gamma)}_{q,h_\gamma}\big]_{ts}
 :=\frac{K_{h_\gamma}\big(\wh\pi^{(\gamma)}_{s,q}-\wh\pi^{(\gamma)}_{t,q}\big)}
        {\sum_{\ell\le T}K_{h_\gamma}\big(\wh\pi^{(\gamma)}_{\ell,q}-\wh\pi^{(\gamma)}_{t,q}\big)}.
\end{equation}
The nonparametric CUH estimator (NPCUH) residualises with these
smoothers,\footnote{Appendix~\ref{app:npcuh} analyses the backfitted form of
this smoother. The product form here is usually numerically very close and is
cheaper to compute.}
\begin{equation}\label{eq:np-sweep}
 \wh{\mathcal M}[Z]:=(I-\wh W^\alpha)Z(I-\wh W^\gamma)',
 \quad
 I-\wh W^\alpha:=\prod_{q\le Q_\pi}
 (I-S^{(\alpha)}_{q,h_\alpha}),
 \quad
 I-\wh W^\gamma:=\prod_{q\le Q_\pi}
 (I-S^{(\gamma)}_{q,h_\gamma}).
\end{equation}
The NPCUH estimator is then constructed as
$\wh\beta=\wh Q^{-1}\wh S$, replacing
$\wh{\mathcal R}$ with $\wh{\mathcal M}$ in \eqref{eq:gram_score}. In words, NPCUH compares observations in terms of their estimated latent heterogeneity and removes the shared systematic variation.

The blended estimator (BCUH) smooths the CUH projection residual, using
$\wh{\mathcal M}\circ\wh{\mathcal R}$ in place of $\wh{\mathcal R}$. Both
maps are linear, so \eqref{eq:score-decomp} holds with either in place of
$\wh{\mathcal R}$.
Sections~\ref{sec:sims} and~\ref{sec:apps} report the cross-fitted versions of
both, built as in Algorithm~\ref{alg:cf} with the smoothers in place.

\section{Weak-factor theory}\label{sec:weak}

This section bounds the estimated singular subspaces of $W$ in
\eqref{eq:W} using a single factor-strength index and a
singular-vector perturbation bound. Write
$\Delta_\beta:=X(\beta_0-\wt\beta)$, so that
$W=G+E+\Delta_\beta$. We maintain the standard noise condition
$\norm{E}_{\rm op}=\Op(\sqrt{N\vee T})$ (\citealp{Bai2009,MoonWeidner2015}), which implies
\[
s_{NT}:=\frac{\norm{E}_{\rm op}}{\sqrt{NT}}
=\Op((N\mw T)^{-1/2}).
\]
The noise edge $s_{NT}$ bounds how weak a factor can be and remain detectable.
Appendix~\ref{app:weak-proofs} provides sufficient conditions
for the noise bound under independence or weak dependence.

\subsection{Factor strength and detectability}

\begin{assumption}[Singular-value strength and detectability]\label{ass:strength}
Let $\sigma_1(G)\ge\sigma_2(G)\ge\ldots$ be the singular values of $G$
under the statistical normalisation \eqref{eq:Gsvd}, and write
$\sigma_{R}:=\sigma_{R}(G)$ for the weakest candidate strength.
(i) (Detectability) $c_{NT}:=\sigma_{R}$ satisfies
\begin{equation}\label{eq:SNR}
  c_{NT}\sqrt{N\mw T}\to\infty,
\end{equation}
equivalently $c_{NT}/s_{NT}\to\infty$ with $s_{NT}$ read as its
rate $(N\mw T)^{-1/2}$ from Lemma~\ref{lem:edge}.
(ii) (Spectral gap)
$\delta_{NT}:=\sigma_{R}(G)-\sigma_{R+1}(G)$ satisfies
\begin{align}
\label{eq:spectral_gap}
\delta_{NT}\asymp c_{NT} .
\end{align}
(iii) (Distinct candidate strengths) There is $c>0$ such that
$|\sigma_r(G)^2-\sigma_s(G)^2|\ge c\,\sigma_r(G)\,\sigma_s(G)$ for all
$r\neq s\le R$.
\end{assumption}

Condition \eqref{eq:SNR} is the detection threshold of the spiked-covariance
literature. A population singular direction appears in the sample spectrum of
$W$ if and only if its singular value exceeds the upper edge of the noise
spectrum \citep{onatski2012,bai2023approximate}.

Part (ii) is the spectral gap that the perturbation argument needs in the
spiked (finite-rank) case. When the spectrum decays the gap vanishes, and
Section~\ref{sec:smooth} treats that case. Part (iii) makes the candidate
strengths asymptotically distinct, so that results hold for individual singular
directions (Proposition~\ref{prop:xi}, Corollary~\ref{cor:select}). At an exact
tie only the span of the tied block is estimable, but the transforms are
invariant to rotations within it, so for slope estimation part (iii) is a
simplification rather than a restriction.

\begin{remark}[Calibrating the singular-value strength]\label{rem:alpha}
Assumption~\ref{ass:strength} restricts the singular values $\sigma_r$ and is
agnostic to whether weakness originates in the loadings, the factors, or both.
Write $G=\Lambda^0F^{0\prime}$, let $\Lambda^0_r\in\R^N$ and $F^0_r\in\R^T$,
$r=1,\ldots,N\wedge T$, be the columns of the true loading and factor matrices,
and let $s_r(G)$ be the ordinary singular values of $G$. Three calibrations of
$c_{NT}=\sigma_{R}$ follow.
\begin{enumerate}[(a)]
\item One-sided weak loadings \citep{bai2023approximate},
$\Lambda^{0\prime}_r\Lambda^0_r/N^{\alpha_r}\to c_r\in(0,\infty)$ with
$\alpha_r\in(0,1]$ and strong factors, so $s_r(G)^2\asymp N^{\alpha_r}T$ and
$c_{NT}\asymp N^{(\alpha_{R}-1)/2}\asymp T^{\kappa(\alpha_{R}-1)/2}$.
\item Symmetric weak loadings and factors,
$\Lambda^{0\prime}_r\Lambda^0_r/N^{\alpha_r}\to c_r\in(0,\infty)$ \emph{and}
$F^{0\prime}_rF^0_r/T^{\tau_r}\to d_r\in(0,\infty)$ with $\alpha_r,\tau_r\in(0,1]$,
so $s_r(G)^2\asymp N^{\alpha_r}T^{\tau_r}$ and
$c_{NT}\asymp N^{(\alpha_{R}-1)/2}T^{(\tau_{R}-1)/2}$. Here $N$ and $T$
enter symmetrically, and case~(a) is $\tau_r\equiv1$.
\item Smooth function or decaying spectrum \citep{freeman2023linear,
freeman2026inference,freeman2026nonparametric}, where the rank is not fixed
and $\sigma_r$ decays in $r$ (Section~\ref{sec:smooth}).
\end{enumerate}
Under (b), detectability \eqref{eq:SNR} reads
\begin{equation}\label{eq:detect-rho}
  \kappa(1-\alpha_{R})+(1-\tau_{R})<\min\{1,\kappa\},
\end{equation}
and under (a), the calibration used in the rest of the paper, it reads
\begin{equation}
    \alpha_{R}>\max\{0,1-1/\kappa\}. \label{eq:detectability}
\end{equation}
\end{remark}

\subsection{Estimation of the singular subspaces}

Define the population covariate projections
$\lambda_{j,r}(\alpha):=\int_{\Omega_\gamma}g_{x,j}(\alpha,c)v_r(c)\pi_\gamma(c)\,dc$
and symmetrically $f_{j,r}(\gamma)$, with feasible counterparts
$\wh\lambda_{j,r},\wh f_{j,r}$ from Section~\ref{sec:cuh-alg}. The same
symbols denote the stacked vectors, such as
$\lambda_{j,r}=(\lambda_{j,r}(\alpha_i))_{i\le N}$ and
$f_{j,r}=(f_{j,r}(\gamma_t))_{t\le T}$. For vectors indexed by units or
periods, write $\|a\|_N^2=N^{-1}\sum_i a_i^2$ and
$\|b\|_T^2=T^{-1}\sum_t b_t^2$.

\begin{assumption}[Regularity]\label{ass:reg}
(i) $\E\eta_{it}^4<\infty$, $\E\varepsilon_{it}^4<\infty$,
$g_{x,j}\in L^4(\pi_\alpha\otimes\pi_\gamma)$,
$u_r\in L^4(\pi_\alpha)$, $v_r\in L^4(\pi_\gamma)$ for every fixed
$j\le d$, $r\in\mathcal I_{R}$, and
$\norm{g}_{L^2(\pi_\alpha\otimes\pi_\gamma)}<\infty$.
(ii) The eigenfunctions are orthonormal,
$\int u_r u_{r'}\pi_\alpha=\mathbbm 1\{r=r'\}$ and
$\int v_r v_{r'}\pi_\gamma=\mathbbm 1\{r=r'\}$.
(iii) For every fixed $r,r'\in\mathcal I_{R}$ and $j\le d$, the
evaluated eigenfunctions and regressors satisfy
$\big|N^{-1}\sum_i u_r(\alpha_i)u_{r'}(\alpha_i)-\mathbbm 1\{r=r'\}\big|
=\Op(N^{-1/2})$,
$\big|T^{-1}\sum_t v_r(\gamma_t)v_{r'}(\gamma_t)-\mathbbm 1\{r=r'\}\big|
=\Op(T^{-1/2})$ and $\|X_j\|_F/\sqrt{NT}=\Op(1)$.
(iv) For every fixed $j\le d$, $r\in\mathcal I_{R}$, the oracle covariate
projections and their energies obey
\begin{align*}
 \|T^{-1}X_jv_r-\lambda_{j,r}\|_N&=\Op(T^{-1/2}),&
 \|N^{-1}X_j'u_r-f_{j,r}\|_T&=\Op(N^{-1/2}),\\
 \big|\|\lambda_{j,r}\|_N^2-\E[\lambda_{j,r}^2(\alpha)]\big|&=\Op(N^{-1/2}),&
 \big|\|f_{j,r}\|_T^2-\E[f_{j,r}^2(\gamma)]\big|&=\Op(T^{-1/2}).
\end{align*}
\end{assumption}

Parts (iii)--(iv) are sampling regularity conditions. They hold under
independent sampling with the displayed moments and under weakly dependent or
mixing designs with the same root-$N$ and root-$T$ laws of large numbers. The
product measure $\pi_\alpha\otimes\pi_\gamma$ reflects averaging over the
rectangular panel, and neither it nor parts (iii)--(iv) require the unit and
time effects to be independent. The first pair of bounds in part (iv)
follows, for example, from bounded average long-run variances of $x_{it,j}v_r(\gamma_t)$
over $t$ and of $x_{it,j}u_r(\alpha_i)$ over $i$. As $d$ and $R$ are fixed in the spiked
regime, all rates hold jointly over $j\le d$ and $r,r'\in\mathcal I_{R}$.

Let $\xi_u,\xi_v$ be the sign-aligned root-mean-square errors of the
estimated singular vectors relative to the evaluated population
eigenfunctions,
\[
  \xi_u^2:=\max_{r\in \mathcal{I}_{R}}\,\min_{s\in\{\pm1\}}\,
  \|u_r^{(g)}-s\,\wh u_r\|_N^2,
  \qquad
  \xi_v^2:=\max_{r\in \mathcal{I}_{R}}\,\min_{s\in\{\pm1\}}\,
  \|v_r^{(g)}-s\,\wh v_r\|_T^2 .
\]
The minimum removes the sign indeterminacy of the SVD.

\begin{proposition}[Subspace singular-vector]\label{prop:xi}
Impose Assumptions~\ref{ass:strength}--\ref{ass:reg}(i)--(iii), the noise edge
$\|E\|_{\rm op} = \Op(\sqrt{N\vee T})$, and
$\|E+\Delta_\beta\|_{\rm op}/\sqrt{NT}=\op(\delta_{NT})$. Then, for every
$\kappa>0$,
\begin{equation}\label{eq:xi}
  \norm{P_{\wh U_{R}}-P_{U_{R}}}_{\rm op}^2,\ \norm{P_{\wh V_{R}}-P_{V_{R}}}_{\rm op}^2,\ \xi_u^2,\ \xi_v^2
  \;=\;\Op\!\Big(\underbrace{\frac{1}{c_{NT}^2\,(N\mw T)}}_{\text{noise}}\Big)
  \;+\;\Op\!\Big(\underbrace{\frac{\norm{X}_{\rm op}^2\,\norm{\beta_0-\wt\beta}^2}{NT\,c_{NT}^2}}_{\text{first-step }\beta}\Big),
\end{equation}
where $U_{R}=(u_r(\alpha_i))_{i\le N,r\in \mathcal{I}_{R}}$ and
$V_{R}=(v_r(\gamma_t))_{t\le T,r\in \mathcal{I}_{R}}$
are the evaluated eigenfunction frames.
\end{proposition}

\begin{remark}[Sharp rate and admissible $\kappa$]\label{rem:xi-rho}
Bound \eqref{eq:xi} is dimension-symmetric and sharp for $\kappa\le1$. Under calibration (a) of Remark~\ref{rem:alpha},
$c_{NT}\asymp T^{\kappa(\alpha_{R}-1)/2}$, so the noise part is
$\xi_v^2\asymp T^{\kappa(1-\alpha_{R})-\min\{1,\kappa\}}$. It vanishes if and
only if $\kappa(1-\alpha_{R})<\min\{1,\kappa\}$, that is, if $\alpha_{R}>0$ for
$\kappa\le1$ and if $\kappa<1/(1-\alpha_{R})$ for $\kappa>1$, the upper bound
of \eqref{eq:detect-rho}. For $\kappa>1$ the symmetric Wedin bound is not
sharp. In an approximate factor model the one-sided rate
$\xi_v^2=\Op(N^{-\alpha_{R}}+(N^{1-\alpha_{R}}/T)^2)$ of
\citet{bai2023approximate} refines it with the same consistency window
$\kappa<1/(1-\alpha_{R})$. The mirror statement in the loading dimension
governs $\xi_u$.
\end{remark}

\begin{theorem}[Projection rates]\label{thm:proj}
Under Assumption~\ref{ass:reg}, with the estimated singular vectors
sign-aligned as in the definitions of $\xi_u,\xi_v$,
\[
 \max_{j\le d,\,r\in\mathcal I_{R}}
 \|\wh\lambda_{j,r}-\lambda_{j,r}\|_N=\Op(\xi_v+T^{-1/2}),
 \qquad
 \max_{j\le d,\,r\in\mathcal I_{R}}
 \|\wh f_{j,r}-f_{j,r}\|_T=\Op(\xi_u+N^{-1/2}).
\]
\end{theorem}

With Proposition~\ref{prop:xi}, Theorem~\ref{thm:proj} gives consistent
covariate projections under the noise edge, detectability and the spectral
gap. The $T^{-1/2}$ and $N^{-1/2}$ terms are the oracle rates for known
population singular functions. Under strong factors $\xi_u$ and $\xi_v$ are
root-$N$ and root-$T$ consistent, and generally slower otherwise.

\subsection{Selection consistency}

Let $\sigma_{\alpha,j,r}^2:=\E[\lambda_{j,r}^2(\alpha)]$ and
$\sigma_{\gamma,j,r}^2:=\E[f_{j,r}^2(\gamma)]$ be the population counterparts
of the projection energies $\wh\sigma_{\alpha,j,r}^2$ and
$\wh\sigma_{\gamma,j,r}^2$ of Section~\ref{sec:cuh-alg}.

\begin{proposition}[Energy expansion]\label{prop:sigma}
Under Assumptions~\ref{ass:strength}--\ref{ass:reg},
uniformly over $r\in \mathcal{I}_{R}$,
\begin{align}
     \wh\sigma_{\alpha,j,r}^2
  =&\sigma_{\alpha,j,r}^2
   +\Op\big(\sigma_{\alpha,j,r}(\xi_v+T^{-1/2})\big)
   +\Op\big(\xi_v^2+(N\mw T)^{-1}\big), \label{eq:sigmahat_alpha} \\
  \wh\sigma_{\gamma,j,r}^2
  =&\sigma_{\gamma,j,r}^2
   +\Op\big(\sigma_{\gamma,j,r}(\xi_u+N^{-1/2})\big)
   +\Op\big(\xi_u^2+(N\mw T)^{-1}\big).\label{eq:sigmahat_gamma}
\end{align}
\end{proposition}

The estimated energies are therefore consistent whenever the covariate
projections are, and their error rates set the minimum $x$-relevance strength
that consistent ranking requires.

\begin{assumption}[Covariate-relevance strength]
\label{ass:Xstrength}
Among the candidates $r\in\mathcal I_R$, the selected directions have energies
$\sigma_{\alpha,j,r}^2\ge\epsilon_{NT}^2$ with
$\epsilon_{NT}^2\gg\xi_v^2+(N\wedge T)^{-1}$, while the non-selected directions
have $\sigma_{\alpha,j,r}^2=\Op((N\wedge T)^{-1})$, and symmetrically for the
time side.
\end{assumption}

\begin{corollary}[Thresholding consistency]\label{cor:select}
Make Assumptions~\ref{ass:strength}--\ref{ass:Xstrength}, and let the smallest
\emph{nonzero} eigenvalues of $\Lambda'\Lambda/N$ and $F'F/T$ be at least
$c\,\epsilon_{NT}^2$ for a fixed $c>0$. Then, with probability approaching one,
ranking by $\wh\sigma_{\alpha,j,r}^2$ (resp.\ $\wh\sigma_{\gamma,j,r}^2$) over
$r\in\mathcal I_{R}$ selects exactly the $x$-relevant common directions. Write
$\Lambda \in \mathbb{R}^{p_\alpha},F \in \mathbb{R}^{p_\gamma}$ for the
population components, and
\begin{equation}\label{eq:zeta}
 \zeta_\Lambda:=(\xi_v+T^{-1/2})/\epsilon_{NT},\qquad
 \zeta_F:=(\xi_u+N^{-1/2})/\epsilon_{NT} .
\end{equation}
Then the estimated components satisfy
\begin{equation}\label{eq:select-angle}
 \norm{M_{\wh\Lambda}P_\Lambda}_{\rm op}
 =\norm{P_\Lambda M_{\wh\Lambda}}_{\rm op}=\Op(\zeta_\Lambda),
 \qquad
 \norm{M_{\wh F}P_F}_{\rm op}
 =\norm{P_FM_{\wh F}}_{\rm op}=\Op(\zeta_F),
\end{equation}
and for every $N\times T$ matrix $H$,
\begin{align}
 \norm{(P_{\wh\Lambda}-P_\Lambda)H}_F
 &\le\sqrt{p_\alpha}\,\norm{M_\Lambda H}_{\rm op}
 +\Op(\zeta_\Lambda)\norm{H}_F, \label{eq:select-energy-L}\\
 \norm{H(P_F-P_{\wh F})}_F
 &\le\sqrt{p_\gamma}\,\norm{HM_F}_{\rm op}
 +\Op(\zeta_F)\norm{H}_F. \label{eq:select-energy-R}
\end{align}
\end{corollary}

The ranking statistic $\wh\sigma_{\alpha,j,r}^2=\norm{\wh\lambda_{j,r}}_N^2$
measures the strength of a common direction in the $x$ equation, not the
singular value $\sigma_r$ of the residualised outcome. Selection is therefore
governed by a second strength index $\alpha^x_r$, generally distinct from the
outcome strength $\alpha_r$ of Remark~\ref{rem:alpha}, so that $\epsilon_{NT}$
depends on $\alpha^x$ while $c_{NT}$ and $\delta_{NT}$ depend on $\alpha$.

\begin{remark}[Factor strength requirements]
\label{rem:strength-consequences}
Let $c_{NT}^2\asymp N^{\alpha_{R}-1}$ and $N\asymp T^\kappa$.
\begin{enumerate}[(i)]
\item By \eqref{eq:detectability}, detection requires
$\alpha_{R}>\max\{0,1-1/\kappa\}$.
\item By \eqref{eq:xi},
$\xi_u^2\vee\xi_v^2=\Op\big(T^{\kappa(1-\alpha_{R})-\min\{1,\kappa\}}\big)$
when the preliminary-estimator term is negligible.
\item Let $\epsilon_{NT}^2\asymp N^{\alpha_{R}^x-1}$ be the strength of the
weakest selected direction in $x$. The condition
$\epsilon_{NT}^2\gg\xi_v^2+(N\mw T)^{-1}$ of Assumption~\ref{ass:Xstrength} is
implied by
$\alpha_{R}+\alpha_{R}^x>2-\min\{1,\kappa\}/\kappa$.
\item Under Assumption~\ref{ass:cf-exact-spike} of
Section~\ref{sec:inference}, $\xi_u+\xi_v+\zeta_\Lambda+\zeta_F$ is of order
$(c_{NT}\epsilon_{NT})^{-1}(N\mw T)^{-1/2}$, and
$\sqrt{NT}\,(\xi_u+\xi_v+\zeta_\Lambda+\zeta_F)^3=o_p(1)$ holds when
\begin{equation}\label{eq:inference-frontier}
 c_{NT}\,\epsilon_{NT}\gg\Big\{\frac{N\vee T}{(N\mw T)^2}\Big\}^{1/6},
 \qquad
 \alpha_{R}+\alpha_{R}^x>\tau(\kappa):=
 \begin{cases}
  \frac43+\frac1{3\kappa}, & \kappa\le1,\\[2pt]
  \frac73-\frac2{3\kappa}, & \kappa\ge1.
 \end{cases}
\end{equation}
By \eqref{eq:zeta}, $\epsilon_{NT}$ must then exceed $\xi_v+T^{-1/2}$ and
$\xi_u+N^{-1/2}$ by a factor larger than $(NT)^{1/6}$, which is stronger than
Assumption~\ref{ass:Xstrength}. At $\kappa=1$, $\tau(1)=5/3$.
\end{enumerate}
\end{remark}

\subsection{The smooth-function regime}\label{sec:smooth}

The spiked calibrations (a)--(b) of Remark~\ref{rem:alpha} keep $R$ fixed.
When $g$ is a smooth nonparametric function, $G$ has infinitely many decaying
singular values and the candidate truncation grows, $R=R_{NT}\to\infty$.
Assume the spectral decay
\begin{equation}\label{eq:decay}
  \sigma_r\asymp r^{-(\rho+\frac12)},
  \qquad\text{so that the truncation error}\quad
  \zeta_{R}^2:=\sum_{r>R}\sigma_r^2
  \asymp R^{-2\rho},
\end{equation}
with smoothness index $\rho=p/\min\{d_\alpha,d_\gamma\}$ for
$g\in H^p(\Omega_\alpha,\Omega_\gamma)$. Applied to the weakest candidate,
$c_{NT}=\sigma_{R}\asymp R^{-(\rho+1/2)}$, detectability \eqref{eq:SNR}
becomes
\begin{equation}\label{eq:detect-R}
  \sigma_{R}\sqrt{N\mw T}\to\infty
  \qquad\Longleftrightarrow\qquad
  R=o\big((N\mw T)^{1/(2\rho+1)}\big),
\end{equation}
the growing-$R$ counterpart of \eqref{eq:detect-rho}.

The relative gap now vanishes,
$\delta_{R}/\sigma_{R}=(\sigma_{R}-\sigma_{R+1})/\sigma_{R}\asymp R^{-1}\to0$,
so Lemma~\ref{lem:wedin}, which needs $\delta_{NT}\asymp c_{NT}$, no longer
gives a useful rate. Instead split $G=G_{R}+G_{>R}$ into its leading-$R$ part
and the tail, with $\norm{G_{>R}}_F/\sqrt{NT}\asymp\zeta_{R}$, and carry the
tail as an additional perturbation. Increasing $R$ shrinks the tail but adds
weaker directions whose proxies are harder to estimate, so $R$ trades
truncation bias against accuracy within the limit
\eqref{eq:detect-R}.

CUH projection leaves the tail $G_{>R}$, which is nonzero for every finite $R$,
and its product with the common block of $X$ left after projection need
not be negligible. CUH and Theorem~\ref{thm:cf} are therefore confined to the
spiked regime. NPCUH (Section~\ref{sec:smoothing}) smooths instead of projecting, and
Appendix~\ref{app:npcuh} develops its theory for the decaying spectrum. Under a uniform immersion of the proxy collection NPCUH can
keep the number of outcome directions fixed, while an almost-everywhere
immersion requires growing $R$.
\section{Inference}\label{sec:inference}

\subsection{Cross-fitted CUH and inference}\label{sec:cf-lead}

Cross-fitting (Section~\ref{sec:cuh-alg}) is widely used in large panels
\citep[e.g.][]{freeman2023linear,beyhum2025inference,freeman2026inference}.
A temporal gap separates same-unit training and evaluation observations, which
allows weak temporal dependence.

\subsubsection{Sample split}\label{sec:sampleSplit}

Split the units into data-independent balanced halves
$\mathcal N_1,\mathcal N_2$, with $-a:=3-a$, and the periods into $K$
adjacent balanced blocks $\mathcal M_1,\ldots,\mathcal M_K$. Let
$\mathcal M_k^{\rm qc}$ collect the periods at distance greater than $q_T$
from $\mathcal M_k$. For $b=(a,k)$, $\mathcal I_b=\mathcal N_a\times\mathcal M_k$
is the evaluation block and $H^{(b)}$ the corresponding submatrix of $H$.

The preliminary estimator $\wt\beta_b$ uses
$\mathcal N_{-a}\times\mathcal M_k^{\rm qc}$. The unit-side transform uses
$\mathcal N_a\times\mathcal M_k^{\rm qc}$, with regressor and error arrays
$X_j^{A,b},E^{A,b}$, and the time-side transform uses
$\mathcal N_{-a}\times\mathcal M_k$, with arrays $X_j^{B,b},E^{B,b}$. Let
$\mathcal F_b$ be the sigma algebra generated by these training data. With
$U_{R,b},\Lambda_b$ and $V_{R,b},F_b$ the population frames restricted to
$\mathcal N_a$ and $\mathcal M_k$, the oracle block transforms and operator are
\[
  A_b:=M_{\Lambda_b}M_{U_{R,b}},
  \qquad
  B_b:=M_{V_{R,b}}M_{F_b},
  \qquad
  \mathcal R_b^0[H^{(b)}]
  :=A_bH^{(b)}B_b.
\]
Their feasible counterparts $\wh A_b,\wh B_b$ are built from the training
data, and $\wh\beta^{\rm cf}$ aggregates the residualised Gram matrices and
moments over the evaluation blocks. As $K$ is fixed and $q_T=o(T)$, every
training rectangle has dimensions proportional to $N$ and $T$.

\subsubsection{Conditions and asymptotic normality}\label{sec:cf-conditions}

Let
$\mathcal C:=\sigma(\{\alpha_i,\gamma_t,\widetilde\alpha_i,
\widetilde\gamma_t\}_{i,t})$ and
$\mathcal C_0:=\sigma(\{\alpha_i,\gamma_t\}_{i,t})$, and write
$\beta_i(q\mid X,\mathcal C)$ for the beta-mixing coefficient at lag $q$ of
$\{\varepsilon_{it}:t\le T\}$ conditional on $(X,\mathcal C)$. Let
$\mathcal V_j:=X_j-\Gamma_{X,j}$, with $\Gamma_{X,j}$ the array of
$g_{x,j}(\alpha_i,\gamma_t)$, so that $\mathcal V_j$ has entries
$\wt g_{x,j}(\wt\alpha_i,\wt\gamma_t)+\eta_{it,j}$ by \eqref{eq:x}. Put
$\mathcal V_{j,b}:=\mathcal V_j^{(b)}$,
$\mathring{\mathcal V}_{j,b}:=
X_{j}^{(b)}-\E[X_{j}^{(b)}\mid\mathcal C_0,\mathcal F_b]$,
\begin{align*}
    \Sigma_{E,b}:=&
\E[\operatorname{vec}(E^{(b)})\operatorname{vec}(E^{(b)})'
   \mid X,\mathcal C],\\
    \Sigma_{\mathring V,j,b}
 :=&
 \E\!\left[
   \operatorname{vec}(\mathring{\mathcal V}_{j,b})
   \operatorname{vec}(\mathring{\mathcal V}_{j,b})'
   \,\middle|\,\mathcal C_0,\mathcal F_b
 \right],
\end{align*}
and the oracle-residualised idiosyncratic-regressor Gram matrix
\[
 Q_{\nu,NT}^0
 :=
 \frac1{NT}\sum_b
 \E\!\left[
 \left\{
 \left\langle
   \mathcal R_b^0[\mathcal V_{j,b}],
   \mathcal R_b^0[\mathcal V_{k,b}]
 \right\rangle_F
 \right\}_{j,k=1}^d
 \right].
\]

\begin{assumption}[Cross-fitting regularity]\label{ass:cf-reg}
For every block $b$ and $j\le d$:
\begin{enumerate}[(i)]
\item Conditional on $(X,\mathcal C)$, the unit processes
$\{\varepsilon_{it}:t\le T\}$ are independent across $i$, centred, and
uniformly geometrically beta mixing over time, with
$\max_{i\le N}\beta_i(q\mid X,\mathcal C)\le C_\beta e^{-c_\beta q}$ for
$q\ge1$ and fixed constants $C_\beta,c_\beta>0$. For some $\delta>0$,
$\max_{i,t}\E[|\varepsilon_{it}|^{4+\delta}\mid X,\mathcal C]=O_p(1)$.
\item $K$ is fixed, $\log N=o(q_T)$, $q_T=o(T)$, and
$\|\Sigma_{E,b}\|_{\rm op}
 +\|\Sigma_{\mathring V,j,b}\|_{\rm op}=O_p(1)$.
\item $\liminf_{N,T\to\infty}
 \lambda_{\min}(Q_{\nu,NT}^0)>0$, with $\lambda_{\min}$ the smallest
 eigenvalue.
\end{enumerate}
\end{assumption}

The bound on $\Sigma_{E,b}$ in part (ii) follows from part (i) by
covariance inequalities for beta-mixing arrays. Part (iii) is no stronger than
the full-rank condition that CCEF places on the remaining idiosyncratic
regressor variation.

\begin{assumption}[Blockwise selected-frame strength]
\label{ass:cf-frame-strength}
For fixed $c>0$, with probability approaching one, every nonzero eigenvalue
of $\Lambda_b'\Lambda_b/N$ and of $F_b'F_b/T$ is at least $c\,\epsilon_{NT}^2$,
for every block $b$.
\end{assumption}

Assumption~\ref{ass:cf-frame-strength}, the block analogue of the eigenvalue
condition of Corollary~\ref{cor:select}, excludes selected columns that are
nearly but not exactly collinear. Lemma~\ref{lem:cf-transform} uses it to
show that the frame truncation of Algorithm~\ref{alg:cf} recovers the ranks of
$\Lambda_b$ and $F_b$. 

We collect estimation errors of the outcome
directions and covariate projections together with the noise to signal ratio into an aggregated nuisance error rate:
\begin{equation}\label{eq:cf-varsigma}
 \varsigma_{NT}:=\xi_u+\xi_v+\zeta_\Lambda+\zeta_F+\frac{1}{c_{NT}\sqrt{N\wedge T}},
\end{equation}
with $\zeta_\Lambda$ and $\zeta_F$ as in \eqref{eq:zeta}.

\begin{assumption}[Cross-fitted regularity]
\label{ass:cf-exact-spike}
\begin{enumerate}[(i)]
\item $G$ has rank $R_0$ and $R=R_0$. Uniformly over training samples, the
block analogues of Assumptions~\ref{ass:strength}--\ref{ass:Xstrength} hold
for $r\in\mathcal I_R$, and, for $s\in\{A,B\}$,
$\norm{E^{s,b}}_{\rm op}=\Op(\sqrt{N\vee T})$ and
$\norm{\operatorname{Cov}\{\operatorname{vec}(E^{s,b})\mid X,\mathcal C\}}_{\rm op}
=\Op(1)$.
\item The blockwise preliminary estimators obey
\begin{equation}\label{eq:cf-initial}
 \max_b\norm{\wt\beta_b-\beta_0}=\Op\big((N\mw T)^{-1/2}\big).
\end{equation}
\item The population frames satisfy
\begin{equation}\label{eq:cf-commute}
 \max_b\min\Big\{\norm{P_{U_{R,b}}P_{\Lambda_b}-P_{\Lambda_b}P_{U_{R,b}}}_{\rm op},\
 \norm{P_{V_{R,b}}P_{F_b}-P_{F_b}P_{V_{R,b}}}_{\rm op}\Big\}
 =\Op(\varsigma_{NT}).
\end{equation}
\item The spectral rates satisfy
\begin{equation}\label{eq:cf-exact-rate}
 \sqrt{NT}\,\varsigma_{NT}^3=o_p(1).
\end{equation}
\end{enumerate}
\end{assumption}

Part (i) adds the noise edge of Proposition~\ref{prop:xi} on each training
rectangle and a covariance bound implied by Assumption~\ref{ass:cf-reg}(i).
Part (ii) holds for the least-squares estimator of \citet{Bai2009} on
$\mathcal N_{-a}\times\mathcal M_k^{\rm qc}$ with any preliminary rank
$R_{\rm pre}\ge R_0$, under the conditions of
\citet[Theorem~4.1]{MoonWeidner2015}, which require no strong factors. Part (iii) restricts the common block alone. It holds with both
commutators zero when $g_x$ is a multiple of $g$, and under the spanning
condition \eqref{eqn:CCEXstrongfactor} of Example~\ref{ex:cce}.
Section~\ref{sec:cf-transform-consistency} shows how to check it from the training frames. Both parts
(i) and (iv) restrict factor strengths. See Remark~\ref{rem:strength-consequences}(iv) for how these rates translate into factor strength. 

\begin{theorem}[Asymptotic normality of cross-fitted CUH]\label{thm:cf}
Let $\wh\beta^{\rm cf}$ be the estimator of Algorithm~\ref{alg:cf}, and
suppose Assumptions~\ref{ass:cf-reg}, \ref{ass:cf-frame-strength},
\ref{ass:cf-exact-spike} and~\ref{ass:cf-score} hold. Then
\[
 \sqrt{NT}\,(\wh\beta^{\rm cf}-\beta_0)
 \Rightarrow\mathcal N\big(0,Q^{-1}\Omega Q^{-1}\big),
\]
where $Q$ and $\Omega$ are the limiting oracle Gram matrix and oracle score
variance of Assumption~\ref{ass:cf-score}.
\end{theorem}

The errors may be heteroskedastic and serially dependent, and
Proposition~\ref{prop:cf-csdep} relaxes their independence across units to
weak cross-sectional dependence of unknown ordering.
Remark~\ref{rem:strength-consequences}(iv) calibrates the factor strength that
Assumption~\ref{ass:cf-exact-spike}(iv) requires. The two-way robust standard
errors of Section~\ref{sec:cf-variance} are consistent by
Theorem~\ref{thm:cf-chs}.

\subsection{Efficiency}\label{sec:eff}

CCEF (Example~\ref{ex:cce}), FK \citep{freeman2026inference} and CUH each
run OLS after a purge that residualises $Y$ and $X$ on an estimate of the
latent structure. Under homoskedastic errors each asymptotic variance is
proportional to the inverse second moment of the regressor variation
$z^e_{it}$ that the purge of estimator $e$ leaves,
\begin{equation}\label{eq:zscores}
  z^{\rm CCEF}_{it}:=\eta_{it},\qquad
  z^{\rm FK}_{it}:=\eta_{it}+\wt g_{x,it},\qquad
  z^{\rm CUH}_{it}:=\eta_{it}+\wt g_{x,it}+w_{it},
\end{equation}
with $\wt g_{x,it}:=\wt g_x(\wt\alpha_i,\wt\gamma_t)$. CCEF removes the
common and $x$-only blocks, FK only the common block, and CUH additionally
keeps $w_{it}$, the part of $g_x(\alpha_i,\gamma_t)$ linearly orthogonal to
$g(\alpha_i,\gamma_t)$. Let $\Sigma_\eta$, $\wt G$ and $\mathcal W$ be the
limiting average second moments of $\eta_{it}$, $\wt g_{x,it}$ and $w_{it}$,
$\sigma^2$ the error variance, and $V_e$ the asymptotic variance of estimator
$e$. Assumption~\ref{ass:homo} in Appendix~\ref{sec:cf-efficiency} states
homoskedasticity and the standard expansions for CCEF and FK.

\begin{proposition}[Variance ordering]\label{prop:eff}
Under Assumptions~\ref{ass:cf-score} and~\ref{ass:homo},
\[
  V_{\rm CUH}
  \;\preceq\;
  \sigma^2\big(\Sigma_\eta+\wt G+\mathcal W\big)^{-1}
  \;\preceq\;
  V_{\rm FK}
  =\sigma^2\big(\Sigma_\eta+\wt G\big)^{-1}
  \;\preceq\;
  V_{\rm CCEF}
  =\sigma^2\Sigma_\eta^{-1}.
\]
The equality \(V_{\rm FK}=V_{\rm CCEF}\) holds
if and only if \(\wt G=0\), i.e. the $x$-only block is negligible.
\end{proposition}

The CUH entry is an inequality because its residual makers need not commute,
so its variance is a full sandwich (Appendix~\ref{sec:cf-efficiency}). Under
heteroskedasticity the ordering can fail.

The factor estimator removes only the outcome directions. As
\[
 A_bZB_b=M_{\Lambda_b}(M_{U_{R,b}}ZM_{V_{R,b}})M_{F_b}
\]
and residual makers are contractions, it retains weakly more regressor variation than CUH, but it
needs strong factors. Removing the covariate projections is the price CUH pays
for validity under weaker factors.

\begin{remark}[The post-purge floor]\label{rem:ppfloor}
Each estimator buys bias protection by removing regressor variation, and the
post-purge floor estimates the share that survives,
\[
  \wh{\mathcal F}
  :=\frac{\lambda_{\min}\big(\wh Q\big)}
         {\lambda_{\min}\big(\tfrac1{NT}\textstyle\sum_{it}x_{it}x_{it}'\big)}
  \qquad\Big(=\frac{\|\wh AX\wh B\|_F^2}{\|X\|_F^2}
  \ \text{ for scalar } x_{it}\Big).
\]
A value near zero means all identifying variation has been purged, which the
CCE rank condition assumes away. Section~\ref{sec:apps}
reports floors for the empirical applications.
\end{remark}

{
\section{Monte Carlo evidence}\label{sec:sims}

\subsection{Design}\label{sec:sim-design}

The design follows \eqref{eq:y}--\eqref{eq:x}. The common effects
$\alpha_i,\gamma_t\in\R$ and the $x$-only effects
$\wt\alpha_i,\wt\gamma_t\in\R^3$ are i.i.d.\ standard Gaussian. The common
block $g_x(\alpha_i,\gamma_t)=\alpha_i\gamma_t$ has rank one and the $x$-only
block $\wt g_x(\wt\alpha_i,\wt\gamma_t)=\wt\alpha_i'\wt\gamma_t$ rank three,
each two-way demeaned and normalised to unit sample variance. The covariate
and outcome are
\begin{align*}
  x_{it}&=g_x(\alpha_i,\gamma_t)
  +\big\{\wt g_x(\wt\alpha_i,\wt\gamma_t)+\eta_{it}/\sqrt{NT}\big\}
  \big/\mathrm{sd}\big(\wt g_x+\eta/\sqrt{NT}\big),\\
  y_{it}&=x_{it}\beta_0+ c_{NT}\, g_x(\alpha_i,\gamma_t)+\varepsilon_{it},
\end{align*}
with $\beta_0=1$, $\varepsilon_{it},\eta_{it}$ i.i.d.\ standard Gaussian, and
common-factor strength $c_{NT}=(N\mw T)^{-w}$ in $y_{it}$, so $w=0$ gives
strong factors and $w=1/2$ the detection boundary of \eqref{eq:SNR}. Both
$x_{it}$ and $y_{it}$ are two-way demeaned before estimation. Scaling
$\eta_{it}$ by $1/\sqrt{NT}$ makes $\Sigma_\eta$ of
Proposition~\ref{prop:eff} negligible, so almost all identifying variation
lies in the $x$-only block. CUH and the factor estimator keep it, while CCEF
projects it out.

The factor estimator is the least-squares estimator of \citet{Bai2009} with
$R=1$. CCEF projects out the outcome direction and the leading four singular
directions of $X$, the rank of $g_x+\wt g_x$. CUH follows Algorithm~\ref{alg:cf} with $R=L=1$,
$K=2$ time halves and a gap $q_T=2$. NPCUH and BCUH are cross-fitted
likewise, with the bandwidths of Remark~\ref{rem:npcuh-bandwidth}. The
bias-aware intervals of
\citet{armstrong2026robust} take the form of Example~\ref{ex:ba}, with their
own standard errors and a worst-case bias bound that allows the factor to be
weak. Table~\ref{tab:mc} adds BM, the cross-fitted two-way grouped estimator
of \citet{beyhum2025inference}, which projects out unit and time group means
formed by k-means on sample averages and is applied before two-way
demeaning. Apart from the bias-aware intervals, standard errors use the
two-way variance estimator of \citet{CHS_Restat_2024} (CHS hereafter) of
Section~\ref{sec:cf-variance}, with the degrees-of-freedom correction of
Remark~\ref{rem:cf-dfc}.

Theorem~\ref{thm:cf} gives valid inference for CUH inside the inference frontier of
Remark~\ref{rem:strength-consequences}(iv). With $T=N^t$ and $\alpha^x_R=1$,
the calibration of Remark~\ref{rem:alpha} gives $\kappa=1/t$ and
$\alpha_R=1-2wt$, so \eqref{eq:inference-frontier} reads
\[
 w<\tfrac13-\tfrac1{6t},
\]
that is $w<0.03$, $0.10$, $0.14$ and $0.17$ at $t=0.55$, $0.7$, $0.85$ and
$1$.

\subsection{Design manifolds against the factor benchmark}
\label{sec:sim-manifold}

Figures~\ref{fig:headline}, \ref{fig:cmp-bias} and \ref{fig:cmp-sd} report
coverage, bias and standard deviation at $N=200$, with
$T=\mathrm{round}(N^{t})$ for $t\in\{0.55,0.7,0.85,1\}$ and
$w\in\{0,0.1,0.25,0.4,0.55,0.7,0.85\}$, and $400$ replications in each of
the $28$ cells. Blue marks performance better than the factor benchmark,
amber within noise and red worse.\footnote{Noise is an unpaired
two-standard-error band,
$1.96\{p_e(1-p_e)+p_f(1-p_f)\}^{1/2}/\sqrt{400}$ for coverage, with $p_e$ and
$p_f$ the coverages of the estimator and the benchmark,
$1.96({\rm sd}_e^2+{\rm sd}_f^2)^{1/2}/\sqrt{400}$ for bias, and the same
expression over $\sqrt{2\cdot399}$ for the standard deviation. The bands
ignore the positive covariance of the paired surfaces and so are
conservative. Dots give the verdict at each grid node, and facet shading
applies the same rule to the average of the four surrounding cells.}

Inside the inference frontier of Section~\ref{sec:sim-design}, CUH covers at
$0.935$ to $0.95$, and near the detection boundary its coverage dips to $0.77$ to
$0.84$ at $w=0.4$ for $T\ge41$. Beyond the boundary all three CUH variants
have smaller absolute bias than the factor benchmark
(Figure~\ref{fig:cmp-bias}) and similar or higher variance
(Figure~\ref{fig:cmp-sd}). BCUH is red nowhere on the coverage surface
(Figure~\ref{fig:headline}). NPCUH also covers near nominal at and beyond the
boundary, but over-covers in some strong-factor cells.

\begin{figure}[!ht]
\centering
\includegraphics[width=0.96\textwidth]{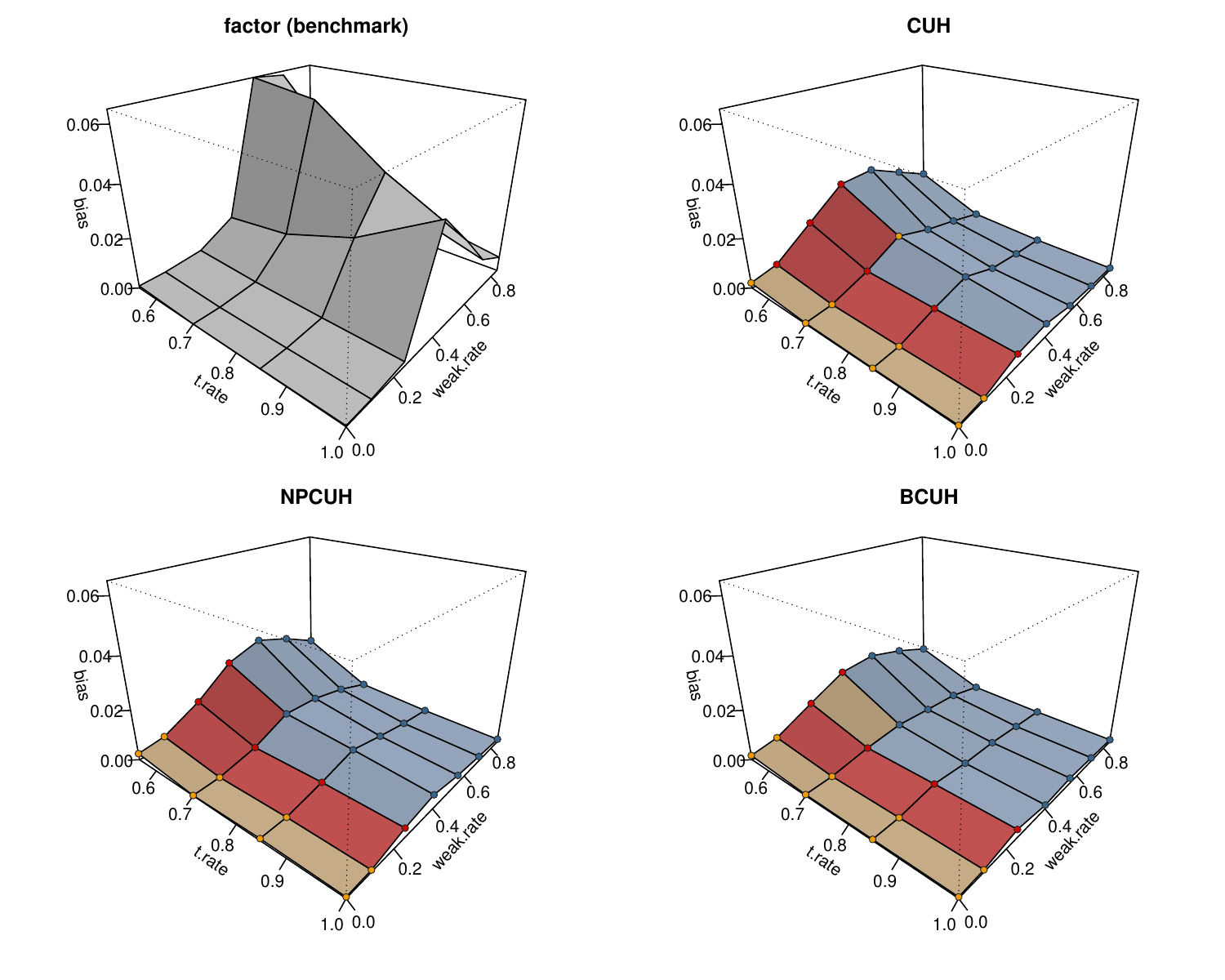}
\caption{Bias surfaces over the design manifold, coloured against the
factor benchmark (grey) as in Section~\ref{sec:sim-manifold}. The
benchmark's bias ridge tracks the detection boundary. CUH-class panels are
red on a band around $w=0.25$ and blue beyond the boundary.}
\label{fig:cmp-bias}
\end{figure}

\begin{figure}[!ht]
\centering
\includegraphics[width=0.96\textwidth]{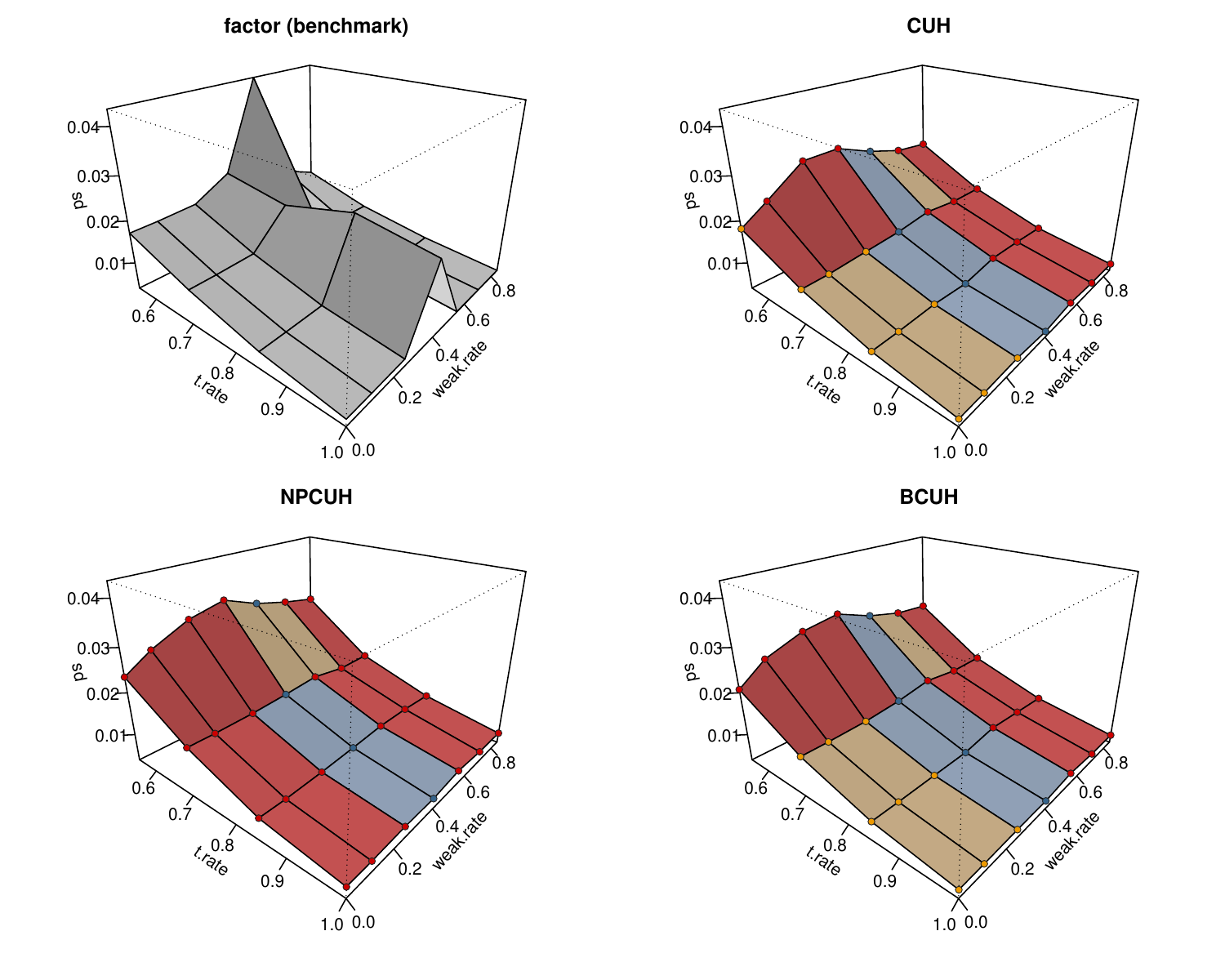}
\caption{Monte Carlo standard deviation surfaces, coloured as in
Figure~\ref{fig:cmp-bias}. NPCUH is red over most of the domain, and CUH and
BCUH are red beyond the detection boundary and along the shortest panel.}
\label{fig:cmp-sd}
\end{figure}

\subsection{Point comparison across sample sizes}\label{sec:sim-table}

Table~\ref{tab:mc} compares the estimators at three sample sizes. Under
strong factors CUH is almost as precise as the factor estimator, while CCEF
is uninformative by design. Under weak factors the factor estimator's
coverage falls from $0.07$ to $0.00$ as the sample grows, NPCUH and BCUH
cover at $0.93$ to $0.96$, CUH at $0.84$ to $0.87$, and the bias-aware
intervals at $1.00$ throughout. BM fails in both regimes. The common block is
two-way demeaned, so the sample averages on which BM forms its groups carry
no signal.

\begin{table}[!ht]
\centering
\caption{Bias, standard deviation and RMSE (all $\times100$) and coverage
of nominal $95\%$ intervals, with $T=\mathrm{round}(N^{0.7})$ and $2{,}000$
replications. The bias-aware row reports the bias, sd and RMSE of its point
estimate.}
\label{tab:mc}
\resizebox{\textwidth}{!}{
\begin{tabular}{l rrrr rrrr rrrr}
\hline\hline
 & \multicolumn{4}{c}{$(N,T)=(100,25)$} & \multicolumn{4}{c}{$(200,41)$} & \multicolumn{4}{c}{$(400,66)$}\\
 & bias & sd & rmse & cov & bias & sd & rmse & cov & bias & sd & rmse & cov\\ \hline
\multicolumn{13}{l}{\emph{Strong factors} ($w=0.00$)}\\
factor & 0.07 & 2.04 & 2.04 & 0.96 & 0.03 & 1.10 & 1.10 & 0.94 & 0.01 & 0.61 & 0.61 & 0.95\\
CCEF & 6.64 & 111.34 & 111.51 & 0.84 & 6.07 & 107.52 & 107.66 & 0.81 & 10.47 & 103.65 & 104.15 & 0.78\\
CUH & 0.22 & 2.12 & 2.13 & 0.96 & 0.07 & 1.12 & 1.12 & 0.93 & 0.02 & 0.62 & 0.62 & 0.95\\
NPCUH & 0.16 & 2.80 & 2.81 & 0.98 & 0.11 & 1.35 & 1.35 & 0.99 & 0.02 & 0.73 & 0.73 & 0.99\\
BCUH & 0.16 & 2.43 & 2.44 & 0.99 & 0.08 & 1.19 & 1.20 & 0.97 & 0.02 & 0.66 & 0.66 & 0.97\\
BiasAware & 0.06 & 4.72 & 4.72 & 1.00 & -0.15 & 3.25 & 3.25 & 1.00 & 0.06 & 2.18 & 2.18 & 1.00\\
BM & 44.97 & 9.55 & 45.97 & 0.00 & 45.16 & 8.91 & 46.03 & 0.00 & 45.24 & 8.53 & 46.04 & 0.00\\
\hline
\multicolumn{13}{l}{\emph{Weak factors} ($w=0.55$)}\\
factor & 7.77 & 2.54 & 8.17 & 0.07 & 6.11 & 1.30 & 6.24 & 0.02 & 4.85 & 0.62 & 4.89 & 0.00\\
CCEF & -0.70 & 94.69 & 94.66 & 0.78 & -1.07 & 88.67 & 88.66 & 0.69 & 2.98 & 83.61 & 83.64 & 0.61\\
CUH & 2.17 & 2.59 & 3.38 & 0.84 & 0.99 & 1.30 & 1.64 & 0.84 & 0.44 & 0.69 & 0.82 & 0.87\\
NPCUH & 2.31 & 3.27 & 4.00 & 0.94 & 1.17 & 1.55 & 1.94 & 0.95 & 0.53 & 0.80 & 0.96 & 0.96\\
BCUH & 1.60 & 2.96 & 3.36 & 0.94 & 0.63 & 1.43 & 1.56 & 0.93 & 0.23 & 0.73 & 0.77 & 0.94\\
BiasAware & 5.32 & 4.54 & 6.99 & 1.00 & 3.94 & 3.10 & 5.02 & 1.00 & 3.30 & 2.06 & 3.89 & 1.00\\
BM & 8.02 & 1.98 & 8.26 & 0.05 & 6.14 & 1.17 & 6.25 & 0.01 & 4.72 & 0.77 & 4.79 & 0.00\\
\hline
\hline
\end{tabular}
}
\end{table}

Table~\ref{tab:mc-re} uses a second design in which BM's sample averages
carry signal. The latent effects are i.i.d.\ uniform on $[0,1]$ without
demeaning, scaled to unit sample variance, and $x_{it}$ and $y_{it}$ each
carry an independent additive effect $a_i+b_t$, with $a_i,b_t$ i.i.d.\
$N(5,5^2)$. Two-way demeaning removes the means and the additive effects, so
for every estimator but BM only the latent distribution differs, and most of
their rows resemble Table~\ref{tab:mc}. BM now selects many more groups but
still fails, which suggests its groups track the additive effects rather than
the interactive confounder. Under weak factors it covers more often than the
factor estimator because its standard errors are larger.

\begin{table}[!ht]
\centering
\caption{As Table~\ref{tab:mc}, in the second design of
Section~\ref{sec:sim-table}, with uniform latent effects that are not
demeaned and two-way additive effects in $x_{it}$ and $y_{it}$.}
\label{tab:mc-re}
\resizebox{\textwidth}{!}{
\begin{tabular}{l rrrr rrrr rrrr}
\hline\hline
 & \multicolumn{4}{c}{$(N,T)=(100,25)$} & \multicolumn{4}{c}{$(200,41)$} & \multicolumn{4}{c}{$(400,66)$}\\
 & bias & sd & rmse & cov & bias & sd & rmse & cov & bias & sd & rmse & cov\\ \hline
\multicolumn{13}{l}{\emph{Strong factors} ($w=0.00$)}\\
factor & 0.04 & 2.09 & 2.09 & 0.90 & -0.03 & 1.11 & 1.11 & 0.94 & -0.00 & 0.63 & 0.63 & 0.94\\
CCEF & 6.24 & 115.08 & 115.22 & 0.84 & 11.47 & 100.59 & 101.22 & 0.81 & 15.94 & 101.12 & 102.34 & 0.79\\
CUH & 0.17 & 2.13 & 2.14 & 0.93 & 0.02 & 1.12 & 1.12 & 0.96 & 0.01 & 0.64 & 0.64 & 0.94\\
NPCUH & 0.17 & 2.64 & 2.65 & 0.98 & 0.01 & 1.29 & 1.29 & 0.99 & 0.01 & 0.70 & 0.70 & 0.98\\
BCUH & 0.16 & 2.34 & 2.35 & 0.97 & 0.01 & 1.18 & 1.18 & 0.98 & 0.00 & 0.66 & 0.66 & 0.97\\
BiasAware & -0.05 & 4.94 & 4.94 & 1.00 & 0.05 & 3.15 & 3.15 & 1.00 & 0.09 & 2.17 & 2.17 & 1.00\\
BM & 42.12 & 9.95 & 43.28 & 0.07 & 41.93 & 7.72 & 42.64 & 0.01 & 42.01 & 6.26 & 42.47 & 0.00\\
\hline
\multicolumn{13}{l}{\emph{Weak factors} ($w=0.55$)}\\
factor & 7.79 & 2.52 & 8.18 & 0.06 & 6.18 & 1.21 & 6.30 & 0.01 & 4.84 & 0.66 & 4.89 & 0.00\\
CCEF & -2.41 & 99.11 & 99.11 & 0.78 & 1.35 & 81.19 & 81.18 & 0.68 & 5.04 & 82.00 & 82.13 & 0.63\\
CUH & 1.86 & 2.50 & 3.12 & 0.84 & 0.80 & 1.28 & 1.51 & 0.89 & 0.38 & 0.69 & 0.79 & 0.89\\
NPCUH & 2.03 & 2.94 & 3.58 & 0.95 & 0.93 & 1.51 & 1.77 & 0.95 & 0.47 & 0.78 & 0.91 & 0.95\\
BCUH & 1.34 & 2.77 & 3.07 & 0.94 & 0.46 & 1.39 & 1.46 & 0.95 & 0.19 & 0.73 & 0.76 & 0.94\\
BiasAware & 5.12 & 4.75 & 6.98 & 1.00 & 4.19 & 2.98 & 5.14 & 1.00 & 3.34 & 2.01 & 3.90 & 1.00\\
BM & 7.95 & 3.08 & 8.52 & 0.58 & 6.01 & 1.75 & 6.26 & 0.32 & 4.64 & 1.03 & 4.75 & 0.11\\
\hline
\hline
\end{tabular}
}
\end{table}

\subsection{The price of robustness in power}

Figure~\ref{fig:power} plots the rate at which each interval excludes
$\beta_0+\xi$, $\xi\in[-0.3,0.3]$, for data generated at $\beta_0$, with
$400$ replications per panel. The height at $\xi=0$ is size and elsewhere
power. The bias-aware intervals hold size in every panel but leave a wide
non-rejection region. CCEF neither holds size nor discriminates.

CUH matches the factor benchmark under strong factors. At $w=0.4$, near the
detection boundary, CUH rejects the truth with probability $0.22$ while BCUH
stays near nominal size. At $w=0.7$ the factor benchmark's non-rejection
region moves off the truth, while CUH and BCUH stay close to it, rejecting
the truth at rates near $0.10$ and $0.06$. NPCUH rejects the truth at or
below nominal size in all three panels, with a wider non-rejection region.
Near and beyond the detection boundary the smoothing variants are therefore
the safer choice.

\begin{figure}[!ht]
\centering
\includegraphics[width=0.32\linewidth]{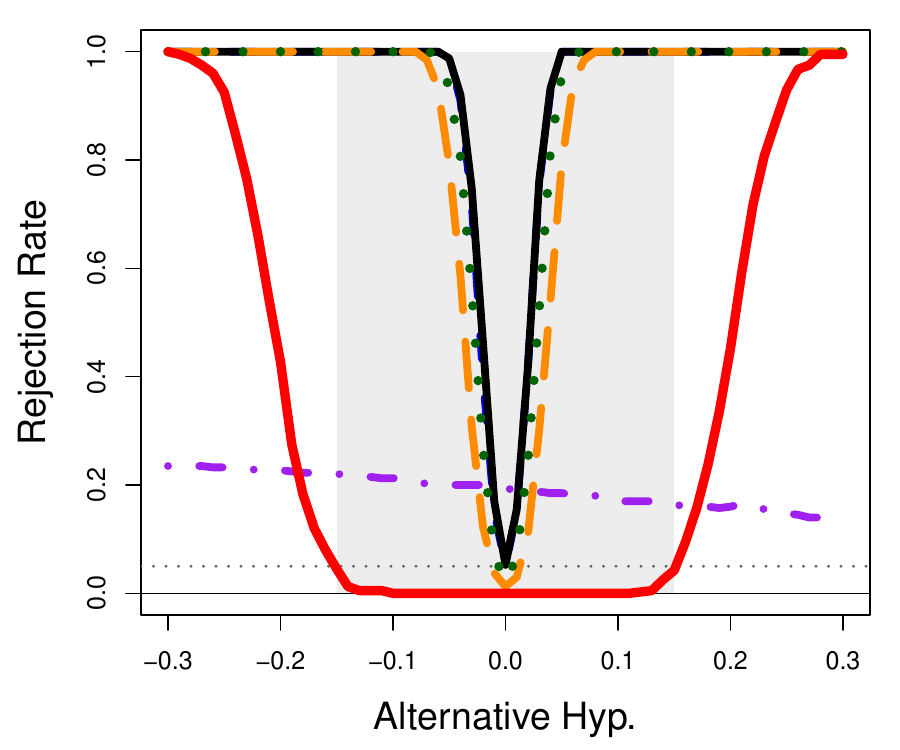}
\includegraphics[width=0.32\linewidth]{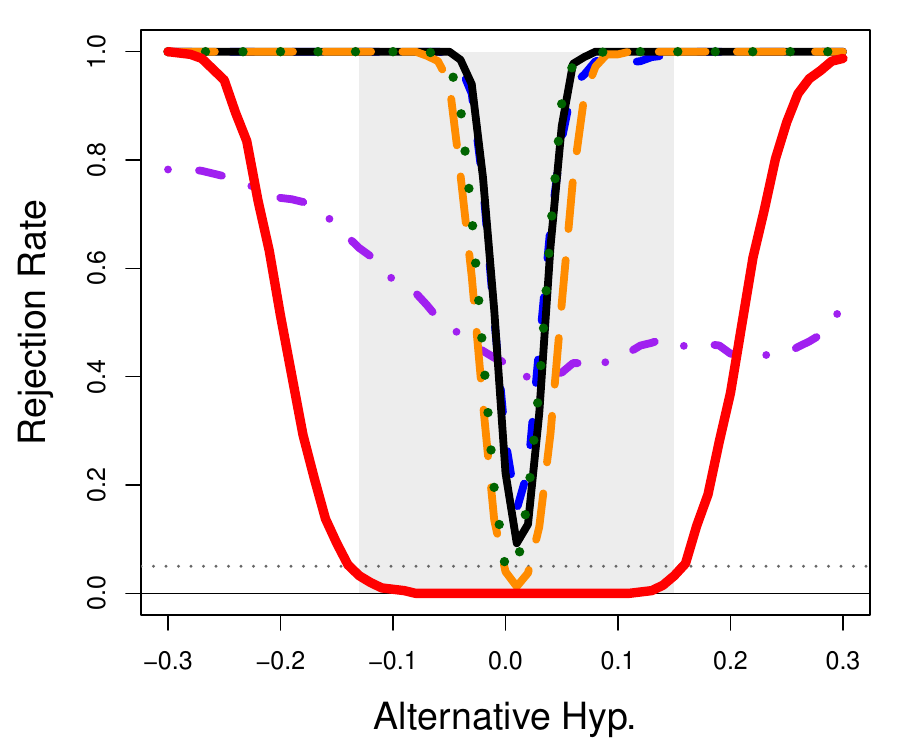}
\includegraphics[width=0.32\linewidth]{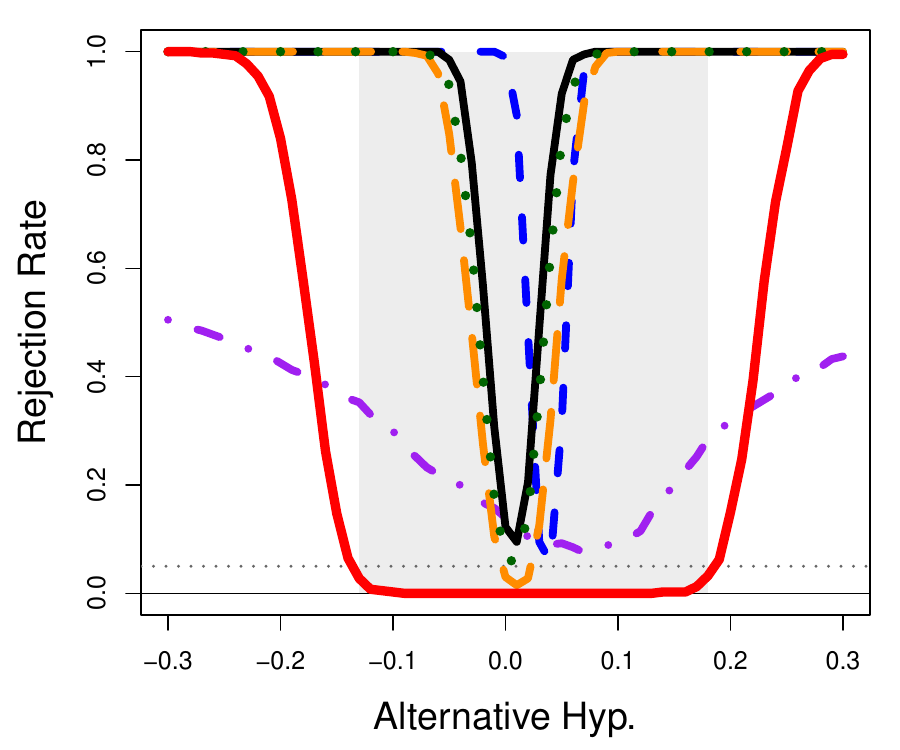}
\caption{Rejection rates at $\beta_0+\xi$, with data generated at $\beta_0$
and $(N,T)=(200,41)$, for $w=0$, $0.4$ and $0.7$. Blue dashed factor, purple
dot-dash CCEF, black solid CUH, orange long-dash NPCUH, green dotted BCUH,
red solid bias-aware, grey band the bias-aware non-rejection region, dotted
line $0.05$.}
\label{fig:power}
\end{figure}
}

{
\section{Empirical applications}\label{sec:apps}

We apply the estimators to the state fiscal panel of \citet{james2015}, also
used by \citet{beyhum2025inference}, and to the state cigarette demand panel
of \citet{baltagilevin1992} and \citet{baltagigriffinxiong2000}, the standard
testbed for interactive fixed effects estimators \citep{badaliebl2014}.
Standard errors, cross-fitting and bandwidths are as in
Section~\ref{sec:sims}, and the table notes give the remaining settings.

\subsection{State fiscal policy and resource revenue}\label{sec:app-fiscal}

\citet{james2015} asks how state governments absorb resource revenue. The
panel covers $N=50$ states over $T=51$ years (1958 to 2008). The regressor is
resource revenue and the five outcomes are non-resource revenue, income tax
revenue, total expenditure, education expenditure and savings, all as shares
of personal income.
\citet{beyhum2025inference} report their grouped estimator alongside two-way
fixed effects, the factor estimator with seven factors and CCE, and these
disagree by an order of magnitude on total expenditure. The post-purge floor
explains why.

After two-way demeaning, a single left singular vector accounts for
ninety-seven percent of the regressor variation, so any residualisation
against an estimated direction leaves only a few percent of it, the
near-zero case of Remark~\ref{rem:ppfloor}. Table~\ref{tab:fiscal} reports
each estimate with its post-purge floor. On four of the five outcomes every
purge-based estimator keeps at most eight percent of the regressor
variation, and NPCUH and BM keep under one percent. On education expenditure
the outcome directions are less aligned with resource revenue, so the
outcome-side purges keep more regressor variation, which the covariate-side
CCEF and BM purges remove.

On total expenditure BM reports a large positive effect, $0.53$ with
standard error $0.26$, while CUH, NPCUH and BCUH lie between $0.09$ and
$0.17$ with standard errors between $0.06$ and $0.09$. On savings the cross-fitted estimators lie between $0.38$ and
$0.61$, while the factor estimator is smaller and sensitive to $R$. At a
floor this low each estimator rests on a different thin slice of variation,
so choosing an estimator is choosing which covariate variation remains.

\begin{table}[!ht]
\centering\scriptsize
\caption{State fiscal policy and resource revenue, one outcome per column, $N=50$, $T=51$}
\label{tab:fiscal}
\setlength{\tabcolsep}{4pt}
\begin{tabular}{l ccccc}
\hline\hline
 & non-resource rev. & income tax & total exp. & education exp. & savings\\
\hline
2WFE & $+0.006$ (0.020) & $+0.018$ (0.010) & $+0.397$ (0.031) & $+0.063$ (0.008) & $+0.609$ (0.040)\\
 & [1.000] & [1.000] & [1.000] & [1.000] & [1.000]\\
factor $R=2$ & $-0.346$ (0.123) & $-0.063$ (0.048) & $-0.032$ (0.060) & $+0.004$ (0.019) & $+0.279$ (0.077)\\
 & [0.061] & [0.053] & [0.078] & [0.217] & [0.030]\\
factor $R=3$ & $-0.335$ (0.133) & $-0.034$ (0.036) & $-0.011$ (0.058) & $+0.012$ (0.018) & $+0.165$ (0.071)\\
 & [0.042] & [0.030] & [0.071] & [0.172] & [0.024]\\
CCEF $R=2$ & $-0.376$ (0.132) & $-0.081$ (0.054) & $+0.121$ (0.175) & $+0.060$ (0.043) & $+0.385$ (0.155)\\
 & [0.023] & [0.020] & [0.022] & [0.021] & [0.016]\\
CUH $R=2$ & $-0.107$ (0.123) & $+0.072$ (0.059) & $+0.174$ (0.064) & $+0.031$ (0.013) & $+0.470$ (0.161)\\
 & [0.031] & [0.070] & [0.058] & [0.090] & [0.020]\\
CUH $R=3$ & $-0.075$ (0.117) & $+0.048$ (0.021) & $+0.113$ (0.068) & $+0.033$ (0.020) & $+0.610$ (0.133)\\
 & [0.031] & [0.030] & [0.035] & [0.114] & [0.020]\\
NPCUH $R=3$ & $-0.275$ (0.206) & $-0.033$ (0.042) & $+0.095$ (0.070) & $+0.020$ (0.038) & $+0.380$ (0.105)\\
 & [0.002] & [0.002] & [0.003] & [0.011] & [0.003]\\
BCUH $R=3$ & $-0.074$ (0.130) & $+0.059$ (0.026) & $+0.085$ (0.085) & $+0.027$ (0.026) & $+0.610$ (0.180)\\
 & [0.021] & [0.023] & [0.026] & [0.046] & [0.014]\\
BM (CHS s.e.) & $-0.478$ (0.161) & $-0.035$ (0.045) & $+0.527$ (0.264) & $-0.041$ (0.083) & $+0.587$ (0.221)\\
 & [0.004] & [0.006] & [0.003] & [0.006] & [0.004]\\
BM, as published & $-0.478$ (0.143) & $-0.035$ (0.048) & $+0.527$ (0.284) & $-0.041$ (0.079) & $+0.587$ (0.228)\\
\hline\hline
\end{tabular}
\begin{minipage}{0.92\textwidth}\vspace{2pt}\footnotesize
Notes: variables are standardised for estimation, with coefficients rescaled as in \citet{beyhum2025inference}. Cells report the estimate (CHS s.e.) and, in brackets, the post-purge floor $\wh{\mathcal F}$ of Remark~\ref{rem:ppfloor}. CCEF purges one covariate principal component and CUH uses one covariate projection. BM uses $2$ to $17$ unit groups and $1$ to $5$ time groups per quadrant, chosen by the authors' rule, and is reconstructed from its k-means labels so that its standard errors use the CHS variance. The last row reproduces the published estimates and Arellano standard errors of \citet[Table 3]{beyhum2025inference}, which the reconstruction matches to three decimals.
\end{minipage}
\end{table}
\subsection{Cigarette demand}\label{sec:app-cigar}

\begin{table}[!ht]
\centering\small
\caption{Cigarette demand elasticities, $N=46$, $T=30$}
\label{tab:cigar}
\begin{tabular}{l ll c}
\hline\hline
 & log price & log income & $\wh{\mathcal F}$\\
\hline
2WFE & $-1.02$ (0.22) [$-1.45$, $-0.60$] & $+0.52$ (0.18) [$+0.17$, $+0.87$] & 1.00\\
factor $R=4$ & $-0.38$ (0.04) [$-0.46$, $-0.31$] & $+0.36$ (0.11) [$+0.15$, $+0.57$] & 0.29\\
CCEF $R=1$ & $-0.40$ (0.05) [$-0.50$, $-0.30$] & $+0.37$ (0.15) [$+0.08$, $+0.66$] & 0.24\\
CUH $R=3$ & $-0.41$ (0.06) [$-0.53$, $-0.28$] & $+0.36$ (0.14) [$+0.09$, $+0.64$] & 0.20\\
NPCUH $R=2$ ($h=0.54$) & $-0.48$ (0.13) [$-0.74$, $-0.23$] & $+0.27$ (0.14) [$-0.00$, $+0.55$] & 0.04\\
BCUH $R=1$ & $-0.41$ (0.11) [$-0.63$, $-0.20$] & $+0.27$ (0.16) [$-0.04$, $+0.57$] & 0.21\\
BM & $-0.55$ (0.07) [$-0.69$, $-0.41$] & $+0.19$ (0.16) [$-0.13$, $+0.50$] & 0.13\\
\hline\hline
\end{tabular}
\begin{minipage}{0.92\textwidth}\vspace{2pt}\footnotesize
Notes: estimate (CHS s.e.) [95\% CI]. $\wh{\mathcal F}$ is the post-purge floor of Remark~\ref{rem:ppfloor}. The small, insignificant neighbouring-price coefficient is omitted. Estimators that depend on $R$ are shown at the $R\le4$ giving the smallest standard error on log price, and their estimates move little across $R$. CCEF purges $R_X=3$ covariate principal components, matching the three covariate projections of CUH with $L=1$. BM uses groups $(9,3)$, $(12,3)$, $(15,5)$ and $(15,6)$ in the four quadrants, and its Arellano standard errors are $0.07$ and $0.13$.
\end{minipage}
\end{table}

}

The panel covers $N=46$ states over $T=30$ years (1963 to 1992). We
regress log per-capita sales on log real price, log real disposable income
and log real minimum price in neighbouring states, all two-way demeaned.
Every purge-based estimator other than BM puts the price elasticity between
$-0.38$ and $-0.48$ and the income elasticity between $0.27$ and $0.37$,
against $-1.02$ and $0.52$ for two-way fixed effects (Table~\ref{tab:cigar}).
The projection estimators keep post-purge floors of $0.20$ to $0.29$, and
Section~\ref{sec:app-overlap} shows that they draw on largely shared
variation. BM is the exception, at $-0.55$ on price and an insignificant
$0.19$ on income, with a floor of $0.13$.

\begin{figure}[!ht]
\centering
\includegraphics[width=0.9\textwidth]{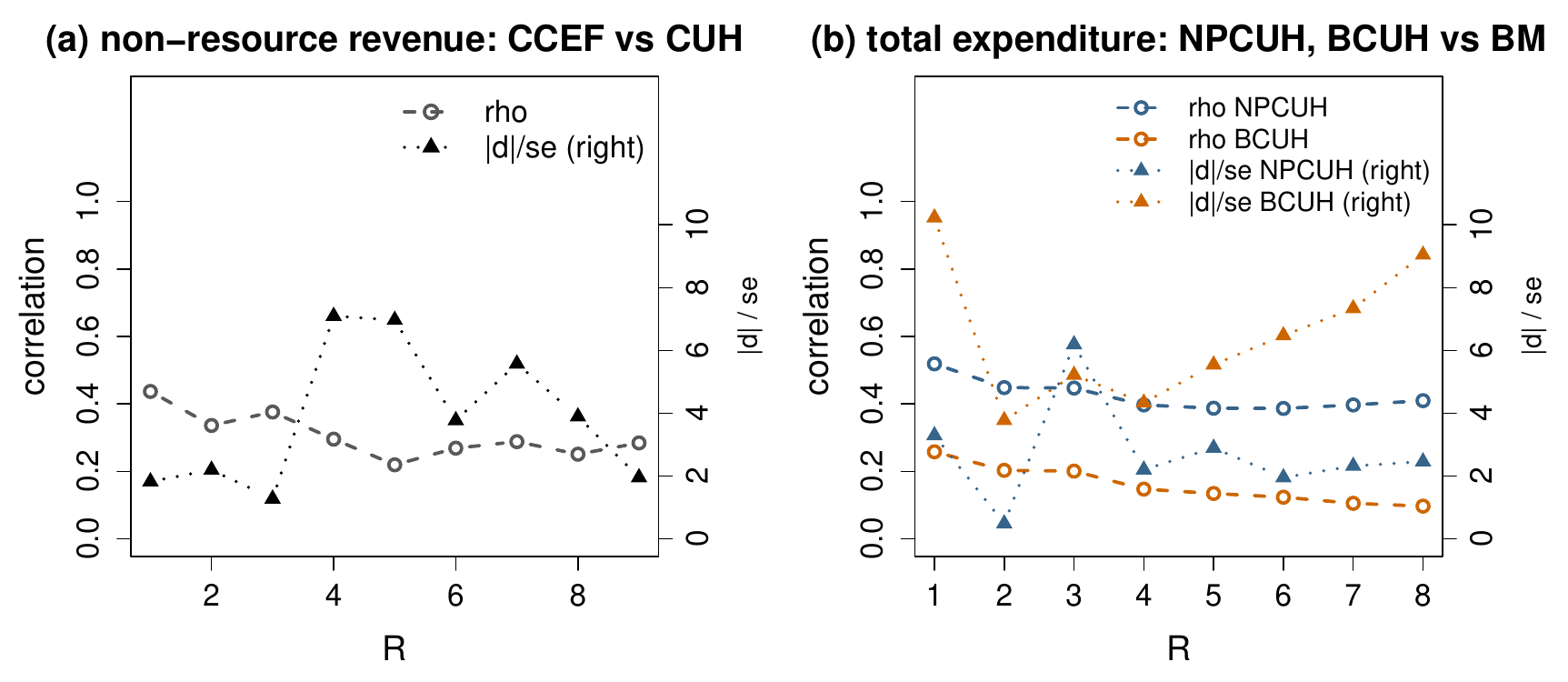}
\caption{Shared identifying variation across the number of purged outcome
directions $R$. Panel (a) compares CCEF with CUH on non-resource revenue,
panel (b) NPCUH and BCUH with BM on total expenditure.}
\label{fig:apps-diag}
\end{figure}

{
\subsection{Shared identifying variation}\label{sec:app-overlap}

When two purge-based estimators disagree, their residualised regressors show
which identifying variation they share. For each $R$,
Table~\ref{tab:overlap} reports the post-purge floors of the factor
estimator, CCEF and CUH, the correlation $\rho$ of CUH's residualised
regressor with that of each of the other two, and the estimate gap
$|d|/{\rm se}$ in units of CUH's standard error. A large gap at high $\rho$
signals conflict on shared variation, and a large gap at low $\rho$ signals
that the two estimators purged different variation.
Figure~\ref{fig:apps-diag} plots both diagnostics.

On cigarette price $\rho$ lies between $0.61$ and $0.84$ and no gap exceeds
$1.7$ standard errors, so the agreement in Table~\ref{tab:cigar} is evidence
of shared identifying variation. On the fiscal outcomes $\rho$ is at most
$0.48$, and the two largest gaps, $12.0$ and $11.3$ standard errors on total
expenditure, come with correlations of $0.12$ and $0.13$, which is evidence
that the estimators disagree because they use different variation.

Figure~\ref{fig:intro-apps}(b) fixes $R=5$ and raises the number of CCEF and
CUH covariate components from one to nine. CUH's estimate stays near
$-0.03$, because the regressor is almost rank one and the frame truncation
of Algorithm~\ref{alg:cf} detects this and keeps the variation. CCEF purges
it indiscriminately, with estimates ranging from $-0.53$ to $0.25$ as its
post-purge floor falls from $0.008$ to $0.0005$.

Table~\ref{tab:overlap-bm} compares the smoothing variants with BM. The
correlations stay below $0.6$ throughout, so BM and the smoothing variants
keep different variation. The gaps stay within $1.4$ standard errors on cigarette
price but reach $6.2$ for NPCUH and $10.2$ for BCUH on total expenditure,
where BM stood apart in Table~\ref{tab:fiscal}.

\begin{table}[!ht]
\centering\footnotesize
\caption{Shared identifying variation and estimate divergence}
\label{tab:overlap}
\setlength{\tabcolsep}{5pt}
\begin{tabular}{l ccc c cc c cc}
\hline\hline
 & \multicolumn{3}{c}{post-purge floor $\wh{\mathcal F}$} &
 & \multicolumn{2}{c}{$\rho$ with CUH} &
 & \multicolumn{2}{c}{$|d|/{\rm se}$ against CUH}\\
\cline{2-4}\cline{6-7}\cline{9-10}
$R$ & factor & CCEF & CUH && factor & CCEF && factor & CCEF\\
\hline
\multicolumn{10}{l}{\emph{Cigarette, log price}}\\
$1$ & 0.74 & 0.35 & 0.47 && 0.78 & 0.78 && 1.7 & 1.3\\
$2$ & 0.50 & 0.33 & 0.35 && 0.80 & 0.84 && 1.1 & 0.5\\
$3$ & 0.40 & 0.33 & 0.27 && 0.77 & 0.77 && 0.3 & 0.4\\
$4$ & 0.35 & 0.29 & 0.20 && 0.68 & 0.70 && 0.2 & 0.2\\
$5$ & 0.30 & 0.26 & 0.15 && 0.64 & 0.65 && 0.1 & 0.2\\
$6$ & 0.24 & 0.21 & 0.13 && 0.61 & 0.62 && 0.4 & 0.5\\\hline
\multicolumn{10}{l}{\emph{Fiscal, non-resource revenue}}\\
$1$ & 0.197 & 0.033 & 0.043 && 0.38 & 0.44 && 0.9 & 1.8\\
$2$ & 0.061 & 0.023 & 0.031 && 0.33 & 0.34 && 2.0 & 2.2\\
$3$ & 0.042 & 0.014 & 0.031 && 0.32 & 0.38 && 2.2 & 1.3\\
$4$ & 0.052 & 0.009 & 0.023 && 0.29 & 0.30 && 0.6 & 7.1\\
$5$ & 0.045 & 0.008 & 0.022 && 0.37 & 0.22 && 0.9 & 7.0\\
$6$ & 0.032 & 0.007 & 0.008 && 0.22 & 0.27 && 0.1 & 3.8\\\hline
\multicolumn{10}{l}{\emph{Fiscal, total expenditure}}\\
$1$ & 0.156 & 0.024 & 0.056 && 0.48 & 0.37 && 1.1 & 5.0\\
$2$ & 0.078 & 0.022 & 0.058 && 0.40 & 0.32 && 3.2 & 0.8\\
$3$ & 0.071 & 0.013 & 0.035 && 0.29 & 0.34 && 1.8 & 2.8\\
$4$ & 0.040 & 0.014 & 0.042 && 0.30 & 0.21 && 1.4 & 0.3\\
$5$ & 0.038 & 0.009 & 0.020 && 0.26 & 0.13 && 0.7 & 2.9\\
$6$ & 0.008 & 0.005 & 0.013 && 0.12 & 0.13 && 12.0 & 11.3\\
\hline\hline
\end{tabular}
\begin{minipage}{0.92\textwidth}\vspace{2pt}\footnotesize
Notes: $\wh{\mathcal F}$ is the share of the named regressor's variation
surviving each purge at $R$ outcome directions, $\rho$ the correlation of the
residualised regressors of CUH and of $e$, the factor estimator or CCEF, and
$|d|/{\rm se}=|\wh\beta_{\rm CUH}-\wh\beta_e|$ over the CHS standard error
of CUH. CUH uses $L=1$, and CCEF $R_X=3$ for cigarette (log price) and
$R_X=1$ for fiscal, with fiscal outcomes standardised as in
Table~\ref{tab:fiscal}.
\end{minipage}
\end{table}

\begin{table}[!ht]
\centering\footnotesize
\caption{Shared identifying variation against the grouped estimator}
\label{tab:overlap-bm}
\setlength{\tabcolsep}{3.4pt}
\begin{tabular}{l ccc c ccc}
\hline\hline
 & \multicolumn{7}{c}{(a) Cigarette, log price}\\
\cline{2-8}
 & \multicolumn{3}{c}{NPCUH} && \multicolumn{3}{c}{BCUH}\\
\cline{2-4}\cline{6-8}
$R$ & $\wh{\mathcal F}$ & $\rho$ & $|d|/{\rm se}$ && $\wh{\mathcal F}$ & $\rho$ & $|d|/{\rm se}$\\
\hline
$1$ & 0.11 & 0.46 & 0.5 && 0.25 & 0.53 & 1.3\\
$2$ & 0.07 & 0.56 & 0.6 && 0.20 & 0.58 & 1.4\\
$3$ & 0.05 & 0.52 & 0.8 && 0.16 & 0.56 & 1.0\\
$4$ & 0.04 & 0.46 & 1.0 && 0.12 & 0.51 & 0.7\\
$5$ & 0.02 & 0.42 & 1.2 && 0.09 & 0.43 & 0.5\\
$6$ & 0.02 & 0.37 & 1.2 && 0.07 & 0.39 & 0.4\\
\hline\hline
\end{tabular}

\vspace{10pt}
\begin{tabular}{l ccc c ccc c ccc c ccc}
\hline\hline
 & \multicolumn{7}{c}{(b) Fiscal, income tax} &
 & \multicolumn{7}{c}{(c) Fiscal, total expenditure}\\
\cline{2-8}\cline{10-16}
 & \multicolumn{3}{c}{NPCUH} && \multicolumn{3}{c}{BCUH}
 && \multicolumn{3}{c}{NPCUH} && \multicolumn{3}{c}{BCUH}\\
\cline{2-4}\cline{6-8}\cline{10-12}\cline{14-16}
$R$ & $\wh{\mathcal F}$ & $\rho$ & $|d|/{\rm se}$ && $\wh{\mathcal F}$ & $\rho$ & $|d|/{\rm se}$ && $\wh{\mathcal F}$ & $\rho$ & $|d|/{\rm se}$ && $\wh{\mathcal F}$ & $\rho$ & $|d|/{\rm se}$\\
\hline
$1$ & 0.001 & 0.13 & 0.4 && 0.005 & 0.12 & 0.8 && 0.005 & 0.52 & 3.3 && 0.041 & 0.26 & 10.2\\
$2$ & 0.003 & 0.26 & 0.3 && 0.048 & 0.03 & 1.6 && 0.004 & 0.45 & 0.5 && 0.041 & 0.20 & 3.8\\
$3$ & 0.002 & 0.26 & 0.0 && 0.023 & 0.05 & 3.6 && 0.003 & 0.45 & 6.2 && 0.026 & 0.20 & 5.2\\
$4$ & 0.002 & 0.26 & 0.2 && 0.008 & 0.11 & 2.4 && 0.002 & 0.40 & 2.2 && 0.028 & 0.15 & 4.3\\
$5$ & 0.002 & 0.24 & 0.3 && 0.006 & 0.13 & 0.4 && 0.002 & 0.39 & 2.9 && 0.016 & 0.13 & 5.5\\
$6$ & 0.001 & 0.22 & 0.1 && 0.005 & 0.12 & 0.5 && 0.001 & 0.39 & 2.0 && 0.011 & 0.12 & 6.5\\
\hline\hline
\end{tabular}
\begin{minipage}{0.96\textwidth}\vspace{2pt}\footnotesize
Notes: as Table~\ref{tab:overlap}, with NPCUH or BCUH in place of CUH and BM
in place of $e$. BM does not vary with $R$. Its floor and
estimate (CHS s.e.) are $0.18$ and $-0.553$ $(0.071)$ for cigarette,
$0.006$ and $-0.035$ $(0.045)$ for income tax, and $0.003$ and $+0.527$
$(0.264)$ for total expenditure.
\end{minipage}
\end{table}
}

\section{Conclusion}

Bias in linear panels with two-way unobserved heterogeneity comes only from
the common block that the outcome shares with the covariates. CUH removes
it through the outcome directions and the covariate projections and keeps the
$x$-only variation. Under our conditions CUH is $\sqrt{NT}$-asymptotically
normal with weak factors, allowing $N$ and $T$ to grow at different rates,
and under homoskedastic errors its asymptotic variance is no larger than
those of FK and CCEF. In simulations it matches the factor estimator under strong factors,
and beyond the detection boundary, where the factor estimator's coverage
collapses, NPCUH and BCUH hold coverage near nominal at the cost of higher
variance. In the applications the post-purge floor and the shared-variation
diagnostics show when the estimators agree and why they diverge. We recommend
reporting this floor with any purge-based estimator.

\setlength{\bibsep}{2pt}
\bibliographystyle{chicago3}
\bibliography{refs}
\appendix

\titleformat{\section}[block]
  {\normalfont\Large\bfseries}
  {\appendixname~\thesection:}
  {0.5em}
  {}

\numberwithin{theorem}{section}
\numberwithin{proposition}{section}
\numberwithin{remark}{section}
\numberwithin{lemma}{section}
\numberwithin{corollary}{section}
\numberwithin{definition}{section}
\numberwithin{example}{section}
\numberwithin{assumption}{section}

\section{Proofs for the weak-factor theory}\label{app:weak-proofs}

\subsection{Noise edge}

\begin{lemma}[Noise edge]\label{lem:edge}
Suppose the idiosyncratic array is separable,
$E=R_N^{1/2}\,\mathcal E_0\,R_T^{1/2}$, where $\mathcal E_0=(e_{it})$ has
independent, mean-zero, unit-variance entries with $\sup_{i,t}\E e_{it}^4<\infty$,
and $R_N$ ($N\times N$) and $R_T$ ($T\times T$) are symmetric positive
semidefinite with $\norm{R_N}_{\rm op}=O(1)$ and $\norm{R_T}_{\rm op}=O(1)$. Then
$\norm{E}_{\rm op}=\Op(\sqrt{\max\{N,T\}})$, so that
\[
  s_{NT}:=\frac{\norm{E}_{\rm op}}{\sqrt{NT}}
  =\Op\!\Big(\sqrt{\tfrac{\max\{N,T\}}{NT}}\Big)
  =\Op\big((N\mw T)^{-1/2}\big).
\]
\end{lemma}

\begin{remark}[Weak dependence]\label{rem:edge-dep}
The separable structure $E=R_N^{1/2}\mathcal E_0R_T^{1/2}$ is the linear-process
error model of \citet{onatski2012}, and its two operator-norm conditions are
long-run-variance conditions in operator form. For a covariance-stationary
series $\norm{R_T}_{\rm op}\to 2\pi\sup_\omega f(\omega)$, so
$\norm{R_T}_{\rm op}=O(1)$ if and only if the spectral density $f$ is bounded,
in particular under absolutely summable autocovariances. Symmetrically,
$\norm{R_N}_{\rm op}=O(1)$ is weak cross-sectional dependence. A bounded
\emph{scalar} long-run variance for a single series is therefore necessary but
not sufficient. The largest eigenvalue of each covariance matrix must stay
bounded as $N,T\to\infty$.
\end{remark}

\begin{proof}[Proof of Lemma~\ref{lem:edge}]
The inequality of \citet{latala2005some} gives
$\E\norm{\mathcal E_0}_{\rm op}\lesssim\sqrt N+\sqrt T+(NT)^{1/4}\lesssim\sqrt{N\vee T}$,
so $\norm{\mathcal E_0}_{\rm op}=\Op(\sqrt{N\vee T})$ by Markov's inequality.
By submultiplicativity,
\[
  \norm{E}_{\rm op}\le\norm{R_N}_{\rm op}^{1/2}\norm{\mathcal E_0}_{\rm op}
  \norm{R_T}_{\rm op}^{1/2}=\Op(\sqrt{N\vee T}),
\]
and dividing by $\sqrt{NT}$ gives the rate of $s_{NT}$.
\end{proof}

\subsection{Subspace recovery}

\begin{lemma}[Singular-subspace perturbation]\label{lem:wedin}
Let $\mathcal M$ and $\wh{\mathcal M}=\mathcal M+\Delta$ be $N\times T$
matrices with SVDs under the statistical normalisation and
leading-$R$ singular-subspace projections
$P_{U_{R}(\mathcal M)},P_{V_{R}(\mathcal M)}$ and
$P_{U_{R}(\wh{\mathcal M})},P_{V_{R}(\wh{\mathcal M})}$. If
$\sigma_{R}(\mathcal M)-\sigma_{R+1}(\wh{\mathcal M})>\ell_{NT}>0$ with
probability approaching one, then
\[
  \max\Big\{
  \norm{P_{U_{R}(\wh{\mathcal M})}-P_{U_{R}(\mathcal M)}}_{\rm op},
  \norm{P_{V_{R}(\wh{\mathcal M})}-P_{V_{R}(\mathcal M)}}_{\rm op}
  \Big\}
  =\Op\!\Big(\frac{\norm{\Delta}_{\rm op}}{\sqrt{NT}\,\ell_{NT}}\Big).
\]
\end{lemma}

\begin{proof}[Proof of Lemma~\ref{lem:wedin}]
Projections are invariant to rescaling the frames, and singular values
under the conventional normalisation are $\sqrt{NT}$ times those under the
statistical normalisation. On the separation event, the $\sin\Theta$ theorem of \citet{Wedin1972}, applied to the unit-norm
leading frames $U_R^*,V_R^*$ of $\mathcal M$ with residuals $\Delta V_R^*$ and
$\Delta'U_R^*$ of operator norm at most $\norm{\Delta}_{\rm op}$ and
separation $\sqrt{NT}\,\ell_{NT}$, gives the bound.
\end{proof}

\begin{proof}[Proof of Proposition~\ref{prop:xi}]
Let $U_{R}^{(G)}$ and $V_{R}^{(G)}$ be the leading-$R$ left and right
singular frames of $G$.

Step 1 (from $W$ to the frames of $G$). As $W=G+E+\Delta_\beta$, Weyl's
inequality and the hypothesis
$\norm{E+\Delta_\beta}_{\rm op}/\sqrt{NT}=\op(\delta_{NT})$ give, with
probability approaching one,
$\sigma_R(G)-\sigma_{R+1}(W)\ge\delta_{NT}-\norm{E+\Delta_\beta}_{\rm op}/\sqrt{NT}>\tfrac12\delta_{NT}$.
Lemma~\ref{lem:wedin} with $\mathcal M=G$, $\Delta=E+\Delta_\beta$ and
$\ell_{NT}=\tfrac12\delta_{NT}\asymp c_{NT}$
(Assumption~\ref{ass:strength}(ii)), together with
$\norm{\Delta_\beta}_{\rm op}\le\norm{X}_{\rm op}\norm{\beta_0-\wt\beta}$,
gives
\begin{align*}
  \norm{P_{\wh U_{R}}-P_{U_{R}^{(G)}}}_{\rm op},\
  \norm{P_{\wh V_{R}}-P_{V_{R}^{(G)}}}_{\rm op}
  &=\Op\!\Big(\frac{\norm{E}_{\rm op}+\norm{\Delta_\beta}_{\rm op}}{\sqrt{NT}\,c_{NT}}\Big)\\
  &=\Op\!\Big(\frac{s_{NT}}{c_{NT}}\Big)
  +\Op\!\Big(\frac{\norm{X}_{\rm op}\norm{\beta_0-\wt\beta}}{\sqrt{NT}\,c_{NT}}\Big).
\end{align*}

Step 2 (from the frames of $G$ to the evaluated eigenfunctions). By
Assumption~\ref{ass:reg}(iii) the evaluated frames have sample Grams
$U_{R}'U_{R}/N=I_{R}+\Op(N^{-1/2})$ and
$V_{R}'V_{R}/T=I_{R}+\Op(T^{-1/2})$. Factor
$U_{R}/\sqrt N=Q(I_{R}+\Theta_A)^{1/2}$ and
$V_{R}/\sqrt T=P(I_{R}+\Theta_B)^{1/2}$
with exactly orthonormal $Q,P$, $\norm{\Theta_A}=\Op(N^{-1/2})$ and
$\norm{\Theta_B}=\Op(T^{-1/2})$. The leading-$R$ singular frames of $G$ are
then $QO_1,PO_2$ up to a head--tail coupling of order
$\Op((N\mw T)^{-1/2}\sigma_{R+1}/\delta_{NT})=\Op((N\mw T)^{-1/2})$ by
Assumption~\ref{ass:strength}(ii), where $O_1,O_2$ are the singular frames of
the $R\times R$ matrix
$K=(I_{R}+\Theta_A)^{1/2}D_{R}(I_{R}+\Theta_B)^{1/2}$ with
$D_{R}={\rm diag}(\sigma_1,\dots,\sigma_{R})$.
Perturbing this $R\times R$ problem under
Assumption~\ref{ass:strength}(iii) gives, after sign alignment,
\[
  \max_{r\in \mathcal{I}_{R}}\ \tfrac1N\,\norm{u_r^{(G)}-u_r}^2
  \ \vee\
  \norm{P_{U_{R}^{(G)}}-P_{U_{R}}}_{\rm op}^2
  =\Op\!\big(N^{-1}+T^{-1}\big),
\]
and symmetrically on the right. As
$c_{NT}\le\sigma_1\le\norm{g}_{L^2(\pi_\alpha\otimes\pi_\gamma)}=O(1)$ by
Assumption~\ref{ass:reg}(i)--(ii), this bridge term is at most a constant
multiple of the noise term in \eqref{eq:xi}.

Step 3 (per-direction errors). By Assumption~\ref{ass:strength}(ii)--(iii)
every gap between the leading $R$ singular values of $G$ is at least of order
$c\,\sigma_{R}\asymp c_{NT}$, and by Weyl's inequality the same holds for $W$
with probability approaching one. Step 1, applied to each
one-dimensional singular block, therefore bounds the sign-aligned per-vector
errors at the same rate. Combining Steps 1--3, squaring and substituting
$s_{NT}=\Op((N\mw T)^{-1/2})$ gives \eqref{eq:xi}.
\end{proof}

\subsection{Projections, energies and selection}

\begin{proof}[Proof of Theorem~\ref{thm:proj}]
Fix $j,r$ and sign-align $\wh v_r$ with $v_r:=(v_r(\gamma_t))_{t\le T}$. Then
$\wh\lambda_{j,r}-\lambda_{j,r}=T^{-1}X_j(\wh v_r-v_r)+(T^{-1}X_jv_r-\lambda_{j,r})$.
By Assumption~\ref{ass:reg}(iii) and $\norm{X_j}_{\rm op}\le\norm{X_j}_F$, the
first term has $\norm{\cdot}_N\le\norm{X_j}_{\rm op}\xi_v/\sqrt{NT}=\Op(\xi_v)$,
and by Assumption~\ref{ass:reg}(iv) the second is $\Op(T^{-1/2})$. Maximising over
the finitely many $j,r$ gives the first bound, and the second is symmetric.
\end{proof}

\begin{proof}[Proof of Proposition~\ref{prop:sigma}]
With $\Delta_i:=\wh\lambda_{j,r}(\alpha_i)-\lambda_{j,r}(\alpha_i)$,
\[
  \wh\sigma_{\alpha,j,r}^2-\sigma_{\alpha,j,r}^2
  =\underbrace{\tfrac1N\textstyle\sum_i\Delta_i^2}_{\Op(\xi_v^2+T^{-1})}
  +\underbrace{\big(\tfrac1N\textstyle\sum_i\lambda_{j,r}^2(\alpha_i)-\sigma_{\alpha,j,r}^2\big)}_{\Op(\sigma_{\alpha,j,r}^2N^{-1/2})}
  +\underbrace{\tfrac2N\textstyle\sum_i\Delta_i\lambda_{j,r}(\alpha_i)}_{\Op(\sigma_{\alpha,j,r}(\xi_v+T^{-1/2}))}.
\]
The three orders follow from Theorem~\ref{thm:proj},
Assumption~\ref{ass:reg}(iv) read as a relative error, and Cauchy--Schwarz with
$(\tfrac1N\sum_i\lambda_{j,r}^2(\alpha_i))^{1/2}=\sigma_{\alpha,j,r}(1+\op(1))$.
Collecting the dominant terms gives \eqref{eq:sigmahat_alpha}, and
\eqref{eq:sigmahat_gamma} is symmetric.
\end{proof}

\begin{proof}[Proof of Corollary~\ref{cor:select}]
By Assumption~\ref{ass:Xstrength} each candidate is either $x$-relevant, with
$\sigma_{\alpha,j,r}^2\ge\epsilon_{NT}^2$, or has
$\sigma_{\alpha,j,r}^2=\Op((N\mw T)^{-1})$, and
$\epsilon_{NT}^2\gg\xi_v^2+(N\mw T)^{-1}$. In the second case
Proposition~\ref{prop:sigma} and $2ab\le a^2+b^2$ give
$\wh\sigma_{\alpha,j,r}^2=\Op(\xi_v^2+(N\mw T)^{-1})=\op(\epsilon_{NT}^2)$. In
the first, dividing \eqref{eq:sigmahat_alpha} by
$\sigma_{\alpha,j,r}^2\ge\epsilon_{NT}^2$ gives
$\wh\sigma_{\alpha,j,r}^2=\sigma_{\alpha,j,r}^2\{1+\op(1)\}$. As $d$ and $R$
are fixed, with probability approaching one every estimated $x$-relevant energy
exceeds $\tfrac12\epsilon_{NT}^2$ and every other estimated energy falls below
it, so the ranking
selects exactly the $x$-relevant directions. The time side is symmetric.

On the selection event $\wh\Lambda$ and $\Lambda$ carry the same index set. Put
$\Delta:=\wh\Lambda-\Lambda$ and let $s$ be the smallest nonzero singular value
of $\Lambda$, at least $\sqrt{cN}\,\epsilon_{NT}$ by hypothesis.
Theorem~\ref{thm:proj} gives
$\norm{\Delta}_{\rm op}\le\norm{\Delta}_F=\sqrt{p_\alpha N}\,\Op(\xi_v+T^{-1/2})$.
A unit vector $a$ in the column span of $\Lambda$ is $a=\Lambda w$ with $w$ in
the row space and $\norm{w}\le s^{-1}$. As $M_{\wh\Lambda}\wh\Lambda=0$ and
$M_{\wh\Lambda}$ is a contraction,
$\norm{M_{\wh\Lambda}a}=\norm{M_{\wh\Lambda}(\Lambda-\wh\Lambda)w}
\le\norm{\Delta}_{\rm op}/s=\Op(\zeta_\Lambda)$, as $p_\alpha$ is fixed, and the
supremum over such $a$ gives \eqref{eq:select-angle}.

For \eqref{eq:select-energy-L} write
$P_{\wh\Lambda}-P_\Lambda=P_{\wh\Lambda}M_\Lambda-M_{\wh\Lambda}P_\Lambda$.
As $\operatorname{rank}(P_{\wh\Lambda}M_\Lambda)\le p_\alpha$ and
$\norm{K}_F\le\sqrt{\operatorname{rank}K}\,\norm{K}_{\rm op}$, the first term
obeys $\norm{P_{\wh\Lambda}M_\Lambda H}_F\le\sqrt{p_\alpha}
\norm{M_\Lambda H}_{\rm op}$, and the second
$\norm{M_{\wh\Lambda}P_\Lambda H}_F\le
\norm{M_{\wh\Lambda}P_\Lambda}_{\rm op}\norm{H}_F$. For
\eqref{eq:select-energy-R} use
$P_F-P_{\wh F}=P_FM_{\wh F}-M_FP_{\wh F}$, the angle bound on the first term
and the rank bound on the second.
\end{proof}

\subsection{Noise edge without separability}\label{sec:primitive-noise-edge}

Write $\mathcal E:=\E[\operatorname{vec}(E)\operatorname{vec}(E)']$ for the
covariance matrix of the stacked errors and, under covariance stationarity in
$t$, $\gamma_{ij}(h):=\E[\varepsilon_{it}\varepsilon_{j,t+h}]$.

\begin{assumption}[Weak dependence]\label{ass:weakdep}
(i) The errors are mean zero and covariance stationary in $t$ with
\[
  \sup_i\sum_{j=1}^N\sum_{h\in\mathbb Z}|\gamma_{ij}(h)|=O(1).
\]
(ii) There is $K=O(\norm{\mathcal E}_{\rm op})$
such that $\E\exp(\theta\,a'Eb)\le\exp(\theta^2 K/2)$ for every $\theta\in\R$ and
all unit vectors $a\in\R^N$, $b\in\R^T$.
\end{assumption}

Condition (ii) holds, for example, when $\varepsilon_{it}$ is an absolutely
summable linear filter of i.i.d.\ sub-Gaussian innovations.

\begin{lemma}[Primitive noise edge]\label{lem:edge-primitive}
Under Assumption~\ref{ass:weakdep}(i), $\norm{\mathcal E}_{\rm op}=O(1)$. If
Assumption~\ref{ass:weakdep}(ii) also holds, then
$\norm{E}_{\rm op}=\Op(\sqrt{\max\{N,T\}})$ and
$s_{NT}=\Op((N\mw T)^{-1/2})$.
\end{lemma}

\begin{proof}[Proof of Lemma~\ref{lem:edge-primitive}]
Step 1. As $\mathcal E$ is symmetric positive semidefinite, Gershgorin's
theorem bounds $\norm{\mathcal E}_{\rm op}=\lambda_{\max}(\mathcal E)$ by its
maximal absolute row sum, and by Assumption~\ref{ass:weakdep}(i) the row
indexed by $(i,t)$ has
$\sum_{j,s}|\gamma_{ij}(s-t)|\le\sup_i\sum_j\sum_h|\gamma_{ij}(h)|=O(1)$.

Step 2. By Step 1 and Assumption~\ref{ass:weakdep}(ii), $a'Eb$ is sub-Gaussian
with proxy $K=O(1)$ uniformly over unit $a$ and $b$, so
$\Pr(|a'Eb|>x)\le2\exp(-x^2/2K)$. Let $\mathcal S_N\subset S^{N-1}$
and $\mathcal S_T\subset S^{T-1}$ be $1/4$-nets with $|\mathcal S_N|\le 9^N$ and
$|\mathcal S_T|\le 9^T$, so that
$\norm{E}_{\rm op}\le 2\max_{a\in\mathcal S_N,\,b\in\mathcal S_T}a'Eb$. A union
bound gives
\[
  \Pr\Big(\max_{a\in\mathcal S_N,\,b\in\mathcal S_T}a'Eb>C\sqrt{N+T}\Big)
  \le 2\cdot 9^{N+T}\exp\!\Big(-\tfrac{C^2(N+T)}{2K}\Big)\to 0
  \quad\text{once }C^2>2K\log 9.
\]
Therefore $\norm{E}_{\rm op}=\Op(\sqrt{N+T})=\Op(\sqrt{\max\{N,T\}})$, and
dividing by $\sqrt{NT}$ gives the rate of $s_{NT}$.
\end{proof}
\section{Inferential results and proofs}\label{app:crossfit}

\subsection{Algorithm}\label{sec:cf-construction}

Algorithm~\ref{alg:cf} uses the block notation of Section~\ref{sec:sampleSplit}.

\begin{algorithm}[h!]
\caption{Cross-fitted CUH}\label{alg:cf}
\begin{algorithmic}[1]
  \STATE For each evaluation block $\mathcal I_b$, $b=(a,k)\in \{1,2\}\times [K]$,
  compute $\wt\beta_b$ on $\mathcal N_{-a}\times \mathcal M_k^{\rm qc}$.
  \begin{enumerate}[(i).]
    \item Unit side. On $\mathcal N_a\times \mathcal M_k^{\rm qc}$, take the
    singular vectors $\wh u^{(A,b)}_{ir}$ and $\wh v^{(A,b)}_{tr}$ of
    $W^{A,b}=Y-X\wt\beta_b$ and the covariate projections
    $\wh\lambda^{(a,k)}_{j,r}(\alpha_i)
    =\frac{1}{|\mathcal M_k^{\rm qc}|}\sum_{t\in \mathcal M_{k}^{\rm qc}}x_{it,j}\,\wh v^{(A,b)}_{tr}$, $i\in\mathcal N_{a}$.
    \item Time side. On $\mathcal N_{-a}\times \mathcal M_k$, take the
    singular vectors $\wh u^{(B,b)}_{ir}$ and $\wh v^{(B,b)}_{tr}$ of
    $W^{B,b}=Y-X\wt\beta_b$ and the covariate projections
    $\wh f^{(a,k)}_{j,r}(\gamma_t)
    =\frac{1}{|\mathcal{N}_{-a}|}\sum_{i\in\mathcal N_{-a}}x_{it,j}\,\wh u^{(B,b)}_{ir}$, $t\in \mathcal M_{k}$.
    \item Selection. Collect the leading $R$ vectors $\wh u^{(A,b)}_{ir}$ in
    $\wh U_{R,b}$ and $\wh v^{(B,b)}_{tr}$ in $\wh V_{R,b}$. For each regressor,
    rank all singular directions of each training rectangle by
    covariate-projection energy, unit side and time side separately, and keep
    the leading $L_j$, without appending $\mathcal I_R$, to form $\wh\Lambda_b$
    and $\wh F_b$.
    \item Frame truncation. Divide each column of $\wh\Lambda_b$ and
    $\wh F_b$ by the root mean square $\wh s_j:=(NT)^{-1/2}\norm{X_j}_F$ of its
    regressor, giving $\wh\Lambda_b^\circ$ and $\wh F_b^\circ$. Let
    $\check\Lambda_b$ collect the leading eigenvector of
    $\wh\Lambda_b^\circ\wh\Lambda_b^{\circ\prime}/|\mathcal N_a|$ and those with
    eigenvalues above $\upsilon_{NT}^2$, where $\upsilon_{NT}:=(N\mw T)^{-1/4}$,
    and $\check F_b$ those of $\wh F_b^\circ\wh F_b^{\circ\prime}/|\mathcal M_k|$
    likewise. Set
    $\wh A_b=M_{\check\Lambda_b}M_{\wh U_{R,b}}$ and
    $\wh B_b=M_{\wh V_{R,b}}M_{\check F_b}$.
  \end{enumerate}
  \STATE Residualise each block by its own transforms and accumulate, for
  $j,\ell=1,\dots,d$,
    \begin{align*}
         [\wh Q]_{j\ell}=& \tfrac1{NT}\sum_{b}
    \fro{\wh A_{b}X_j^{(b)}\wh B_{b},
         \wh A_{b}X_\ell^{(b)}\wh B_{b}},\\
         [\wh S]_j =&  \tfrac1{NT}\sum_{b}
    \fro{\wh A_{b}X_j^{(b)}\wh B_{b},
         \wh A_{b}Y^{(b)}\wh B_{b}} .
    \end{align*}
  Return $\wh\beta^{\rm cf}=\wh Q^{-1}\wh S$.
\end{algorithmic}
\end{algorithm}

The estimate $\wt\beta_b$ and the transforms $\wh A_b,\wh B_b$ are
$\mathcal F_b$-measurable, and each transform, a product of residual makers,
is a contraction. The unit side excludes $\mathcal N_{-a}$ because its errors
at the times $\mathcal M_k^{\rm qc}$ need not be independent of the time-side
frames built from $\mathcal N_{-a}\times\mathcal M_k$, and the time side is
restricted likewise.

\subsection{Oracle score}\label{sec:cf-oracle}

On block $b$ write $\wh{\mathcal R}_b[H]:=\wh A_bH\wh B_b$,
$q_{j,b}:=\mathcal R_b^0[X_j^{(b)}]=A_bX_j^{(b)}B_b$,
$\wh q_{j,b}:=\wh{\mathcal R}_b[X_j^{(b)}]$ and
$e_b^0:=\mathcal R_b^0[E^{(b)}]$, and define the oracle weights
$z_{j,b}:=A_b'A_bX_j^{(b)}B_bB_b'$. For $(i,t)\in\mathcal I_b$ let $q_{it}$
and $z_{it}$ be the $d$-vectors of the $(i,t)$ entries of $q_{j,b}$ and
$z_{j,b}$, and let $\psi_{it}:=q_{it}e^0_{it}$ be the oracle score. The adjoint
identity gives
\begin{equation}\label{eq:cf-oracle-score-identity}
\begin{split}
 \sum_{(i,t)\in\mathcal I_b}\psi_{it,j}
 =\langle A_bX_{j}^{(b)}B_b,A_bE^{(b)}B_b\rangle_F
 =\langle A_b'A_bX_{j}^{(b)}B_bB_b',E^{(b)}\rangle_F
 =\sum_{(i,t)\in\mathcal I_b}z_{it,j}\varepsilon_{it}.
\end{split}
\end{equation}
By linearity of $\wh{\mathcal R}_b$, the feasible score $NT[\wh S-\wh Q\beta_0]_j$ is
the blockwise form of \eqref{eq:score-decomp},
\begin{equation}\label{eq:cf-linear-score}
 \sum_b\langle\widehat{\mathcal R}_b[X_{j}^{(b)}],\widehat{\mathcal R}_b[G^{(b)}]\rangle_F
 +\sum_b\langle\widehat{\mathcal R}_b[X_{j}^{(b)}],\widehat{\mathcal R}_b[E^{(b)}]\rangle_F .
\end{equation}

\begin{assumption}[Oracle score limits]\label{ass:cf-score}
(i) The matrix
\[
 \overline Q_{NT}:=\frac1{NT}\sum_b
 \big\{\langle\mathcal R_b^0[X_{j}^{(b)}],
 \mathcal R_b^0[X_{\ell}^{(b)}]\rangle_F\big\}_{j,\ell}
\]
satisfies $\overline Q_{NT}\to_p Q$ with $Q$ nonsingular.
(ii) $\frac1{\sqrt{NT}}\sum_{t=1}^T\sum_{i=1}^N\psi_{it}
 \Rightarrow\mathcal N(0,\Omega).$
\end{assumption}

\subsection{Coupling, transform consistency and score replacement}\label{sec:cf-transform-consistency}

\begin{lemma}[Coupling across the temporal gap]
\label{lem:cf-coupling}
Let $\mathcal Z_b^{\rm tr}$ be the training data of block $b$, so that
$\mathcal F_b=\sigma(\mathcal Z_b^{\rm tr})$.
(a) Under Assumption~\ref{ass:cf-reg}(i)--(ii), on an enriched probability space
there are copies $(\widetilde E^{(b)},\widetilde{\mathcal Z}_b^{\rm tr})$ with the
conditional marginal laws of $E^{(b)}$ and $\mathcal Z_b^{\rm tr}$, conditionally
independent of each other given $(X,\mathcal C)$, such that
\begin{equation}\label{eq:cf-coupling}
 \Pr\!\left(
 \text{some }(E^{(b)},\mathcal Z_b^{\rm tr})\ne
 (\widetilde E^{(b)},\widetilde{\mathcal Z}_b^{\rm tr})
 \mid X,\mathcal C\right)
 \le CKN e^{-c q_T}=o(1).
\end{equation}
(b) Let $\mathcal H\supseteq\sigma(X,\mathcal C)$ be a sigma algebra, $\bar E$
an error array with $\E[\bar E\mid\mathcal H]=0$ and
$c_{\mathcal H}:=\norm{\operatorname{Cov}\{\operatorname{vec}(\bar E)\mid\mathcal H\}}_{\rm op}$,
and $H_1,\dots,H_m$ $\mathcal H$-measurable arrays of the same dimensions. Then
\[
 \E\Big[\sum_{l\le m}\fro{H_l,\bar E}^2\,\Big|\,\mathcal H\Big]
 \le c_{\mathcal H}\sum_{l\le m}\norm{H_l}_F^2,
\]
so $\sum_l\fro{H_l,\bar E}^2=\Op(\sum_l\norm{H_l}_F^2)$ when
$c_{\mathcal H}=\Op(1)$. Under Assumption~\ref{ass:cf-reg} this conclusion
also holds with $\bar E=E^{(b)}$ when every $H_l$ is a function of
$(X,\mathcal C)$ and $\mathcal Z_b^{\rm tr}$.
\end{lemma}

\begin{proof}[Proof of Lemma~\ref{lem:cf-coupling}]
(a) With $\mathcal G=\sigma(X,\mathcal C)$, the conditional
absolute-regularity coefficient of random elements $Z_1,Z_2$ is
$\gamma_{\mathcal G}(Z_1,Z_2):=\frac12\|P_{(Z_1,Z_2)\mid\mathcal G}
-P_{Z_1\mid\mathcal G}\otimes P_{Z_2\mid\mathcal G}\|_{\rm TV}$. Given $(X,\mathcal C)$, $X\beta_0+G$ is fixed, and by
Assumption~\ref{ass:cf-reg}(i) the errors of $\mathcal N_{-a}$ are independent
of those of $\mathcal N_a$. It therefore suffices to couple
$E_{a,k}:=(\varepsilon_{it})_{i\in\mathcal N_a,t\in\mathcal M_k}$ with
$E^{\rm qc}_{a,k}:=(\varepsilon_{it})_{i\in\mathcal N_a,t\in\mathcal M_k^{\rm qc}}$,
with unit rows $E_{i,k}$ and $E^{\rm qc}_{i,k}$. The $q_T$ periods omitted on
both sides of $\mathcal M_k$ give
$\gamma_{\mathcal G}(E_{i,k},E^{\rm qc}_{i,k})\le2\beta_i(q_T\mid X,\mathcal C)
\le2C_\beta e^{-c_\beta q_T}$, and independence across units with the product
total-variation inequality gives
$\gamma_{\mathcal G}(E_{a,k},E^{\rm qc}_{a,k})\le C|\mathcal N_a|e^{-cq_T}$.
The conditional version of \citet[Lemma~3.1]{semenova2023inference} gives copies
with product conditional law that differ from the originals with conditional
probability $\gamma_{\mathcal G}(E_{a,k},E^{\rm qc}_{a,k})$. A union bound
over the $2K$ blocks gives \eqref{eq:cf-coupling}, which is $o(1)$ as
$\log N=o(q_T)$.

(b) Given $\mathcal H$, each $\fro{H_l,\bar E}$ is centred with variance
$\operatorname{vec}(H_l)'\operatorname{Cov}\{\operatorname{vec}(\bar E)\mid\mathcal H\}\operatorname{vec}(H_l)
\le c_{\mathcal H}\norm{H_l}_F^2$, and conditional Markov gives the rate. For the second claim let $\wt H_l$ be
$H_l$ recomputed from $\wt{\mathcal Z}_b^{\rm tr}$. By part (a),
$\wt E^{(b)}$ has the conditional law of $E^{(b)}$, centred by
Assumption~\ref{ass:cf-reg}(i), and is conditionally independent of
$\wt{\mathcal Z}_b^{\rm tr}$ given $(X,\mathcal C)$. The first claim therefore
applies with $\mathcal H=\sigma(X,\mathcal C,\wt{\mathcal Z}_b^{\rm tr})$ and
$c_{\mathcal H}=\norm{\Sigma_{E,b}}_{\rm op}=\Op(1)$ by
Assumption~\ref{ass:cf-reg}(ii). As $\wt{\mathcal Z}_b^{\rm tr}$ has the
conditional law of $\mathcal Z_b^{\rm tr}$, the $\norm{\wt H_l}_F$ have the
joint law of the $\norm{H_l}_F$, and by \eqref{eq:cf-coupling} the original
and coupled quantities coincide outside an event of probability $o(1)$.
\end{proof}

Write $X^{s,b}\delta:=\sum_jX^{s,b}_j\delta_j$ for $\delta\in\R^d$ and
$s\in\{A,B\}$. Frame truncation, step (iv) of Algorithm~\ref{alg:cf}, matters
when $\Lambda_b$ is rank deficient, as $\wh\Lambda_b$ then generically spans a
direction that $\Lambda_b$ lacks, and a transform built from it would remove
variation of $X_j^{(b)}$ that the oracle transform keeps.

\begin{lemma}[Blockwise transform consistency]\label{lem:cf-transform}
Under Assumptions~\ref{ass:cf-reg}, \ref{ass:cf-frame-strength},
\ref{ass:cf-exact-spike}(i), (ii), (iv) and~\ref{ass:cf-score}(i), with
probability approaching one
$\operatorname{rank}\check\Lambda_b=\operatorname{rank}\Lambda_b$ and
$\operatorname{rank}\check F_b=\operatorname{rank}F_b$ for every $b$,
\begin{equation}\label{eq:cf-frame-consistency}
 \max_b\norm{P_{\check\Lambda_b}-P_{\Lambda_b}}_{\rm op}=\Op(\zeta_\Lambda),
 \qquad
 \max_b\norm{P_{\check F_b}-P_{F_b}}_{\rm op}=\Op(\zeta_F),
\end{equation}
and
\begin{equation}\label{eq:cf-operator-consistency}
 \max_b\big\{\norm{\wh A_b-A_b}_{\rm op}+\norm{\wh B_b-B_b}_{\rm op}\big\}
 =\Op(\varsigma_{NT})=\op(1).
\end{equation}
Moreover, with $\Delta_{A,b}:=\wh A_b-A_b$ and $\Delta_{B,b}:=\wh B_b-B_b$,
every array $H$ on $\mathcal I_b$ satisfies
\begin{equation}\label{eq:cf-D-bound}
 \norm{\wh A_bH\wh B_b-A_bHB_b}_F\vee
 \norm{\wh A_b'\wh A_bH\wh B_b\wh B_b'-A_b'A_bHB_bB_b'}_F
 \le2\big(\norm{\Delta_{A,b}}_{\rm op}+\norm{\Delta_{B,b}}_{\rm op}\big)\norm{H}_F.
\end{equation}
\end{lemma}

\begin{proof}[Proof of Lemma~\ref{lem:cf-transform}]
Fix $b$. A maximum over the fixed $2K$ blocks preserves $\Op$ orders. On each training rectangle the noise edge of Assumption~\ref{ass:cf-exact-spike}(i),
$\norm{X^{s,b}_j}_{\rm op}\le\norm{X_j}_F=\Op(\sqrt{NT})$ and
\eqref{eq:cf-initial} give
$\norm{E^{s,b}+X^{s,b}(\beta_0-\wt\beta_b)}_{\rm op}/\sqrt{NT}
=\Op((N\mw T)^{-1/2})=\op(c_{NT})$ by detectability, the edge hypothesis of
Proposition~\ref{prop:xi}. Hence
$\norm{P_{\wh U_{R,b}}-P_{U_{R,b}}}_{\rm op}=\Op(\xi_u)$ and
$\norm{P_{\wh V_{R,b}}-P_{V_{R,b}}}_{\rm op}=\Op(\xi_v)$. On the selection
event of Corollary~\ref{cor:select}, read blockwise, $\wh\Lambda_b$ and
$\Lambda_b$ carry the same columns, and sign alignment changes neither the
spans nor the frame truncation, so, as in the proof of
Corollary~\ref{cor:select},
$\norm{\wh\Lambda_b-\Lambda_b}_{\rm op}=\sqrt{p_\alpha N}\,\Op(\xi_v+T^{-1/2})$.

Below, singular values are those of the unnormalised matrices, and each claim
holds with probability approaching one. Let $\Lambda_b^\circ$ be $\Lambda_b$ with each column divided by the $\wh s_j$
of step (iv). Here $\wh s_j=\Op(1)$, and $\wh s_j^2\ge[\overline Q_{NT}]_{jj}$
because the blocks partition the panel and the oracle transforms are
contractions, so $\wh s_j^{-1}=\Op(1)$ by Assumption~\ref{ass:cf-score}(i).
Column scaling preserves the span and rank of $\Lambda_b$, and the rank is at
least one as each selected column has energy at least $\epsilon_{NT}^2$ by
Assumption~\ref{ass:Xstrength}. By Assumption~\ref{ass:cf-frame-strength} the
nonzero singular values of $\Lambda_b^\circ$ are at least
$c^{1/2}\sqrt N\epsilon_{NT}/\max_j\wh s_j$, and by Weyl's inequality those of
$\wh\Lambda_b^\circ$ lie within
$\omega_b:=\norm{\wh\Lambda_b^\circ-\Lambda_b^\circ}_{\rm op}
=\sqrt N\,\Op(\xi_v+T^{-1/2})$ of them. As
$\varsigma_{NT}\ge\zeta_\Lambda+\zeta_F\ge(N\mw T)^{-1/2}/\epsilon_{NT}$,
\eqref{eq:cf-exact-rate} forces the deterministic sequence
$\sqrt{NT}(N\mw T)^{-3/2}\epsilon_{NT}^{-3}$ to vanish, so
$\epsilon_{NT}\gg(NT)^{1/6}(N\mw T)^{-1/2}\ge(N\mw T)^{-1/6}$ and
$\upsilon_{NT}=o(\epsilon_{NT})$. It also gives
$\xi_v\le\varsigma_{NT}=\op((NT)^{-1/6})=\op(\upsilon_{NT})$, and
$T^{-1/2}=o(\upsilon_{NT})$. Hence the eigenvalues of
$\wh\Lambda_b^\circ\wh\Lambda_b^{\circ\prime}/|\mathcal N_a|$ above
$\upsilon_{NT}^2$ are exactly the $\operatorname{rank}\Lambda_b$ leading ones,
and $\operatorname{rank}\check\Lambda_b=\operatorname{rank}\Lambda_b$.

A unit vector $a$ in the span of $\Lambda_b$ is $a=\Lambda_b^\circ w$ with
$\norm w\le\max_j\wh s_j/(c^{1/2}\sqrt N\epsilon_{NT})$. The singular values
of $\wh\Lambda_b^\circ$ that frame truncation discards are at most
$\omega_b$, so
$\norm{M_{\check\Lambda_b}a}\le\norm{M_{\check\Lambda_b}\wh\Lambda_b^\circ w}
+\omega_b\norm w\le2\omega_b\norm w=\Op(\zeta_\Lambda)$ uniformly in $a$. The
supremum over $a$ is
$\norm{M_{\check\Lambda_b}P_{\Lambda_b}}_{\rm op}
=\norm{P_{\check\Lambda_b}-P_{\Lambda_b}}_{\rm op}$, as the two projectors
have equal finite rank, giving the first half of
\eqref{eq:cf-frame-consistency}. The identity
\[
 \wh A_b-A_b=(P_{\Lambda_b}-P_{\check\Lambda_b})M_{\wh U_{R,b}}
 +M_{\Lambda_b}(P_{U_{R,b}}-P_{\wh U_{R,b}})
\]
and the contraction property give
$\norm{\wh A_b-A_b}_{\rm op}=\Op(\zeta_\Lambda+\xi_u)$. The time side,
with $\wh B_b-B_b=(P_{V_{R,b}}-P_{\wh V_{R,b}})M_{\check F_b}
+M_{V_{R,b}}(P_{F_b}-P_{\check F_b})$, is symmetric and gives
$\Op(\zeta_F+\xi_v)$. Finally $\varsigma_{NT}=\op(1)$ by
\eqref{eq:cf-exact-rate}.

The bound \eqref{eq:cf-D-bound} follows from the contraction property and the
identities
\[
\begin{aligned}
 \wh A_bH\wh B_b-A_bHB_b&=\Delta_{A,b}H\wh B_b+A_bH\Delta_{B,b},\\
 \wh A_b'\wh A_bH\wh B_b\wh B_b'-A_b'A_bHB_bB_b'
 &=(\wh A_b'\Delta_{A,b}+\Delta_{A,b}'A_b)H\wh B_b\wh B_b'
 +A_b'A_bH(\wh B_b\Delta_{B,b}'+\Delta_{B,b}B_b').
\end{aligned}
\]
\end{proof}

Assumption~\ref{ass:cf-exact-spike}(iii) can be checked from the training
frames. At the equal ranks of Lemma~\ref{lem:cf-transform}, each estimated
commutator, such as
$\norm{P_{\wh U_{R,b}}P_{\check\Lambda_b}-P_{\check\Lambda_b}P_{\wh U_{R,b}}}_{\rm op}$,
is within $\Op(\varsigma_{NT})$ of its population counterpart.

\begin{lemma}[Score replacement]\label{lem:cf-score-replacement}
Let $D_{j,b}:=\wh A_b'\wh A_bX_j^{(b)}\wh B_b\wh B_b'-A_b'A_bX_j^{(b)}B_bB_b'$.
Under the conditions of Lemma~\ref{lem:cf-transform},
$(NT)^{-1/2}\sum_b\fro{D_{j,b},E^{(b)}}=\Op(\varsigma_{NT})=\op(1)$ for
$j=1,\ldots,d$.
\end{lemma}

\begin{proof}[Proof of Lemma~\ref{lem:cf-score-replacement}]
By \eqref{eq:cf-D-bound} at $H=X_j^{(b)}$, with
$\norm{X_j^{(b)}}_F\le\norm{X_j}_F=\Op(\sqrt{NT})$, and
Lemma~\ref{lem:cf-transform}, both $\norm{\wh q_{j,b}-q_{j,b}}_F$ and
$\norm{D_{j,b}}_F$ are $\Op(\varsigma_{NT}\sqrt{NT})$. As $D_{j,b}$ is a
function of $(X,\mathcal C)$ and $\mathcal Z_b^{\rm tr}$,
Lemma~\ref{lem:cf-coupling}(b) with $m=1$ gives
$(NT)^{-1/2}\fro{D_{j,b},E^{(b)}}=\Op(\varsigma_{NT})$ on each of the fixed
$2K$ blocks.
\end{proof}

\begin{lemma}[Gram consistency]\label{lem:cf-gram}
Under the conditions of Lemma~\ref{lem:cf-transform}, the Gram matrix of
Algorithm~\ref{alg:cf} satisfies
$\widehat Q-\overline Q_{NT}=\Op(\varsigma_{NT})$ and $\widehat Q\to_pQ\succ0$.
\end{lemma}

\begin{proof}[Proof of Lemma~\ref{lem:cf-gram}]
By the adjoint identity,
\[
 \fro{\wh A_bX_{j}^{(b)}\wh B_b,\wh A_bX_{\ell}^{(b)}\wh B_b}
 -\fro{A_bX_{j}^{(b)}B_b,A_bX_{\ell}^{(b)}B_b}=\fro{X_{j}^{(b)},D_{\ell,b}},
\]
so $|[\wh Q-\overline Q_{NT}]_{j\ell}|\le(NT)^{-1}\sum_b\norm{X_{j}^{(b)}}_F
\norm{D_{\ell,b}}_F=\Op(\varsigma_{NT})$ by the proof of
Lemma~\ref{lem:cf-score-replacement}. Assumption~\ref{ass:cf-score}(i)
completes the proof.
\end{proof}

\subsection{Proof of Theorem~\ref{thm:cf}}\label{sec:cf-normality-proof}

\begin{proof}[Proof of Theorem~\ref{thm:cf}]
By \eqref{eq:cf-linear-score},
\[
 \sqrt{NT}(\wh\beta^{\rm cf}-\beta_0)
 =\wh Q^{-1}(NT)^{-1/2}\sum_b\Big\{\fro{\wh q_{j,b},\wh{\mathcal R}_b[G^{(b)}]}
 +\fro{\wh q_{j,b},\wh{\mathcal R}_b[E^{(b)}]}\Big\}_{j\le d}.
\]
Lemma~\ref{lem:cf-gram} gives $\wh Q\to_pQ\succ0$. By the adjoint identity,
Lemma~\ref{lem:cf-score-replacement} and
\eqref{eq:cf-oracle-score-identity}, the error term equals the oracle score
up to $\op(1)$, and Assumption~\ref{ass:cf-score}(ii) gives its normal limit.
By Slutsky's theorem it remains to show that the heterogeneity term
$(NT)^{-1/2}\sum_b\fro{\wh q_{j,b},\wh{\mathcal R}_b[G^{(b)}]}$ is $\op(1)$ for
each $j$. As $K$ is fixed, it suffices to do so block by block.

On block $b$, exact rank and $R=R_0$ give
\[
 G^{(b)}=U_{R_0,b}\Sigma V_{R_0,b}',
 \qquad A_bU_{R_0,b}=0,
 \qquad V_{R_0,b}'B_b=0,
\]
by the definitions of $A_b$ and $B_b$. Hence $A_bG^{(b)}=G^{(b)}B_b=0$, so
$\mathcal R_b^0[G^{(b)}]=0$ and, with $\Delta_{A,b},\Delta_{B,b}$ as in
Lemma~\ref{lem:cf-transform},
$\wh{\mathcal R}_b[G^{(b)}]=\Delta_{A,b}G^{(b)}\Delta_{B,b}
=(\widehat A_bU_{R_0,b})\Sigma(V_{R_0,b}'\widehat B_b)$. The heterogeneity
term of block $b$ therefore splits into a leading and a second pairing,
\[
 \frac1{\sqrt{NT}}\fro{\wh q_{j,b},\wh{\mathcal R}_b[G^{(b)}]}
 =\frac1{\sqrt{NT}}\fro{q_{j,b},\Delta_{A,b}G^{(b)}\Delta_{B,b}}
 +\frac1{\sqrt{NT}}\fro{\wh q_{j,b}-q_{j,b},\Delta_{A,b}G^{(b)}\Delta_{B,b}}.
\]
Since $\widehat A_b=\widehat A_bM_{\widehat U_{R,b}}$ and $R=R_0$,
\[
 \|\widehat A_bU_{R_0,b}\|_{\rm op}
 \le\|M_{\widehat U_{R,b}}P_{U_{R,b}}\|_{\rm op}\norm{U_{R_0,b}}_{\rm op}
 =\Op(\xi_u\sqrt N)
\]
by Proposition~\ref{prop:xi} and Assumption~\ref{ass:reg}(iii), and
symmetrically $\|V_{R_0,b}'\widehat B_b\|_{\rm op}=\Op(\xi_v\sqrt T)$. The
product has rank at most $R_0$ and $\norm{\Sigma}_{\rm op}=\sigma_1=O(1)$, so
$\norm{\Delta_{A,b}G^{(b)}\Delta_{B,b}}_F
=\Op(\sqrt{R_0}\,\xi_u\xi_v\sqrt{NT})$. With
$\norm{\wh q_{j,b}-q_{j,b}}_F=\Op(\varsigma_{NT}\sqrt{NT})$ from the proof of
Lemma~\ref{lem:cf-score-replacement}, Cauchy--Schwarz and
$\xi_u\vee\xi_v\le\varsigma_{NT}$ bound the second pairing by
$\Op(\sqrt{NT R_0}\,\varsigma_{NT}^3)=\op(1)$ under \eqref{eq:cf-exact-rate}.

The Cauchy--Schwarz bound $\Op(\sqrt{R_0}\,\xi_u\xi_v\sqrt{NT})$ on the
leading pairing need not vanish, so the pairing is centred instead through the
exact leak of each training rectangle. Put
$\delta_b:=\beta_0-\wt\beta_b$, and let $V_{{\rm qc},b}$ be the rows of
$V_{R_0}$ in $\mathcal M_k^{\rm qc}$ and $U_{-a}$ the rows of $U_{R_0}$ in
$\mathcal N_{-a}$, so that
$W^{A,b}=U_{R_0,b}\Sigma V_{{\rm qc},b}'+E^{A,b}+X^{A,b}\delta_b$ and
$W^{B,b}=U_{-a}\Sigma V_{R_0,b}'+E^{B,b}+X^{B,b}\delta_b$. Let $\wh V^A$ and
$\wh U^B$ collect the $R$ leading right singular vectors of $W^{A,b}$ and left
singular vectors of $W^{B,b}$, and put
\begin{align*}
 \wh\Pi_b&:=\wh V^A(V_{{\rm qc},b}'\wh V^A)^{-1},&
 \Pi_b&:=V_{{\rm qc},b}(V_{{\rm qc},b}'V_{{\rm qc},b})^{-1},\\
 \wh\Psi_b&:=\wh U^B(U_{-a}'\wh U^B)^{-1},&
 \Psi_b&:=U_{-a}(U_{-a}'U_{-a})^{-1}.
\end{align*}
By Proposition~\ref{prop:xi} and Assumption~\ref{ass:reg}(iii), with
probability approaching one the inverses exist,
$\norm{\wh\Pi_b}_{\rm op}+\norm{\Pi_b}_{\rm op}=\Op(T^{-1/2})$ and
$\norm{\wh\Pi_b-\Pi_b}_{\rm op}=\norm{M_{V_{{\rm qc},b}}\wh\Pi_b}_{\rm op}
=\Op(\xi_vT^{-1/2})$, and symmetrically for $\wh\Psi_b,\Psi_b$ with $N$ and
$\xi_u$. Since $M_{\wh U_{R,b}}W^{A,b}\wh V^A=0$ and
$\wh U^{B\prime}W^{B,b}M_{\wh V_{R,b}}=0$,
\begin{equation}\label{eq:cf-leak}
 \wh A_bU_{R_0,b}\Sigma=-\wh A_b(E^{A,b}+X^{A,b}\delta_b)\wh\Pi_b,\qquad
 \Sigma V_{R_0,b}'\wh B_b=-\wh\Psi_b'(E^{B,b}+X^{B,b}\delta_b)\wh B_b,
\end{equation}
and the leading pairing equals $J_1+J_2+J_3$, where
\begin{align*}
 J_1&:=-\tfrac1{\sqrt{NT}}\fro{q_{j,b},\,(\wh A_bU_{R_0,b})\wh\Psi_b'E^{B,b}\wh B_b},\\
 J_2&:=\tfrac1{\sqrt{NT}}\fro{q_{j,b},\,\wh A_bE^{A,b}\wh\Pi_b\Sigma^{-1}\wh\Psi_b'X^{B,b}\delta_b\wh B_b},\\
 J_3&:=\tfrac1{\sqrt{NT}}\fro{q_{j,b},\,\wh A_bX^{A,b}\delta_b\wh\Pi_b\Sigma^{-1}\wh\Psi_b'X^{B,b}\delta_b\wh B_b}.
\end{align*}
Each pairing below is of $q_{j,b}$, with
$\norm{q_{j,b}}_{\rm op}\le\norm{X_j}_F=\Op(\sqrt{NT})$, against an array of
rank at most $R_0$, and $\norm{\Sigma^{-1}}_{\rm op}=1/c_{NT}$ with
$c_{NT}\le\sigma_1=O(1)$.  Two facts reduce the remainders to
$\Op(\sqrt{NT}\,\varsigma_{NT}^3)$.
First, by \eqref{eq:cf-varsigma}, $\varsigma_{NT}
\ge\frac{1}{c_{NT}\sqrt{N\wedge T}}$, so
$ \sqrt{N\vee T}\,\varsigma_{NT}^2 \le c_{NT}\sqrt{NT}\,\varsigma_{NT}^3=o_p(1)$. 
Second, \eqref{eq:cf-initial} and
\eqref{eq:cf-varsigma} imply $ \norm{\delta_b}
=\Op((N\wedge T)^{-1/2}) =\Op(c_{NT}\varsigma_{NT})$.

In $J_1$, replacing $\wh\Psi_b'E^{B,b}\wh B_b$ by $\Psi_b'E^{B,b}B_b$ costs
$\Op(R_0\,\xi_u\varsigma_{NT}\sqrt{N\vee T})=\Op(\sqrt{NT}\,\varsigma_{NT}^3)$,
by $\norm{\wh A_bU_{R_0,b}}_{\rm op}=\Op(\xi_u\sqrt N)$, the noise edge and
Lemma~\ref{lem:cf-transform}. The remainder
$-(NT)^{-1/2}\fro{\Psi_b(\wh A_bU_{R_0,b})'q_{j,b}B_b',E^{B,b}}$ pairs
$E^{B,b}=E^{(-a,k)}$ with an array of Frobenius norm at most
$\norm{\Psi_b}_{\rm op}\norm{\wh A_bU_{R_0,b}}_{\rm op}\norm{X_j}_F
=\Op(\xi_u\sqrt{NT})$ that depends only on $(X,\mathcal C)$ and the data on
$\mathcal N_a\times\mathcal M_k^{\rm qc}$ and
$\mathcal N_{-a}\times\mathcal M_k^{\rm qc}$, training data of block
$(-a,k)$. Lemma~\ref{lem:cf-coupling}(b) for that block bounds it by
$\Op(\xi_u)$, so $J_1=\Op(\xi_u+\sqrt{NT}\,\varsigma_{NT}^3)$.

In $J_2$, replacing $\wh A_bE^{A,b}\wh\Pi_b$ by $A_bE^{A,b}\Pi_b$ costs
$\Op(\varsigma_{NT}\sqrt{(N\vee T)/T})$ in operator norm, by
Lemma~\ref{lem:cf-transform} and the noise edge. As
$\norm{\wh\Psi_b'X^{B,b}\delta_b\wh B_b}_{\rm op}=\Op(\sqrt T\norm{\delta_b})$,
$J_2$ changes by
$\Op(R_0\,\varsigma_{NT}\norm{\delta_b}\sqrt{N\vee T}/c_{NT})$, which is
$\Op(\sqrt{NT}\,\varsigma_{NT}^3)$. The remainder
$(NT)^{-1/2}\fro{A_b'q_{j,b}K_b'\Pi_b',E^{A,b}}$, with
$K_b:=\Sigma^{-1}\wh\Psi_b'X^{B,b}\delta_b\wh B_b$, pairs $E^{A,b}$ with an
array of Frobenius norm $\Op(\sqrt{NT}\norm{\delta_b}/c_{NT})$, since
$\norm{q_{j,b}}_F\le\norm{X_j}_F$,
$\norm{K_b}_{\rm op}=\Op(\sqrt T\norm{\delta_b}/c_{NT})$ and
$\norm{\Pi_b}_{\rm op}=\Op(T^{-1/2})$. The array is measurable with respect
to $\mathcal H:=\sigma(X,\mathcal C,\text{data of the units }\mathcal N_{-a})$.
By Assumption~\ref{ass:cf-reg}(i), $E^{A,b}$ is centred and independent of
$\mathcal H$ given $(X,\mathcal C)$, so its conditional mean and covariance
given $\mathcal H$ are those given $(X,\mathcal C)$.
Lemma~\ref{lem:cf-coupling}(b), with the covariance bound of
Assumption~\ref{ass:cf-exact-spike}(i), gives
$J_2=\Op(\norm{\delta_b}/c_{NT}+\sqrt{NT}\,\varsigma_{NT}^3)$.

The term $J_3$ is bounded directly, without centring. The columns of
$X^{A,b}_j\wh V^A$ are $|\mathcal M_k^{\rm qc}|$ times the covariate
projections $\wh\lambda^{(a,k)}_{j,r}$, $r\in\mathcal I_R$, of
Algorithm~\ref{alg:cf}. For a selected covariate projection $\wh\lambda$, a
column of $\wh\Lambda_b$, write $\wh A_b\wh\lambda=M_{\check\Lambda_b}\wh\lambda
-M_{\check\Lambda_b}P_{\wh U_{R,b}}\wh\lambda$. Let $\lambda$ be its
population counterpart, a column of $\Lambda_b$, so $\lambda=P_{\Lambda_b}\lambda$,
and $\kappa_{A,b}:=\norm{P_{U_{R,b}}P_{\Lambda_b}-P_{\Lambda_b}P_{U_{R,b}}}_{\rm op}$.
Since $M_{\check\Lambda_b}\lambda=M_{\check\Lambda_b}P_{\Lambda_b}\lambda$ and
$M_{\check\Lambda_b}P_{U_{R,b}}P_{\Lambda_b}
=M_{\check\Lambda_b}P_{\Lambda_b}P_{U_{R,b}}
+M_{\check\Lambda_b}(P_{U_{R,b}}P_{\Lambda_b}-P_{\Lambda_b}P_{U_{R,b}})$,
\[
 \norm{\wh A_b\wh\lambda}
 \le2\norm{\wh\lambda-\lambda}
 +\big(\norm{P_{\wh U_{R,b}}-P_{U_{R,b}}}_{\rm op}
 +2\norm{M_{\check\Lambda_b}P_{\Lambda_b}}_{\rm op}+\kappa_{A,b}\big)\norm{\lambda}.
\]
An unselected covariate projection has root-mean-square norm
$\Op(\xi_v+T^{-1/2}+(N\mw T)^{-1/2})$ by Assumption~\ref{ass:Xstrength} and
Theorem~\ref{thm:proj}. With Proposition~\ref{prop:xi},
\eqref{eq:cf-frame-consistency} and Theorem~\ref{thm:proj} this gives
$\norm{\wh A_bX^{A,b}_j\wh\Pi_b}_{\rm op}
=\Op\{\sqrt N(\varsigma_{NT}+\kappa_{A,b})\}$, and the mirror argument gives
$\norm{\wh\Psi_b'X^{B,b}_j\wh B_b}_{\rm op}
=\Op\{\sqrt T(\varsigma_{NT}+\kappa_{B,b})\}$, with $\kappa_{B,b}$ the time-side
commutator of \eqref{eq:cf-commute}. Hence
\[
 |J_3|=\Op\Big(R_0\sqrt{NT}\,\norm{\delta_b}^2
 (\varsigma_{NT}+\kappa_{A,b})(\varsigma_{NT}+\kappa_{B,b})/c_{NT}\Big)
 =\Op\big(R_0\,c_{NT}\sqrt{NT}\,\varsigma_{NT}^3\big)
\]
by \eqref{eq:cf-initial} and \eqref{eq:cf-commute}. The leading pairing is
therefore
$\Op(\xi_u+c_{NT}^{-1}\max_b\norm{\delta_b}+\sqrt{NT}\,\varsigma_{NT}^3)=\op(1)$,
which completes the proof.
\end{proof}
\subsection{Weak cross-sectional dependence}\label{sec:cf-csdep}

We relax the independence across units in Assumption~\ref{ass:cf-reg}(i).
No temporal gap separates the training rectangle
$\mathcal N_{-a}\times\mathcal M_k$, which the time-side covariate
projections need, from $E^{(b)}$, and Lemma~\ref{lem:cf-coupling} decouples
the two only through independence across units. Following
\citet{freeman2026inference}, we keep the rectangle and centre the error
at its fold-conditional mean.

\begin{assumption}[Weak cross-sectional dependence]\label{ass:cf-csdep}
Conditional on $(X,\mathcal C)$, the errors admit the linear-array
representation
\[
 \varepsilon_{it}=\sum_{m=1}^{M_N}a_{im}u_{mt},
\]
where the shock processes $\{u_{mt}:t\le T\}$ are centred, independent
across $m$, uniformly geometrically beta mixing over time with the
constants of Assumption~\ref{ass:cf-reg}(i),
$\max_{m,t}\E[|u_{mt}|^{4+\delta}\mid X,\mathcal C]=O_p(1)$ for the same
$\delta>0$, and $\log M_N=o(q_T)$. The loadings satisfy
\[
 \sup_i\sum_{m=1}^{M_N}|a_{im}|
 +\sup_m\sum_{i=1}^N|a_{im}|\le C_a<\infty .
\]
\end{assumption}

The summability condition needs no ordering of the units, so it covers
spatial autoregressions and moving averages on an unobserved lattice or
network, and $M_N=N$, $a_{im}=\mathbbm 1\{m=i\}$ recover
Assumption~\ref{ass:cf-reg}(i).

Define the fold-conditional mean and the centred errors
\[
 \mu_b:=\E\big[E^{(b)}\,\big|\,X,\mathcal C,\mathcal F_b\big],\qquad
 \mathring E^{(b)}:=E^{(b)}-\mu_b .
\]
Splitting $E^{(b)}=\mathring E^{(b)}+\mu_b$ in the score replacement of
Lemma~\ref{lem:cf-score-replacement} adds the term
$(NT)^{-1/2}\sum_b\fro{D_{j,b},\mu_b}$, which no conditioning centres, as
$D_{j,b}$ and $\mu_b$ are both $\mathcal F_b$-measurable. It is the bias
created by cross-sectional dependence, and we require
\begin{equation}\label{eq:cf-csdep-bias}
 \frac1{\sqrt{NT}}\sum_b\fro{D_{j,b},\mu_b}=\op(1),\qquad
 j=1,\ldots,d .
\end{equation}

\begin{proposition}[Cross-fitted CUH under cross-sectional dependence]
\label{prop:cf-csdep}
Suppose the conditions of Theorem~\ref{thm:cf} hold with
Assumption~\ref{ass:cf-reg}(i) replaced by
Assumption~\ref{ass:cf-csdep}, that the time-side commutator of
\eqref{eq:cf-commute} is of order $\varsigma_{NT}$ on every block, and that
\eqref{eq:cf-csdep-bias} holds. Then the conclusions of
Theorem~\ref{thm:cf} hold, with $\Omega$ the long-run variance of the
aggregated score $N^{-1/2}\sum_{i=1}^N\psi_{it}$ in
Assumption~\ref{ass:cf-score}(ii).
\end{proposition}

\begin{proof}[Proof of Proposition~\ref{prop:cf-csdep}]
Beta-mixing coefficients are subadditive across the independent shock
processes \citep[Theorem~6.2]{bradley2007mixing}, so the vector error
process $\{\varepsilon_{\cdot t}\}_{t\le T}$ is geometrically beta mixing
with coefficient at most $CM_Ne^{-c_\beta q}$. This bound replaces the
reduction to the units of $\mathcal N_a$ and the product bound over them in
the proof of Lemma~\ref{lem:cf-coupling}(a), so part (a) couples $E^{(b)}$ with the data
at the times $\mathcal M_k^{\rm qc}$, with failure probability
$CKM_Ne^{-cq_T}=o(1)$ as $\log M_N=o(q_T)$. Schur's test with the loading
summability gives $\norm{\Sigma_{E,b}}_{\rm op}=\Op(1)$ and
\[
 \norm{\operatorname{Cov}\{\operatorname{vec}(E^{(b)})\mid X,\mathcal C,\mathcal F_b\}}_{\rm op}=\Op(1).
\]
With the coupling and the centring of the shocks, the first bound gives the
last conclusion of Lemma~\ref{lem:cf-coupling}(b) with $\bar E=E^{(b)}$ for
arrays that are functions of $(X,\mathcal C)$ and the data at the times
$\mathcal M_k^{\rm qc}$.

The proof of Lemma~\ref{lem:cf-transform} does not use
Assumption~\ref{ass:cf-reg}(i). As $D_{j,b}$ is $\mathcal F_b$-measurable and $\mathring E^{(b)}$ is centred
given $\mathcal H:=\sigma(X,\mathcal C,\mathcal F_b)$, the proof of
Lemma~\ref{lem:cf-score-replacement}, with this $\mathcal H$ in
Lemma~\ref{lem:cf-coupling}(b) and the displayed covariance bound, gives
$(NT)^{-1/2}\sum_b\fro{D_{j,b},\mathring E^{(b)}}=\Op(\varsigma_{NT})$. The
fold-mean part is $\op(1)$ by \eqref{eq:cf-csdep-bias}. The oracle weights
$z_{j,b}$ are $(X,\mathcal C)$-measurable, so the oracle score stays
centred, and Lemma~\ref{lem:cf-gram} involves no error centring.

In the proof of Theorem~\ref{thm:cf}, $J_1$ is centred across the temporal
gap alone, which the vector mixing bound preserves, and $J_3$ is bounded
directly. Only $J_2$ used independence across units. With the time-side
commutator of order $\varsigma_{NT}$, the bound on
$\norm{\wh\Psi_b'X^{B,b}_j\wh B_b}_{\rm op}$ there is
$\Op(\sqrt T\varsigma_{NT})$, so
$|J_2|=\Op(R_0\,\varsigma_{NT}\norm{\delta_b}\sqrt{N\vee T}/c_{NT})
=\Op(\sqrt{NT}\,\varsigma_{NT}^3)$ without centring.
\end{proof}

\subsection{Variance estimation}\label{sec:cf-variance}

The CUH transforms induce cross-sectional dependence in the scores even when
the errors are independent across units, so standard errors use the two-way
variance estimator of \citet{CHS_Restat_2024}. On each block $b$ define the
residual
$\wh e_b:=\wh{\mathcal R}_b[Y^{(b)}-\sum_jX_j^{(b)}\wh\beta_j^{\rm cf}]$
and the scores $\wh\psi_{it}:=\wh q_{it}\wh e_{it}$ for
$(i,t)\in\mathcal I_b$, where $\wh q_{it}$ stacks the $(i,t)$ entries of the
$\wh q_{j,b}$ of Section~\ref{sec:cf-oracle}. Write
$\widehat R_i:=\sum_{t=1}^T\widehat\psi_{it}$ and
$\widehat S_t:=\sum_{i=1}^N\widehat\psi_{it}$ for the unit and time sums,
and $w_h=\big(1-\frac{h}{M+1}\big)_+$ for the Bartlett kernel with bandwidth
$M$. Following \citet{chen2024fixed}, the CHS estimator with the
cross-fitted score is
\begin{equation}\label{eq:cf-chs-hac}
 \widehat\Omega_{\rm CHS}
 :=\widehat\Omega_A
  +\widehat\Omega_{\rm DK}
  -\widehat\Omega_{\rm NW},
\end{equation}
where
\[
\begin{aligned}
 \widehat\Omega_A
 &:=\frac1{NT}\sum_{i=1}^N\widehat R_i\widehat R_i',
  &&\text{(cluster by individuals)}\\
 \widehat\Omega_{\rm DK}
 &:=\frac1{NT}\sum_{t=1}^{T}\sum_{s=1}^{T}w_{|t-s|}\,
   \widehat S_t\widehat S_s',
  &&\text{(Driscoll--Kraay)}\\
 \widehat\Omega_{\rm NW}
 &:=\frac1{NT}\sum_{i=1}^N\sum_{t=1}^{T}\sum_{s=1}^{T}w_{|t-s|}\,
   \widehat\psi_{it}\widehat\psi_{is}',
  &&\text{(within-unit Newey--West)},
\end{aligned}
\]
and the sandwich is
$\widehat V_{\rm CHS}
 :=\widehat Q^{-1}\widehat\Omega_{\rm CHS}\widehat Q^{-1}$.

\begin{assumption}[HAC regularity]\label{ass:cf-hac}
\begin{enumerate}[(i)]
\item The conditional variance of the oracle score satisfies
\[
 \Omega_{NT}
 :=
 \frac1{NT}\operatorname{Var}\left(
 \sum_{i=1}^N\sum_{t=1}^T\psi_{it}
 \,\middle|\,X,\mathcal C
 \right)
 \to_p\Omega,
\]
where $\Omega$ is the variance in
Assumption~\ref{ass:cf-score}(ii).

\item For some fixed $p>4$, the oracle transforms satisfy
\begin{equation}\label{eq:cf-chs-score-energy}
 \frac1{NT}
 \sum_b\sum_{j=1}^d\sum_{(i,t)\in\mathcal I_b}
 \left\{
 \left|(A_bX_j^{(b)}B_b)_{it}\right|^p
 +
 \left|(A_b'A_bX_j^{(b)}B_b)_{it}\right|^p
 \right\}
 =O_p(1).
\end{equation}

\item The deterministic bandwidth satisfies $M\to\infty$,
$M/T\to0$, and
\begin{equation}\label{eq:cf-chs-feasible-rates}
 \{(M+1)N+T\}
 \{(N+T)^{2/p}+(N\vee T)\varsigma_{NT}^2\}
 \left(\varsigma_{NT}^4+\frac1{NT}\right)
 =o_p(1).
\end{equation}
\end{enumerate}
\end{assumption}

Condition~\eqref{eq:cf-chs-feasible-rates} makes the feasible score
replacement negligible. When $N\asymp T$ and $\varsigma_{NT}=O_p(T^{-1/2})$
it holds if $M=o(T^{1-2/p})$.

\begin{theorem}[Consistency of the CHS variance estimator]\label{thm:cf-chs}
Under the conditions of Theorem~\ref{thm:cf} and
Assumption~\ref{ass:cf-hac},
\[
 \widehat\Omega_{\rm CHS}\to_p\Omega,\qquad
 \widehat V_{\rm CHS}
 \to_p Q^{-1}\Omega Q^{-1}.
\]
\end{theorem}

\subsection{Proof of Theorem~\ref{thm:cf-chs}}\label{sec:cf-chs-proof}

Write $\E_c$, $\operatorname{Var}_c$ and $\operatorname{Cov}_c$ for moments
conditional on $(X,\mathcal C)$, $\Omega^0_{\rm CHS}$ for
\eqref{eq:cf-chs-hac} at the oracle scores $\psi_{it}$, and
$R_i:=\sum_t\psi_{it}$ and $S_t:=\sum_i\psi_{it}$. The oracle frames are
restricted to $\mathcal N_a$ and $\mathcal M_k$, so we write $A_a$ and $B_k$
for $A_{(a,k)}$ and $B_{(a,k)}$. Put
$\mathbb B:=\operatorname{diag}(B_1,\dots,B_K)$ and, within $\mathcal N_a$,
$q_j:=A_aX_j\mathbb B$, $\breve q_j:=A_a'q_j$ and $e^0:=A_aE\mathbb B$, so
that $\psi_{it}=q_{it}e^0_{it}$. Let
$\Gamma_l:=\operatorname{Cov}_c((\varepsilon_{lt})_{t\le T})$ and
$\gamma_l:=\mathbb B'\Gamma_l\mathbb B$. By the covariance inequality for
mixing sequences, Assumption~\ref{ass:cf-reg}(i) gives
$C_\Gamma:=\max_l\sup_t\sum_s|\Gamma_l(t,s)|=\Op(1)$. Schur's test then
gives $\norm{\Gamma_l}_{\rm op},\norm{\gamma_l}_{\rm op}\le C_\Gamma$, and
$\norm{\Sigma_E}_{\rm op}\le C_\Gamma$ for the conditional covariance
$\Sigma_E$ of the stacked errors, as $\norm{\mathbb B}_{\rm op}\le1$ and the
units are conditionally independent. Finally $p>4$ in
\eqref{eq:cf-chs-score-energy} gives
\begin{equation}\label{eq:cf-chs-fourth-energy}
 \sum_{i,t}\{\|q_{it}\|^4+\|\breve q_{it}\|^4\}=O_p(NT).
\end{equation}

\begin{proof}[Proof of Theorem~\ref{thm:cf-chs}]
Decompose
\begin{equation*}
 \widehat\Omega_{\rm CHS}-\Omega
 = (\widehat\Omega_{\rm CHS}-\Omega^0_{\rm CHS})
  +(\Omega^0_{\rm CHS}-\E_c[\Omega^0_{\rm CHS}])-(\Omega_{NT}-\E_c[\Omega^0_{\rm CHS}])
  +(\Omega_{NT}-\Omega).
\end{equation*}
The four terms are $o_p(1)$ by Lemmas~\ref{lem:cf-chs-feasible},
\ref{lem:cf-chs-fluctuation} and~\ref{lem:cf-chs-bias} and
Assumption~\ref{ass:cf-hac}(i), in that order. With
$\widehat Q\to_pQ\succ0$ from Lemma~\ref{lem:cf-gram}, continuity of
inversion gives the claim for $\widehat V_{\rm CHS}$.
\end{proof}

\begin{lemma}[Oracle CHS truncation bias]\label{lem:cf-chs-bias}
Put $\mathfrak B_{NT}:=\Omega_{NT}-\E_c[\Omega^0_{\rm CHS}]$ and
$\vartheta_M:=\max_l\sup_t\sum_s
(1-w_{|t-s|})|\Gamma_l(t,s)|$.
Under the conditions of Theorem~\ref{thm:cf} and Assumption~\ref{ass:cf-hac},
\[
 \|\mathfrak B_{NT}\|
 =O_p\!\left(\varsigma_{NT}+N^{-1/2}+
 \vartheta_M+\sqrt{(M+1)/T}\right)=o_p(1).
\]
\end{lemma}

\begin{proof}
Bias identity. In the Driscoll--Kraay term, the $i=i'$ terms of
$S_tS_s'=\sum_{i,i'}\psi_{it}\psi_{i's}'$ form the Newey--West term, so
\begin{equation}\label{eq:cf-chs-oracle-rewrite}
 \Omega^0_{\rm CHS}
 =\frac1{NT}\left\{\sum_iR_iR_i'
 +\sum_{t,s}w_{|t-s|}\sum_{i\ne i'}
        \psi_{it}\psi_{i's}'\right\}.
\end{equation}
As the oracle scores are conditionally centred,
$\Omega_{NT}=(NT)^{-1}\sum_{i,i'}\sum_{t,s}\E_c[\psi_{it}\psi_{i's}']$, and
$\E_c[R_iR_i']=\sum_{t,s}\E_c[\psi_{it}\psi_{is}']$ cancels its $i=i'$ part,
so
\begin{equation}\label{eq:cf-chs-bias-cross}
 \mathfrak B_{NT}
 =\frac1{NT}\sum_{t,s}(1-w_{|t-s|})
     \sum_{i\ne i'}\E_c[\psi_{it}\psi_{i's}'].
\end{equation}
As the error rows are conditionally independent,
$\E_c[e^0_{it}e^0_{i's}]=\sum_{l\in\mathcal N_a}(A_a)_{il}(A_a)_{i'l}\gamma_l(t,s)$
for $i,i'\in\mathcal N_a$, and the covariance is zero across halves. Writing
the sum over $i\ne i'$ as the double sum less its diagonal and using
$\sum_i(A_a)_{il}q_{it}=\breve q_{lt}$ gives
\begin{equation}\label{eq:cf-chs-oracle-bias}
 \mathfrak B_{NT}
 =\frac1{NT}\sum_{a=1}^2\sum_{l\in\mathcal N_a}\sum_{t,s}
 (1-w_{|t-s|})\gamma_l(t,s)
 \left\{\breve q_{lt}\breve q_{ls}'
    -\sum_{i\in\mathcal N_a}(A_a)_{il}^{2}q_{it}q_{is}'\right\}.
\end{equation}
By dominated convergence $\vartheta_M=o_p(1)$, as $M\to\infty$ and the
conditional covariance tails are summable. We bound
\eqref{eq:cf-chs-oracle-bias} in each of the two cases of
\eqref{eq:cf-commute}.

Unit-side bound. Fix $a$ and two score coordinates, and treat the rows
$q_i,\breve q_l$ as time-indexed row vectors. Let $W=(w_{|t-s|})_{t,s}$ and
$\Xi_l=(\mathbf1\mathbf1'-W)\odot\gamma_l$, with $\odot$ the entrywise
product. As the Bartlett matrix $W$ is positive semidefinite with unit
diagonal, the Schur-multiplier inequality gives
\begin{equation}\label{eq:cf-chs-Kbound}
 \|\Xi_l\|_{\rm op}
 \le\|\gamma_l\|_{\rm op}
    +\|W\odot\gamma_l\|_{\rm op}
 \le2C_\Gamma.
\end{equation}
Set $\bar A_a=I-A_a$. The oracle frames have fixed rank, so
$\operatorname{rank}(\bar A_a)=O(1)$, $\|\bar A_a\|_{\rm op}\le2$
and $\|\bar A_a\|_F=O(1)$. As $A_a=M_{\Lambda_a}M_{U_{R,a}}$,
$\breve q=A_a'q=M_{U_{R,a}}q$, so with
$\kappa_{A,a}=\|[P_{U_{R,a}},P_{\Lambda_a}]\|_{\rm op}$,
\begin{equation}\label{eq:cf-chs-unit-comm}
 \|\breve q_j-q_j\|_F
 =\|P_{U_{R,a}}M_{\Lambda_a}M_{U_{R,a}}
                   X_j\mathbb B\|_F
 \le\kappa_{A,a}\|X_j\|_F.
\end{equation}
For this half, \eqref{eq:cf-chs-oracle-bias} rearranges as
\[
 NT\,\mathfrak B_{NT,a}=\sum_l(\breve q_l\Xi_l\breve q_l'-q_l\Xi_lq_l')
 +\sum_{i,l}\{\mathbf1(i=l)-(A_a)_{il}^2\}q_i\Xi_lq_i'.
\]
By \eqref{eq:cf-chs-Kbound}, Cauchy--Schwarz and
\eqref{eq:cf-chs-unit-comm}, the first sum is $O_p(NT\kappa_{A,a})$. In the
second, $A_{il}=\mathbf1(i=l)-(\bar A_a)_{il}$ gives
$|\mathbf1(i=l)-A_{il}^2|\le2\mathbf1(i=l)|(\bar A_a)_{ii}|+(\bar A_a)_{il}^2$,
so the sum is at most
$2C_\Gamma\sum_i\{2|(\bar A_a)_{ii}|+\sum_l(\bar A_a)_{il}^2\}\|q_i\|^2$ in
absolute value. By the norm bounds on $\bar A_a$, $\sum_i(\bar A_a)_{ii}^2$ and
$\sum_i(\sum_l(\bar A_a)_{il}^2)^2$ are $O(1)$, and
$\sum_i\|q_i\|^4\le T\sum_{i,t}\|q_{it}\|^4=O_p(NT^2)$ by
\eqref{eq:cf-chs-fourth-energy}, so Cauchy--Schwarz bounds the second sum by
$O_p(T\sqrt N)$. Dividing by $NT$,
\begin{equation}\label{eq:cf-chs-bias-unit}
 \|\mathfrak B_{NT,a}\|
       =O_p(\kappa_{A,a}+N^{-1/2}).
\end{equation}

Time-side bound. Put $\bar{\mathbb B}=I-\mathbb B$, of fixed rank with
$\|\bar{\mathbb B}\|_F=O(1)$, and $\kappa_{B,a}:=\max_k\kappa_{B,(a,k)}$
with $\kappa_{B,(a,k)}:=\|[P_{V_{R,k}},P_{F_k}]\|_{\rm op}$. As
$B_k-B_kB_k'=-M_{V_{R,k}}P_{F_k}P_{V_{R,k}}$ has norm at most
$\kappa_{B,(a,k)}$,
$\|q_j\bar{\mathbb B}'\|_F=\|A_aX_j(\mathbb B-\mathbb B\mathbb B')\|_F\le\kappa_{B,a}\|X_j\|_F$,
and $\|\breve q_j\bar{\mathbb B}'\|_F\le\|q_j\bar{\mathbb B}'\|_F$. For row
vectors $x,y$, let $\mathsf H_{xy}=\operatorname{diag}(x)W\operatorname{diag}(y)$,
so that
\[
 \sum_{t,s}(1-w_{|t-s|})\gamma_l(t,s)x_ty_s
 =x\mathbb B'\Gamma_l\mathbb B y'
       -\operatorname{tr}(\mathbb B \mathsf H_{xy}
                               \mathbb B'\Gamma_l).
\]
After adding and subtracting $x\Gamma_ly'$ and
$\operatorname{tr}(\mathsf H_{xy}\Gamma_l)$, Schur's test bounds the
untransformed term by $\vartheta_M\|x\|\|y\|$. Expanding
$\mathbb B=I-\bar{\mathbb B}$ bounds the difference in the quadratic forms by
$C_\Gamma\{\|x\bar{\mathbb B}'\|\|y\|+\|x\|\|y\bar{\mathbb B}'\|
+\|x\bar{\mathbb B}'\|\|y\bar{\mathbb B}'\|\}$ and the trace difference by
$C C_\Gamma\|\bar{\mathbb B}\|_F\|\mathsf H_{xy}\|_F$, where
$\|\mathsf H_{xy}\|_F^2\le(2M+1)\sum_t(x_t^4+y_t^4)$ as the Bartlett weights
vanish for $|t-s|>M$. Apply these bounds in \eqref{eq:cf-chs-oracle-bias}
with $(x,y)=(\breve q_{l,j},\breve q_{l,k})$, and with
$(x,y)=(q_{i,j},q_{i,k})$ weighted by $(A_a)_{il}^2$. As
$\sum_l(A_a)_{il}^2\le1$ and $\sum_i(A_a)_{il}^2\le1$, Cauchy--Schwarz and
\eqref{eq:cf-chs-fourth-energy} show that, after division by $NT$, the
untransformed and commutator terms contribute $O_p(\vartheta_M)$ and
$O_p(\kappa_{B,a})$, and the finite-rank terms at most
\[
 C\sqrt{M+1}\Big\{N\sum_{i,t}\big(\|q_{it}\|^4+\|\breve q_{it}\|^4\big)\Big\}^{1/2}\Big/(NT)
 =O_p\big(\sqrt{(M+1)/T}\big).
\]
Hence
\begin{equation}\label{eq:cf-chs-bias-time}
 \|\mathfrak B_{NT,a}\|
 =O_p\!\left(\vartheta_M+\kappa_{B,a}
                     +\sqrt{\frac{M+1}{T}}\right).
\end{equation}
As $\kappa_{A,a}$ is common to the time blocks,
\[
 \min\{\kappa_{A,a},\max_k\kappa_{B,(a,k)}\}
 =\max_k\min\{\kappa_{A,a},\kappa_{B,(a,k)}\},
\]
which is $O_p(\varsigma_{NT})$ by
Assumption~\ref{ass:cf-exact-spike}(iii). The smaller of
\eqref{eq:cf-chs-bias-unit} and \eqref{eq:cf-chs-bias-time} for each half
gives the stated rate, which is $\op(1)$ as $\varsigma_{NT},\vartheta_M=\op(1)$
and $M/T\to0$.
\end{proof}

\begin{lemma}[Oracle CHS sampling error]\label{lem:cf-chs-fluctuation}
Under the conditions of Theorem~\ref{thm:cf} and Assumption~\ref{ass:cf-hac},
for every score-coordinate pair $(j,k)$,
\[
\operatorname{Var}_c([\Omega^0_{\rm CHS}]_{jk})
 =O_p\!\left(N^{-1}+(M+1)/T\right)=o_p(1),
\]
and consequently
$\Omega^0_{\rm CHS}-\E_c[\Omega^0_{\rm CHS}]=o_p(1)$.
\end{lemma}

\begin{proof}
Each entry of $\Omega^0_{\rm CHS}$ is a quadratic form in
$\mathsf E=\operatorname{vec}(E)$. Let $\mathcal T$ be the block-diagonal
linear map that sends $E$ to its two oracle-transformed halves, so
$\|\mathcal T\|_{\rm op}\le1$ as every $A_a$ and $B_k$ is a contraction. Let
the symmetric cell-indexed matrix $\mathbf K$ have
$\mathbf K_{(i,t),(i',s)}=1$ if $i=i'$ and $w_{|t-s|}$ otherwise, and let
$D_j$ be diagonal with entries $q_{it,j}$. By
\eqref{eq:cf-chs-oracle-rewrite},
$[\Omega^0_{\rm CHS}]_{jk}=\mathsf E'\mathcal Q_{jk}\mathsf E$ with the
symmetric matrix
$\mathcal Q_{jk}=(2NT)^{-1}\mathcal T'(D_j\mathbf K D_k+D_k\mathbf K D_j)\mathcal T$,
and
\begin{align*}
 (NT)^2\|\mathcal Q_{jk}\|_F^2
 &\le C\|D_j\mathbf K D_k\|_F^2\\
 &\le C\left\{\sum_i\left(\sum_t\|q_{it}\|^2\right)^2
   +\sum_{t,s}w_{|t-s|}^2
      \sum_i\|q_{it}\|^2\sum_{i'}\|q_{i's}\|^2\right\}\\
 &\le C\{T+(2M+1)N\}\sum_{i,t}\|q_{it}\|^4,
\end{align*}
by Cauchy--Schwarz over the dates of a unit and over the units at a date,
and $\sum_s w_{|t-s|}^2\le2M+1$. With \eqref{eq:cf-chs-fourth-energy},
\begin{equation}\label{eq:cf-chs-quad-F}
 \|\mathcal Q_{jk}\|_F^2
 =O_p\!\left(\frac1N+\frac{M+1}{T}\right).
\end{equation}
The mixing and moment conditions of Assumption~\ref{ass:cf-reg}(i) give
\[
 C_4:=\max_l\sup_{t_1,t_2}\sum_{t_3,t_4}
 |\operatorname{cum}_c(\varepsilon_{lt_1},\varepsilon_{lt_2},\varepsilon_{lt_3},\varepsilon_{lt_4})|
 =O_p(1).
\]
The conditional variance of $\mathsf E'\mathcal Q_{jk}\mathsf E$ is
$2\operatorname{tr}(\mathcal Q_{jk}\Sigma_E\mathcal Q_{jk}\Sigma_E)$ plus a
fourth-cumulant term. By conditional independence across units the
cumulants vanish unless all four errors share a unit, so, as
$\norm{\Sigma_E}_{\rm op}\le C_\Gamma$, the variance is at most
$(2C_\Gamma^2+C_4)\|\mathcal Q_{jk}\|_F^2$. With
\eqref{eq:cf-chs-quad-F} and the conditional Chebyshev inequality this
proves the lemma.
\end{proof}

\begin{lemma}[Feasible CHS score replacement]\label{lem:cf-chs-feasible}
Under the conditions of Theorem~\ref{thm:cf} and Assumption~\ref{ass:cf-hac},
\[
\widehat\Omega_{\rm CHS}-\Omega^0_{\rm CHS}=o_p(1).
\]
\end{lemma}

\begin{proof}
Let $\Delta\psi_{it}=\widehat\psi_{it}-\psi_{it}$,
$\Delta_t=N^{-1/2}\sum_i\Delta\psi_{it}$,
$\Delta R_i=\sum_t\Delta\psi_{it}$, and
\begin{equation}\label{eq:cf-chs-replacement-errors}
 r_{NT}=\frac1T\sum_t\|\Delta_t\|^2
    +\frac1{NT}\sum_{i,t}\|\Delta\psi_{it}\|^2,
 \qquad
 u_{NT}=\frac1{NT}\sum_i\|\Delta R_i\|^2.
\end{equation}
On block $b$, put $\delta_\beta=\widehat\beta^{\rm cf}-\beta_0$ and
$h_b=\widehat{\mathcal R}_b[G^{(b)}-\sum_\ell X_\ell^{(b)}\delta_{\beta,\ell}]$.
By the definitions of the two scores,
\[
 \widehat\psi_{j,b}-\psi_{j,b}
 =\underbrace{(\widehat q_{j,b}-q_{j,b})\odot e_b^0
    +\widehat q_{j,b}\odot
       (\widehat{\mathcal R}_b-\mathcal R_b^0)[E^{(b)}]}_{\text{error-transform part}}
   +\underbrace{\widehat q_{j,b}\odot h_b}_{\text{residual part}}.
\]

Error-transform part. Put
$\wh\varsigma_b:=\norm{\Delta_{A,b}}_{\rm op}+\norm{\Delta_{B,b}}_{\rm op}$,
with $\Delta_{A,b}$ and $\Delta_{B,b}$ as in Lemma~\ref{lem:cf-transform},
so $\wh\varsigma_b=\Op(\varsigma_{NT})$ by
\eqref{eq:cf-operator-consistency}. In coordinate $j$, each entry of this
part, and each date or unit sum of its entries, is a linear form in
$E^{(b)}$ with coefficient array a function of $(X,\mathcal C)$ and the
training data of block $b$. With $j,b$ suppressed and $e_i,e_t$ standard
basis vectors, the array for cell $(i,t)$ is
\[
 C_{it}=(\widehat q-q)_{it}A_b'e_ie_t'B_b'
 +\widehat q_{it}\big\{
     \widehat A_b'e_ie_t'\widehat B_b'
                      -A_b'e_ie_t'B_b'\big\}.
\]
As the transforms are contractions,
$\|C_{it}\|_F\le|(\widehat q-q)_{it}|+2\widehat\varsigma_b|\widehat q_{it}|$.
The array of a date or unit sum has the same form with the date or unit
slices of $\widehat q-q$ and $\widehat q$ in place of their entries, and
obeys the same bound with the slice norms. By
\eqref{eq:cf-D-bound},
$\norm{\wh q_{j,b}-q_{j,b}}_F\le2\wh\varsigma_b\norm{X_j^{(b)}}_F$ and
$\norm{\wh q_{j,b}}_F\le\norm{X_j^{(b)}}_F$, so the squared norms summed over
cells, dates or units are at most $C\wh\varsigma_b^2\norm{X_j^{(b)}}_F^2$.
Lemma~\ref{lem:cf-coupling}(b), Cauchy--Schwarz over the two halves meeting
at a date and the $K$ blocks meeting in a unit, and
$\sum_b\norm{X_j^{(b)}}_F^2=\Op(NT)$ give
\begin{equation}\label{eq:cf-chs-replace-E}
 r_{NT}^{E}+u_{NT}^{E}=O_p(\varsigma_{NT}^2)
\end{equation}
for the contributions of this part.

Residual part. By the proof of Theorem~\ref{thm:cf},
$\wh{\mathcal R}_b[G^{(b)}]=\Delta_{A,b}G^{(b)}\Delta_{B,b}$, so
$\sum_b\|\widehat{\mathcal R}_b[G^{(b)}]\|_F^2=O_p(NT\varsigma_{NT}^4)$.
With $\|\delta_\beta\|=O_p((NT)^{-1/2})$ by Theorem~\ref{thm:cf} and
$\sum_b\|X_j^{(b)}\|_F^2=O_p(NT)$,
\begin{equation}\label{eq:cf-chs-h-bound}
 \mathfrak h_{NT}:=\frac1{NT}\sum_b\|h_b\|_F^2
 =O_p\!\left(\varsigma_{NT}^4+\frac1{NT}\right).
\end{equation}
Let the maximal slice energy $\widehat m_{NT}$ be the maximum, over $j\le d$,
of $\max_t N^{-1}\sum_i\widehat q_{it,j}^2$ and
$\max_i T^{-1}\sum_t\widehat q_{it,j}^2$. H\"older's inequality and
\eqref{eq:cf-chs-score-energy} bound the oracle analogues by $O_p(T^{2/p})$
and $O_p(N^{2/p})$, and
$\sum_{i,t}|\widehat q_{it,j}-q_{it,j}|^2=O_p(NT\varsigma_{NT}^2)$, so
\begin{equation}\label{eq:cf-chs-m-bound}
 \widehat m_{NT}
 =O_p\!\left((N+T)^{2/p}
                     +(N\vee T)\varsigma_{NT}^2\right).
\end{equation}
Cauchy--Schwarz over the units at a date and over the dates of a unit gives
$r_{NT}^{G}\le CN\widehat m_{NT}\mathfrak h_{NT}$ and
$u_{NT}^{G}\le CT\widehat m_{NT}\mathfrak h_{NT}$ for the contributions of
this part. By
\eqref{eq:cf-chs-replace-E}--\eqref{eq:cf-chs-m-bound},
$(M+1)r_{NT}+u_{NT}$ is $\Op(1)$ times the left-hand side of
\eqref{eq:cf-chs-feasible-rates}, which dominates $(M+1)\varsigma_{NT}^2$
as $N(N\vee T)\ge NT$, so Assumption~\ref{ass:cf-hac}(iii) gives
\begin{equation}\label{eq:cf-chs-replacement-rate}
 (M+1)r_{NT}+u_{NT}=o_p(1).
\end{equation}

It remains to bound the three CHS differences. For $v=1,\ldots,T+M$ let
$\mathcal J_v=\{v-M,\ldots,v\}\cap\{1,\ldots,T\}$. Each date lies in $M+1$
windows and dates $t,s$ lie together in $(M+1)w_{|t-s|}$ of them, so, with
$z(\mathcal J_v)=\sum_{t\in\mathcal J_v}z_t$,
\begin{equation}\label{eq:cf-chs-windows}
 \sum_{t,s}w_{|t-s|}z_ty_s'
 =\frac1{M+1}\sum_v z(\mathcal J_v)y(\mathcal J_v)',
 \quad
 \sum_v\|z(\mathcal J_v)\|^2
 \le(M+1)^2\sum_t\|z_t\|^2.
\end{equation}
Put $\phi_t=N^{-1/2}\sum_i\psi_{it}$ and
$\phi(\mathcal J_v)=\sum_{t\in\mathcal J_v}\phi_t$. In coordinate $j$ and
half $\mathcal N_a$, each oracle window sum and unit sum is a linear form
$\fro{A_a'D\mathbb B',E}$, where $D$ holds the weights $q_{it,j}$ of the
summed cells and $\norm{A_a'D\mathbb B'}_F\le\norm{D}_F$. So
$\norm{\Sigma_E}_{\rm op}\le C_\Gamma$ and $\sum_{i,t}\|q_{it}\|^2=O_p(NT)$
give
$\{T(M+1)\}^{-1}\sum_v\|\phi(\mathcal J_v)\|^2$,
$\{NT(M+1)\}^{-1}\sum_{i,v}\|\psi_i(\mathcal J_v)\|^2$ and
$(NT)^{-1}\sum_i\|R_i\|^2$ all $O_p(1)$. By \eqref{eq:cf-chs-windows} and
\eqref{eq:cf-chs-replacement-errors},
$\{T(M+1)\}^{-1}\sum_v\|\Delta(\mathcal J_v)\|^2\le(M+1)r_{NT}$ and
$\{NT(M+1)\}^{-1}\sum_{i,v}\|\Delta\psi_i(\mathcal J_v)\|^2\le(M+1)r_{NT}$.
The identity $aa'-bb'=(a-b)b'+b(a-b)'+(a-b)(a-b)'$ and Cauchy--Schwarz bound
the $\widehat\Omega_A$ difference by $O_p(\sqrt{u_{NT}}+u_{NT})$ and the
$\widehat\Omega_{\rm DK}$ and $\widehat\Omega_{\rm NW}$ differences each by
$O_p(\sqrt{(M+1)r_{NT}}+(M+1)r_{NT})$. All three vanish by
\eqref{eq:cf-chs-replacement-rate}.
\end{proof}

\subsection{Efficiency}\label{sec:cf-efficiency}

\begin{assumption}[Homoskedastic benchmark]
\label{ass:homo}
\begin{enumerate}[(i)]
\item The limits
\[
 \Sigma_\eta:=\lim_{N,T}\frac1{NT}\sum_{i,t}\E[\eta_{it}\eta_{it}'],\enspace
 \wt G:=\lim_{N,T}\frac1{NT}\sum_{i,t}\E[\wt g_{x,it}\wt g_{x,it}'],\enspace
 \mathcal W:=\lim_{N,T}\frac1{NT}\sum_{i,t}\E[w_{it}w_{it}']
\]
exist and are finite, with \(\Sigma_\eta\succ0\) and \(w_{it}\) as in
Section~\ref{sec:eff}, and the Gram matrix of the two-sided residuals on the
population frames satisfies
\[
\begin{aligned}
 \overline Q^{\perp}_{NT}
 &:=\frac1{NT}\sum_b
 \Big\{\big\langle M_{(\Lambda_b,U_{R,b})}X_j^{(b)}M_{(F_b,V_{R,b})},\,
 M_{(\Lambda_b,U_{R,b})}X_\ell^{(b)}M_{(F_b,V_{R,b})}\big\rangle_F\Big\}_{j,\ell}\\
 &\phantom{:}\to_p\Sigma_\eta+\wt G+\mathcal W .
\end{aligned}
\]
\item The errors satisfy
\(\E[\varepsilon_{it}\mid X,\mathcal C]=0\) and
\(\E[\varepsilon_{it}\varepsilon_{i't'}\mid X,\mathcal C]
 =\sigma^2\mathbbm 1\{(i,t)=(i',t')\}\).
\item For \(e\in\{{\rm CCEF},{\rm FK}\}\), with
\(Q_e:=\lim_{N,T}(NT)^{-1}\sum_{i,t}\E[z^e_{it}z^{e\prime}_{it}]\),
\[
 \sqrt{NT}(\wh\beta_e-\beta_0)
 =Q_e^{-1}\frac1{\sqrt{NT}}\sum_{i,t}z^e_{it}\varepsilon_{it}+\op(1),
 \qquad
 \frac1{\sqrt{NT}}\sum_{i,t}z^e_{it}\varepsilon_{it}
 \Rightarrow\mathcal N(0,\Omega_e).
\]
\end{enumerate}
\end{assumption}

Write \(V_e:=Q_e^{-1}\Omega_eQ_e^{-1}\) and
\(V_{\rm CUH}:=Q^{-1}\Omega Q^{-1}\), the variance of Theorem~\ref{thm:cf}.
For FK, part~(iii) is Theorem~2 of
\citet{freeman2026inference}, and for CCEF it follows from
\citet{westerlund2015cross} under \eqref{eqn:CCEXstrongfactor}. The last
clause of part~(i) requires the outcome directions and covariate
projections of each evaluation block to absorb none of the $x$-only block in
the limit.

\begin{proof}[Proof of Proposition~\ref{prop:eff}]
Step 1 (CCEF and FK). As \(\wt g_{x,it}\) is a function of the conditioning
variables in \eqref{eq:x}, the normalisation there gives
\(\E[\eta_{it}\wt g_{x,it}']=0\), so
\eqref{eq:zscores} and
Assumption~\ref{ass:homo}(i) and~(iii) give \(Q_{\rm CCEF}=\Sigma_\eta\) and
\(Q_{\rm FK}=\Sigma_\eta+\wt G\), both positive definite. As the weights
\(z^e_{it}\) are \((X,\mathcal C)\)-measurable,
Assumption~\ref{ass:homo}(ii) gives \(\Omega_e=\sigma^2Q_e\), so
\(V_e=\sigma^2Q_e^{-1}\).

Step 2 (CUH variance). Fix \(c\in\R^d\), put \(Z_b:=\sum_jc_jX_j^{(b)}\) on
each block \(b\) and \(\mathcal S_b[Z]:=A_b'A_bZB_bB_b'\), so that the
oracle weights of Section~\ref{sec:cf-oracle} are
\(z_{j,b}=\mathcal S_b[X_j^{(b)}]\) and
\[
 c'\overline Q_{NT}c=\frac1{NT}\sum_b\langle Z_b,\mathcal S_b[Z_b]\rangle_F,
 \qquad
 \frac1{NT}\sum_{i,t}(c'z_{it})^2=\frac1{NT}\sum_b\norm{\mathcal S_b[Z_b]}_F^2 .
\]
On the Frobenius space \(\mathcal S_b\) acts through the symmetric matrix
\(B_bB_b'\otimes A_b'A_b\), whose spectrum lies in \([0,1]\) as \(A_b\) and
\(B_b\) are contractions. So \(\mathcal S_b^2\preceq\mathcal S_b\) and
\((NT)^{-1}\sum_{i,t}(c'z_{it})^2\le c'\overline Q_{NT}c\). For \(\delta>0\) let \(\mathcal B_{NT}\) be the event
\(c'\overline Q_{NT}c\le c'Qc+\delta\), of probability approaching one by
Assumption~\ref{ass:cf-score}(i), and put
\(\chi_{NT}:=(NT)^{-1/2}\sum_{i,t}c'z_{it}\varepsilon_{it}\). As the oracle
weights and \(\mathcal B_{NT}\) are \((X,\mathcal C)\)-measurable,
Assumption~\ref{ass:homo}(ii) gives
\[
 \E\big[\chi_{NT}^2\mathbbm 1_{\mathcal B_{NT}}\big]
 =\sigma^2\E\Big[\mathbbm 1_{\mathcal B_{NT}}\frac1{NT}\sum_{i,t}(c'z_{it})^2\Big]
 \le\sigma^2\E\big[\mathbbm 1_{\mathcal B_{NT}}\,c'\overline Q_{NT}c\big]
 \le\sigma^2(c'Qc+\delta).
\]
By \eqref{eq:cf-oracle-score-identity} and Assumption~\ref{ass:cf-score}(ii),
\(\chi_{NT}\mathbbm 1_{\mathcal B_{NT}}\Rightarrow\mathcal N(0,c'\Omega c)\),
so Fatou's lemma gives \(c'\Omega c\le\sigma^2(c'Qc+\delta)\). As \(c\) and
\(\delta\) are arbitrary, \(\Omega\preceq\sigma^2Q\) and
\(V_{\rm CUH}=Q^{-1}\Omega Q^{-1}\preceq\sigma^2Q^{-1}\).

Step 3 (lower bound on \(Q\)). Fix a block and let \(\mathcal K_L\) and
\(\mathcal K_R\) be the column spans of \((\Lambda_b,U_{R,b})\) and
\((F_b,V_{R,b})\). The map
\(Z\mapsto M_{(\Lambda_b,U_{R,b})}ZM_{(F_b,V_{R,b})}\) is the
Frobenius-orthogonal projection whose kernel \(\mathcal U\) consists of the
sums of a matrix with columns in \(\mathcal K_L\) and one with rows in
\(\mathcal K_R\). Expanding the four residual makers in \(A_bZB_b\) shows
\(Z-A_bZB_b\in\mathcal U\), so the minimal-norm property of orthogonal
projection gives
\[
 \norm{A_bZB_b}_F\ge\norm{M_{(\Lambda_b,U_{R,b})}ZM_{(F_b,V_{R,b})}}_F .
\]
At \(Z=Z_b\), summed over the blocks, this is
\(c'\overline Q_{NT}c\ge c'\overline Q^{\perp}_{NT}c\), and Assumption~\ref{ass:cf-score}(i) with the last clause of
Assumption~\ref{ass:homo}(i) gives
\(c'Qc\ge c'(\Sigma_\eta+\wt G+\mathcal W)c\). Inverting, with Step 2,
\[
 V_{\rm CUH}\preceq\sigma^2Q^{-1}
 \preceq\sigma^2\big(\Sigma_\eta+\wt G+\mathcal W\big)^{-1}.
\]

Step 4 (ordering). Inverting
\(\Sigma_\eta+\wt G+\mathcal W\succeq\Sigma_\eta+\wt G\succeq\Sigma_\eta\succ0\)
and using Step 1 gives the rest of the ordering, and
\(V_{\rm FK}=V_{\rm CCEF}\) if and only if
\((\Sigma_\eta+\wt G)^{-1}=\Sigma_\eta^{-1}\), that is, \(\wt G=0\).
\end{proof}
\section{Theory for NPCUH}\label{app:npcuh}

\subsection{Cross-fitted NPCUH}\label{sec:npcuh-construction}

On block $b=(a,k)$, every coordinate of the proxy vectors
$\wh\pi_i^{(\alpha)}$ and $\wh\pi_t^{(\gamma)}$ of Section~\ref{sec:smoothing}
is estimated on the training rectangles of Algorithm~\ref{alg:cf}, with
$\mathcal J_j^\alpha$ and $\mathcal J_j^\gamma$ the $L_j$ directions selected
in its step (iii). Let $S^{(\alpha,b)}_{q,h_\alpha}$ and
$S^{(\gamma,b)}_{q,h_\gamma}$ be the smoothers \eqref{eq:np-smoother} built
from these proxies over the $n:=|\mathcal N_a|$ units of $\mathcal N_a$ and the
times of $\mathcal M_k$. Let $W^{(\alpha,b)}_q$ be the converged backfitting
smoother \citep{opsomer2000asymptotic} for the $q$th component of the additive
model in $\wh\pi^{(\alpha)}_{i,1},\dots,\wh\pi^{(\alpha)}_{i,Q_\pi}$, built
from the centred smoothers
$(I-n^{-1}\mathbf 1\mathbf 1')S^{(\alpha,b)}_{q,h_\alpha}$ with $\mathbf 1$ a
vector of ones, and set
\[
 W^{(\alpha,b)}:=n^{-1}\mathbf 1\mathbf 1'+\sum_{q\le Q_\pi}W^{(\alpha,b)}_q.
\]
Define $W^{(\gamma,b)}$ likewise, with $|\mathcal M_k|$ in place of $n$. The
NPCUH residualisation on block $b$ is the backfitted form of
\eqref{eq:np-sweep},
\[
 \wh{\mathcal M}_b[Z]:=(I-W^{(\alpha,b)})Z(I-W^{(\gamma,b)})'.
\]
NPCUH is Algorithm~\ref{alg:cf} with $\wh{\mathcal M}_b$, a fixed linear map
given $\mathcal F_b$, in place of $\wh{\mathcal R}_b$, so
\eqref{eq:cf-linear-score} holds exactly.

The population proxy vectors $\pi_i^{(\alpha)}$ and $\pi_t^{(\gamma)}$
replace each estimated coordinate by its population counterpart
$u_r(\alpha_i)$, $\lambda_{j,r}(\alpha_i)$, $v_r(\gamma_t)$ or
$f_{j,r}(\gamma_t)$.
Assumption~\ref{ass:kernel}(iii) requires the common conditional mean
$\Gamma_Z$ of each $Z\in\{Y,X_1,\dots,X_d\}$ to be additive across pairs of
proxy coordinates,
\begin{equation}\label{eq:pair-additive}
 \Gamma_{Z,it}=\mu_Z+\sum_{q,q'}\phi_{Z,qq'}
 \!\left(\pi_{i,q}^{(\alpha)},\pi_{t,q'}^{(\gamma)}\right),
\end{equation}
with both sums over the $Q_\pi$ retained coordinates.

\subsection{Injectivity of the proxy collection}

Statements are for the unit side, and the time side is symmetric. The
population proxy maps are the outcome eigenfunctions $u_r$ of \eqref{eq:Gsvd}
and the covariate projections $\lambda_{j,r}$ of Section~\ref{sec:weak}. For
truncation orders $(S_1,S_2)$ the \emph{stacked} map is
\[
 P_{S}(\alpha)
 :=\big(\sigma_1u_1(\alpha),\dots,\sigma_{S_1}u_{S_1}(\alpha);\;
 \{\lambda_{j,r}(\alpha)\}_{j\le d,\,r\le S_2}\big)\in\R^{S},
 \qquad S=S_1+dS_2.
\]
Let $\Pi_V$ be the $L^2(\pi_\gamma)$ projection onto the closed span of
$\{v_r\}$, applied to $\gamma$-sections,
$(\Pi_Vg_{x,j})(\alpha,\cdot):=\Pi_V\{g_{x,j}(\alpha,\cdot)\}$. The stacked
separation is
\begin{align*}
 \Delta^{\rm stack}(\alpha_1,\alpha_2)
 :={}&\norm{g(\alpha_1,\cdot)-g(\alpha_2,\cdot)}^2_{L^2(\pi_\gamma)}\\
 &+\sum_{j\le d}
 \norm{\Pi_V\{g_{x,j}(\alpha_1,\cdot)-g_{x,j}(\alpha_2,\cdot)\}}^2_{L^2(\pi_\gamma)},
\end{align*}
and the stacked gradient information is
$M^{\rm stack}_\infty:=M^{g}_\infty+M^{g_x}_\infty$, with $M^{g}_\infty$ the
gradient information matrix of \citet{freeman2026nonparametric} for $g$ and
$M^{g_x}_\infty$ the sum over $j\le d$ of those for the $\Pi_Vg_{x,j}$.

\begin{assumption}[Stacked reduction and immersion]\label{ass:inject}
(i) $\Delta^{\rm stack}(\alpha_1,\alpha_2)>0$ whenever
$\alpha_1\neq\alpha_2$. Types with identical outcome \emph{and} projected
covariate responses are identified, the observational-equivalence reduction
for the stacked kernel $(g,\Pi_Vg_x)$. (ii) Either
$M^{\rm stack}_\infty(\alpha)\succ0$ for every $\alpha\in\Omega_\alpha$
(uniform immersion), or $\pi_\alpha\{\alpha:\det
M^{\rm stack}_\infty(\alpha)=0\}=0$ (almost-everywhere immersion). (iii)
The domain and smoothness conditions of
\citet{freeman2026nonparametric} hold for $g$ and each $\Pi_Vg_{x,j}$.
\end{assumption}

Condition (ii) holds wherever $M^{g}_\infty$ or $M^{g_x}_\infty$ has full
rank, and $M^{g_x}_\infty$ does not depend on the outcome factor strength
$c_{NT}$.

\begin{proposition}[Population injectivity]
\label{prop:inject-pop}
Under Assumption~\ref{ass:inject}(i), (iii) and the uniform immersion in
(ii), there exist finite $(S_1^*,S_2^*)$ such that $P_S$ is injective on
$\Omega_\alpha$ for all $S_1\ge S_1^*$, $S_2\ge S_2^*$. Under the
almost-everywhere immersion, the same holds on every compact subset of
$\Omega_\alpha$ off the degeneracy set, hence off a set of types of
arbitrarily small $\pi_\alpha$-mass.
\end{proposition}

\begin{proof}
The finite-truncation argument of \citet{freeman2026nonparametric} applies to
the stacked objects. Their truncated separation functions are monotone in each
truncation order with continuous limit $\Delta^{\rm stack}$, and the immersion
floor on $M^{\rm stack}_\infty$ gives near-diagonal separation by their
near-diagonal lemma. Dini's theorem makes the convergence uniform away from
the diagonal, where $\Delta^{\rm stack}>0$ by Assumption~\ref{ass:inject}(i),
and their injectivity criterion applies. Their compact exhaustion and blend
corollary, with $M^{\rm stack}_\infty$ in place of $M_\infty$, give the
almost-everywhere and small-mass versions.
\end{proof}

For $(i,t)\in\mathcal I_b$, each retained coordinate estimated out of block
satisfies
\begin{equation}\label{eq:proxy-image}
 \wh\pi^{(\alpha)}_{i,q}
 =\sum_{s\le S}C_{qs}\,P_{S,s}(\alpha_i)+e_{i,q},
\end{equation}
where the weights $C_{qs}$ are realised inner products of the
training-sample singular vectors with the population directions (for instance
$C_{qs}=T^{-1}v_s'\wh v_{r(q)}$ up to scale), and $e_{i,q}$ is a
type-independent contamination from the idiosyncratic components, mean zero
given the types and the training sample. Under weak factors the outcome rows of
$C$ shrink with $c_{NT}$ and the contamination is not negligible, while the
covariate rows stay of order one, so the covariate coordinates carry
injectivity when the outcome coordinates degrade.

\begin{proposition}[Injectivity of generic linear images]
\label{prop:inject-image}
Let $P:\Omega_\alpha\to\R^S$ be an injective $C^1$ immersion on a compact
$\Omega_\alpha$, and let $m\ge2d_\alpha+1$. Then the set of matrices
$C\in\R^{m\times S}$ for which $\alpha\mapsto CP(\alpha)$ fails to be an
injective immersion has Lebesgue measure zero. Consequently, if the
realised weight matrix in \eqref{eq:proxy-image} has a density on
$\R^{m\times S}$ and $R+\sum_jL_j\ge2d_\alpha+1$, the type-measurable part of
the estimated proxy collection is an injective immersion of the types with
probability one.
\end{proposition}

\begin{proof}
By compactness and the immersion property, the closure $\mathcal K$ of the
normalised secants $(P(\alpha_1)-P(\alpha_2))/\norm{P(\alpha_1)-P(\alpha_2)}$,
$\alpha_1\neq\alpha_2$, and of the unit tangent directions of $P$ lies in a
locally Lipschitz image of a set of dimension at most $2d_\alpha$. A matrix $C$
fails the conclusion only if its kernel meets $\mathcal K$. As
$\{C:Cv=0\}$ has dimension $m(S-1)$ for each unit vector $v$, such $C$ form a
set of dimension at most $2d_\alpha+m(S-1)<mS$ when $m\ge2d_\alpha+1$, and
this set is Lebesgue null by the general-position step of the Whitney embedding theorem
\citep{hirsch1976differential}. A density gives a null set probability zero.
\end{proof}

\subsection{Conditions and proxy rate}

\begin{assumption}[Local NPCUH first-step conditions]\label{ass:npcuh}
Let $m_{NT}=N\mw T$.
\begin{enumerate}[(i)]
\item The spectral decay \eqref{eq:decay} holds with $\rho>2$, and the proxy
count is in one of two regimes. In regime (F), $R$ and the $L_j$ are fixed,
$\sigma_1>\cdots>\sigma_{R+1}$, and $R+\sum_jL_j$ is large enough for the
proxy collection to be injective, as Propositions~\ref{prop:inject-pop}
and~\ref{prop:inject-image} give under the uniform immersion of
Assumption~\ref{ass:inject}(ii). In regime (G), $R\to\infty$ with
$R^{2\rho+3}=o(m_{NT})$, so that the eigengaps $\sigma_r^2-\sigma_{r+1}^2$,
$r\le R$, exceed the noise level, and the population eigenfunctions obey
$\sup_\alpha|u_r(\alpha)|\vee\sup_\gamma|v_r(\gamma)|\le Cr^{b}$ for some
$b<\rho-1/2$. Regime (G) serves the almost-everywhere immersion of
Assumption~\ref{ass:inject}(ii).
\item Assumption~\ref{ass:reg}(i) and (iii) hold for every $j\le d$ and
$r,r'\in\mathcal I_R$ with uniformly bounded fourth moments, and the oracle
projection-sampling error of Assumption~\ref{ass:reg}(iv),
$q_{{\rm proj},NT}:=\max_{j\le d,\,r\in\mathcal I_R}
\{\norm{T^{-1}X_jv_r-\lambda_{j,r}}_N+\norm{N^{-1}X_j'u_r-f_{j,r}}_T\}$, is
$\Op(m_{NT}^{-1/2})$. Moreover $\norm{E}_{\rm op}=\Op(\sqrt{N\vee T})$, the
conditional covariance operator of $\operatorname{vec}(E)$ is bounded, and
the block preliminary estimators obey
\begin{align*}
 \max_b\|\widetilde\beta_b-\beta_0\|
 &=\op\big(m_{NT}^{-1/4}\big)&&\text{in regime (F)},\\
 \max_b\|\widetilde\beta_b-\beta_0\|
 &=\op\big(m_{NT}^{-(7\rho+7)/(12\rho)}\big)&&\text{in regime (G)}.
\end{align*}
\item In regime (G), let $\Lambda:=(\sigma_ru_r^{(g)})_{r\in\mathcal I_R}$ and
$F:=(v_r^{(g)})_{r\in\mathcal I_R}$, so that $G_R=\Lambda F'$ is the
leading-$R$ part of $G$ in Section~\ref{sec:smooth}. For some
$\delta_\lambda\in(0,1]$, uniformly over the candidate coordinates,
$N^{-\delta_\lambda}\Lambda'\Lambda\to\Sigma_\Lambda\succ0$ and
$T^{-1}F'F\to\Sigma_F\succ0$, and the smallest candidate ordinary singular
value clears the noise edge.
\item In regime (G), the error--factor products of the factor expansion are
uniformly bounded. With $e_t=E_{\cdot t}$ and $e_i'=E_{i\cdot}$,
\[
 \sup_t\E\|N^{-\delta_\lambda/2}\Lambda'e_t\|^2=O(1),
 \quad
 \sup_i\E\|T^{-1/2}F'e_i\|^2=O(1),
\]
\[
 \max_{t,s}\E\left[
 N^{-1/2}\sum_i\{\varepsilon_{it}\varepsilon_{is}
 -\E(\varepsilon_{it}\varepsilon_{is})\}\right]^2=O(1),
 \quad
 \|\Lambda'EF\|_F=\Op(\sqrt{N^{\delta_\lambda}T}),
\]
and, uniformly in $t$ and $i$,
\[
 (NT)^{-1}\|e_t'EF\|=\Op(m_{NT}^{-2}),\qquad
 (NT)^{-1}\|e_i'E'\Lambda\|=\Op(N^{-\delta_\lambda}).
\]
\end{enumerate}
\end{assumption}

\begin{lemma}[NPCUH proxy rate]\label{lem:npcuh}
Under Assumption~\ref{ass:npcuh}(i)--(iv), adding the fixed number
$\sum_jL_j$ of covariate projections does not change the proxy rate,
\begin{equation}\label{eq:npcuh-proxy-rate}
 \frac1N\sum_i\|\widehat\pi_i^{(\alpha)}-\pi_i^{(\alpha)}\|^2
 \ \vee\
 \frac1T\sum_t\|\widehat\pi_t^{(\gamma)}-\pi_t^{(\gamma)}\|^2
 =\Op\big(\xi_\pi^2\big),
\end{equation}
where $\xi_\pi^2:=\max_b\|\widetilde\beta_b-\beta_0\|^2+m_{NT}^{-1}$ in regime
(F), the constant depending on $R$, and
$\xi_\pi^2:=R^{2\rho+2}m_{NT}^{-1}+R^{4\rho+5}\max_b\|\widetilde\beta_b-\beta_0\|^2$
in regime (G).
\end{lemma}

\begin{proof}
In regime (F) the fixed $\sigma_1>\cdots>\sigma_{R+1}$ give
Assumption~\ref{ass:strength} with $c_{NT}\asymp1$. On each training
rectangle Proposition~\ref{prop:xi}, with its noise edge from
Assumption~\ref{ass:npcuh}(ii) and its perturbation condition from
$\norm{X}_{\rm op}^2/(NT)=\Op(1)$ (Assumption~\ref{ass:reg}(iii)) and the
preliminary rate, gives eigenfunction errors
$\Op(m_{NT}^{-1}+\max_b\norm{\wt\beta_b-\beta_0}^2)$. In regime (G) the factor
expansion of \citet{freeman2026inference} for a decaying spectrum gives
$\xi_\pi^2$. Its first term sums the cost
$\sigma_r^{-2}m_{NT}^{-1}\asymp r^{2\rho+1}m_{NT}^{-1}$ of the $r$th
eigenfunction over $r\le R$, and its second is the plug-in perturbation
$X(\widetilde\beta_b-\beta_0)$ set against the eigengaps of
Assumption~\ref{ass:npcuh}(i). Truncation at $R$ adds no approximation term.
The leading block of the sample singular value decomposition is orthogonal to
its tail, which enters only through the eigengaps, and the eigenfunction bound
of Assumption~\ref{ass:npcuh}(i) makes the sample non-orthogonality of the
population tail dominated by the first term. In
either regime the covariate projections inherit the eigenfunction error plus
$q_{{\rm proj},NT}$ by the argument of Theorem~\ref{thm:proj}, which is
conditional on the singular-vector error and needs no gap, and $q_{{\rm proj},NT}^2=\Op(m_{NT}^{-1})$ is dominated by $\xi_\pi^2$.
\end{proof}

\begin{assumption}[Kernel, design and representation]\label{ass:kernel}
\begin{enumerate}[(i)]
\item The univariate kernel $K$ of \eqref{eq:np-smoother} is bounded,
symmetric and Lipschitz, integrates to one, has finite positive second
moment, and has compact support or exponentially decaying tails.
\item Each population proxy coordinate has compact support and a density
bounded and bounded away from zero, or estimation is trimmed to a fixed
interior set with these properties.
\item For every $Z\in\{Y,X_1,\dots,X_d\}$, $\Gamma_Z$ has the representation
\eqref{eq:pair-additive}, with components twice continuously differentiable in
both arguments and derivatives bounded uniformly in $(q,q')$, on the trimmed
interior set of (ii), taken inside a compact set on which
Proposition~\ref{prop:inject-pop} makes the proxy collection injective, and
inside its $\gamma$-side analogue.
\item The local sample sizes diverge, $Nh_\alpha/\log N\to\infty$ and
$Th_\gamma/\log T\to\infty$.
\item On every block the backfitting algorithms defining $W^{(\alpha,b)}$ and
$W^{(\gamma,b)}$ converge to unique limits \citep[sufficient conditions are
in][]{opsomer2000asymptotic}.
\end{enumerate}
\end{assumption}

Condition (iii) holds when each $g_Z$ is additive across latent coordinates
and, on each side, every latent coordinate has a retained proxy that is an
injective function of it alone.

Write $\varpi:=h_\alpha^2\vee h_\gamma^2$ and $\varrho:=\varpi+\xi_\pi^2$,
with $\xi_\pi$ from Lemma~\ref{lem:npcuh}.

\begin{assumption}[Nonparametric remainders]\label{ass:sweep}
(i) $\|W^{(\alpha,b)}\|_{\rm op}\vee\|W^{(\gamma,b)}\|_{\rm op}=\Op(Q_\pi)$
uniformly in $b$.
(ii) For each $j\le d$ and $\xi_{\Gamma X}:=(NT)^{-1}\sum_b
\fro{\wh{\mathcal M}_b[X_j^{(b)}],\wh{\mathcal M}_b[G^{(b)}]}$,
\begin{align*}
 \xi_{\Gamma X}&=Q_\pi^4\Op(\varrho^2)
 +\tfrac{1}{\sqrt{NT}}\Op\Bigl[Q_\pi^3\xi_\pi^2
 +Q_\pi^2\{(Nh_\alpha)^{-1/2}+(Th_\gamma)^{-1/2}\}
 +Q_\pi^4\Bigl\{\varpi+\tfrac{1}{\sqrt{NTh_\alpha h_\gamma}}\Bigr\}\Bigr]\\
 &\quad+Q_\pi^3\Op(\varrho\,m_{NT}^{-1/2}),
\end{align*}
the term $Q_\pi^3\Op(\varrho\,m_{NT}^{-1/2})$ absent under block-independent
errors.
(iii) Assumptions~\ref{ass:cf-reg}(iii) and \ref{ass:cf-score} hold with the
oracle residualisation built as $\wh{\mathcal M}_b$ from the population
proxies, with limits $Q_{\rm np},\Omega_{\rm np}$, and feasible and oracle
scores agree to $\op(\sqrt{NT})$.
\end{assumption}

\subsection{Asymptotic normality}

\begin{corollary}[Asymptotic normality of NPCUH]
\label{cor:npcuh-normal}
Let $N\asymp T$, the errors be block independent,
Assumptions~\ref{ass:cf-reg}(i)--(ii), \ref{ass:npcuh}, \ref{ass:kernel}
and \ref{ass:sweep} hold, and
$h_\alpha=c_\alpha N^{-\tau_\alpha}$, $h_\gamma=c_\gamma T^{-\tau_\gamma}$.
In regime (F) let $\tau_\alpha,\tau_\gamma\in(\frac14,1)$, and in regime (G)
let $\rho>4$, $R\asymp m_{NT}^{1/(6\rho)}$ and
\[
 \tau_\alpha,\tau_\gamma\in\left(\tfrac14+\tfrac1{6\rho},
 \tfrac{3\rho-2}{3\rho}\right).
\]
Then the cross-fitted NPCUH estimator with the backfitted smoother of
Section~\ref{sec:npcuh-construction} satisfies
$\sqrt{NT}(\widehat\beta^{\rm NPCUH}-\beta_0)\Rightarrow
\mathcal N(0,Q_{\rm np}^{-1}\Omega_{\rm np}Q_{\rm np}^{-1})$.
\end{corollary}

\begin{proof}
Since $W^{(\alpha,b)}$ reproduces constants, $I-W^{(\alpha,b)}$ removes
$\mu_Z$. The unit index enters
$\phi_{Z,qq'}(\wh\pi_{i,q}^{(\alpha)},\wh\pi_{t,q'}^{(\gamma)})$ only through
$\wh\pi_{i,q}^{(\alpha)}$, the proxy on which $W^{(\alpha,b)}_q$ conditions,
so at fixed $t$ and for every $q'$ this term is part of the $q$th component of
an additive model in the unit-side proxies, which backfitting removes up to a
bias $O(h_\alpha^2)$ in sup norm under Assumption~\ref{ass:kernel}. The
$\gamma$ side is symmetric. One component of the unit-side smoother therefore
absorbs a whole row of \eqref{eq:pair-additive} to bias order $\varpi$, and
one component of the time-side smoother a whole column. Lipschitz components
leave a proxy error $\Op(\xi_\pi)$, and Assumption~\ref{ass:sweep}(ii) records
its propagation by Assumption~\ref{ass:sweep}(i) across the $Q_\pi^2$
components.

Write $\underline\tau=\tau_\alpha\wedge\tau_\gamma$,
$\bar\tau=\tau_\alpha\vee\tau_\gamma$ and $Q_\pi\asymp m_{NT}^{\iota}$, with
$\iota=0$ in regime (F) and $\iota=1/(6\rho)$ in regime (G). The bandwidth
terms of Assumption~\ref{ass:sweep}(ii) are
\begin{gather*}
 Q_\pi^4\varpi^2\asymp m_{NT}^{4\iota-4\underline\tau},
 \qquad
 Q_\pi^2(NT)^{-1/2}(Nh_\alpha)^{-1/2}\asymp m_{NT}^{2\iota+(\tau_\alpha-3)/2},\\
 Q_\pi^4(NT)^{-1}(h_\alpha h_\gamma)^{-1/2}=O\big(m_{NT}^{4\iota+\bar\tau-2}\big),
\end{gather*}
and the $\gamma$-side twin of the second, all $\op(m_{NT}^{-1})$ exactly when
$\frac14+\iota<\tau_\alpha,\tau_\gamma<1-4\iota$, the bandwidth range of the
corollary. For the proxy terms, in regime (F) the preliminary rate of
Assumption~\ref{ass:npcuh}(ii) gives
$\xi_\pi^4=\Op(\max_b\|\wt\beta_b-\beta_0\|^4+m_{NT}^{-2})=\op(m_{NT}^{-1})$
and $(NT)^{-1/2}\xi_\pi^2=\op(m_{NT}^{-1})$. In regime (G),
Lemma~\ref{lem:npcuh} with the preliminary rate gives
$\xi_\pi^2=\Op(m_{NT}^{1/(3\rho)-2/3})+\op(m_{NT}^{-1/2-1/(3\rho)})$. Hence
$Q_\pi^4\xi_\pi^4=\Op(m_{NT}^{4/(3\rho)-4/3})+\op(m_{NT}^{-1})$, which is
$\op(m_{NT}^{-1})$ exactly when $\rho>4$, and
$Q_\pi^3(NT)^{-1/2}\xi_\pi^2=\Op(m_{NT}^{(2\rho+5)/(6\rho)-2})+\op(m_{NT}^{-1})$,
which is $\op(m_{NT}^{-1})$ when $\rho>5/4$. The remaining terms of
Assumption~\ref{ass:sweep}(ii) are of smaller order, in regime (G) once
$\rho>4$.

In either regime $N\asymp T$ then gives
$\xi_{\Gamma X}=\op(m_{NT}^{-1})=\op((NT)^{-1/2})$, so the $G$ term of
\eqref{eq:cf-linear-score}, with $\wh{\mathcal M}_b$ in place of
$\wh{\mathcal R}_b$, is $\op(\sqrt{NT})$. The operators are
$\mathcal F_b$-measurable, so Lemma~\ref{lem:cf-coupling} and
Assumption~\ref{ass:sweep}(iii) handle the $E$ term, and Slutsky's theorem
concludes.
\end{proof}

\section{Implementation details}\label{app:sim-details}

The cross-fit splits the units into first and second halves, after a random
permutation in the simulations and in data order in the applications. In both, $\wt\beta_b$ is
the CUH fixed point on $\mathcal N_{-a}\times\mathcal M_k^{\rm qc}$, iterated
from the least-squares estimator of \citet{Bai2009}. The CHS variance
estimator of Section~\ref{sec:cf-variance} uses the Bartlett bandwidth
$M=\max\{1,\lfloor4(T/100)^{2/9}\rfloor\}$.

\begin{remark}[Finite-sample degrees of freedom]\label{rem:cf-dfc}
The standard error of the $j$th coefficient is
$\{[\widehat V_{\rm CHS}]_{jj}/(NT)\}^{1/2}$, computed on the estimator's own
residualised arrays, times $\{NT/\sum_bd_b\}^{1/2}$, a per-block version of
the degrees-of-freedom correction of \citet{freeman2026inference}. Let
$\mathrm{edf}^{\alpha}_b:=\sum_{q\le Q_\pi}\{\tr(S^{(\alpha,b)}_{q,h_\alpha})-1\}$,
the effective degrees of freedom
\citep[Section~7.6]{hastie2009introduction} of the component smoothers of
\eqref{eq:np-sweep} net of the constant each reproduces, and define
$\mathrm{edf}^{\gamma}_b$ likewise. Following \citet{freeman2026inference},
who adapt \citet{rueda2013degrees}, let
\[
 e^\alpha_b:=\min\{1.5\,\mathrm{edf}^\alpha_b,|\mathcal N_a|/2\},
 \qquad
 e^\gamma_b:=\min\{1.5\,\mathrm{edf}^\gamma_b,|\mathcal M_k|/2\}.
\]
Then
\[
 d_b=\begin{cases}
 (|\mathcal N_a|-R-\operatorname{rank}\check\Lambda_b)
 (|\mathcal M_k|-R-\operatorname{rank}\check F_b) & \text{for CUH},\\[2pt]
 (|\mathcal N_a|-e^\alpha_b)(|\mathcal M_k|-e^\gamma_b) & \text{for NPCUH},\\[2pt]
 (|\mathcal N_a|-R-\operatorname{rank}\check\Lambda_b-e^\alpha_b)
 (|\mathcal M_k|-R-\operatorname{rank}\check F_b-e^\gamma_b) & \text{for BCUH},
 \end{cases}
\]
with each factor floored at two. The CUH case is the analogue of the
factor-model correction of \citet{freeman2023linear}. The other estimators use
the dimension their purge leaves, $(N-1)(T-1)$ for two-way fixed effects,
$(N-R)(T-R)$ for the factor estimator and $(N-R-R_X)(T-R-R_X)$ for CCEF with
$R_X$ covariate principal components, the last two with the same floor, and
for the grouped estimator the sum over its four quadrants of the units less the
unit groups times the periods less the time groups.
Theorem~\ref{thm:cf-chs} uses $NT$ in place of $\sum_bd_b$, which is
asymptotically equivalent for CUH with the $L_j$ fixed, since
$\sum_bd_b/(NT)\to1$ as the $2K$ blocks partition the panel, $R$ and $K$ are
fixed and
$\operatorname{rank}\check\Lambda_b,\operatorname{rank}\check F_b\le\sum_jL_j$.
\end{remark}

In Tables~\ref{tab:mc} and~\ref{tab:mc-re} and Figure~\ref{fig:power} the
factor and CUH standard errors are each multiplied by their own
$\wh c_{\rm boot}:=\max\{1,\mathrm{avg}(\mathrm{se}_{\rm boot}/\mathrm{se}_{\rm CHS})\}$,
and the BCUH standard errors by the CUH multiplier. The wild bootstrap behind
$\mathrm{se}_{\rm boot}$ redraws $Y^*=X\wh\beta+\wh G+\omega\odot\wh e$, with
i.i.d.\ Rademacher $\omega$, $\wh G$ the rank-$R$ approximation of
$Y-X\wh\beta$ and $\wh e$ its remainder, and re-runs the estimator on each
draw \citep{goncalvesperron2014}. The multiplier is estimated once per design
cell, from its first $10$ replications with $40$ draws each, and applied to all
its replications, as in \citet{giacominipolitiswhite2013}. In a design variant
at $(N,T)=(300,31)$ with $w=0$, unscaled $\eta_{it}$, and $\eta_{it}$ and
$\varepsilon_{it}$ autoregressive with coefficient $0.6$ over time and $0.3$
across units, the CUH ratio $\mathrm{se}_{\rm boot}/\mathrm{se}_{\rm CHS}$
averages $0.74$, so the floor at one binds and CUH covers at $0.93$, against
$0.85$ with a bootstrap standard error in every replication.

\begin{remark}[Bandwidth in practice]\label{rem:npcuh-bandwidth}
Corollary~\ref{cor:npcuh-normal} fixes the bandwidth rates but not the
constants, so the implementation estimates the constant and imposes the rate,
as in \citet{linton1995second} and \citet{calonico2020optimal}.
Let $\wh h_{\rm cv}$ be the leave-one-out cross-validated bandwidth of the
smoother, computed once per regression on the full two-way demeaned outcome
panel, with its leading $R+\sum_jL_j$ singular vectors as proxies and one
bandwidth for both sides, as in \citet{freeman2026inference}. NPCUH uses
$h_\alpha=h_\gamma=\wh h_{\rm cv}\,m_{NT}^{-1/5}$ in every block, and BCUH
uses $\wh h_{\rm cv}$. Cross-validation targets prediction risk, whose optimal
bandwidth rate $m_{NT}^{-1/5}$ for a second-order kernel
\citep{hardle1985optimal,li2004cross} lies outside the interval of
Corollary~\ref{cor:npcuh-normal}, so its smoothing bias is not negligible at
the $\sqrt{NT}$ rate. The factor $m_{NT}^{-1/5}$ moves the rate to
$m_{NT}^{-2/5}$, inside the interval in both regimes.
\end{remark}

\end{document}